\documentclass[sigconf]{acmart}
\AtBeginDocument{%
  }

\copyrightyear{2026}
\acmYear{2026}
\setcopyright{none}
\acmConference[CHI '26]{Proceedings of the 2026 CHI Conference on Human Factors in Computing Systems}{April 13--17, 2026}{Barcelona, Spain}
\acmBooktitle{Proceedings of the 2026 CHI Conference on Human Factors in Computing Systems (CHI '26), April 13--17, 2026, Barcelona, Spain}

\author{Dominique Geissler}
\email{d.geissler@lmu.de}
\affiliation{%
  \institution{LMU Munich \& Munich Center for Machine Learning (MCML)}
  \city{Munich}
  \country{Germany}
}

\author{Claire Robertson}
\affiliation{%
  \institution{New York University}
  \city{New York}
  \country{U.S.}}
  \affiliation{%
  \institution{Colby College}
  \city{Waterville}
  \country{U.S.}}

\author{Stefan Feuerriegel}
\affiliation{%
  \institution{LMU Munich \& Munich Center for Machine Learning (MCML)}
  \city{Munich}
  \country{Germany}
}

\usepackage{longtable}
\usepackage{setspace}
\usepackage{dirtytalk}
\usepackage{lscape}
\usepackage{subcaption}
\usepackage{pdflscape}
\usepackage{enumitem}
\usepackage{tabularx}
\usepackage{amsmath}
\usepackage{hyperref}
\usepackage{xurl}

\usepackage{eso-pic}   
\usepackage{xcolor}    
\definecolor{DarkGreen}{RGB}{34,139,34}
\newcommand{\firstpagebottomnotice}{%
  \AddToShipoutPictureFG*{%
    \put(0,0){%
      \parbox[b][30mm][c]{\paperwidth}{%
        \centering\bfseries\Huge\color{red}
        This is a preprint. Accepted at ACM CHI Conference 2026. 
      }%
    }%
  }%
}

\begin{document}
\firstpagebottomnotice

\title{Designing Effective Digital Literacy Interventions for Boosting Deepfake Discernment}

\begin{abstract}
Deepfakes images can erode trust in institutions and compromise election outcomes, as people often struggle to discern real images from deepfake images. Improving digital literacy can help address these challenges. Here, we compare the efficacy of five digital literacy interventions to boost people's ability to discern deepfakes: (1)~textual guidance on common indicators of deepfakes; (2)~visual demonstrations of these indicators; (3)~a gamified exercise for identifying deepfakes; (4)~implicit learning through repeated exposure and feedback; and (5)~explanations of how deepfakes are generated with the help of AI. We conducted an experiment with $N=1,200$ participants from the United States to test the immediate and long-term effectiveness of our interventions. Our results show that our lightweight, easy-to-understand interventions can boost {deepfake image discernment} by up to 13 percentage points while maintaining trust in real images. 
\end{abstract}


\maketitle
\section{Introduction}

{AI-generated or manipulated materials that falsely appear authentic, commonly called ``deepfakes'' \cite{EU.2025}, are nowadays common in online environments and are increasingly being misused to spread disinformation \cite{Goldstein.2023}. While deepfakes encompass various modalities including video, audio, and images, AI-generated images have become particularly prevalent due to the rapid proliferation of accessible AI-image generators that require no technical expertise \cite{Feuerriegel.2023, Nightingale.2022b}. Critically, deepfake images may be a uniquely potent form of mis- and disinformation, as people struggle more to discern deepfake images than deepfake audios or videos \cite{Diel.2024}, are more likely to believe mis- and disinformation when a false claim is presented with an image \cite{Hameleers.2020}, and are both understudied and relatively unregulated \cite{Dan.2021}.}  

{Hence, AI-generated images, which we refer to as \emph{deepfake images} throughout this paper, are emerging as a uniquely troubling form of misinformation requiring effective, empirically tested interventions to combat. These deepfake images pose significant challenges for user trust and information authenticity \cite{Spitale.2023, Zhou.2023} as people may come to believe that any image online could be fabricated \cite{Vaccari.2020, Hancock.2021}. Notable examples include a deepfake image showing the arrest of U.S. President Donald Trump \cite{BBC.2023.Trump} and a deepfake image of Pope Francis in a fashionable white puffer jacket \cite{Forbes.2023}, both of which fooled millions online. These cases illustrate the potential of AI-generated disinformation to manipulate public opinion, influence electoral outcomes, or provoke social instability \cite{Dobber.2019, Ternovski.2022}, and thereby highlight the immediate need for effective strategies to mitigate the susceptibility of people towards deepfake images.} 

Prior human-computer interaction (HCI) research emphasizes that people often struggle to discern authentic from AI-generated content, largely due to cognitive biases and reliance on flawed heuristic processing when evaluating digital information \cite{Jakesch.2023, Kobis.2021, Kasra.2018, Frank.2024, Nightingale.2022, Groh.2024}. {However, being able to discern AI-generated content accurately is an important step in decreasing people's vulnerability to the effects of such content \cite{Groh.2022, Ruffin.2023}.} In this context, digital literacy, i.e., the ability to analyze and evaluate online information, has emerged within HCI research as a promising strategy to build resilience against disinformation \cite{Kozyreva.2024, Costello.2024, Arechar.2023, Maertens.2025, Huang.2024}, {and where discernment accuracy is a common indicator for intervention efficacy \cite{Huang.2024}}. Previous studies have demonstrated that digital literacy interventions effectively improve users' abilities to discern textual misinformation \cite{Guess.2020, Moore.2022, Panizza.2022}, but interventions targeting visual misinformation, specifically {deepfake images}, remain scarce.

{Some digital literacy interventions educate people on verification strategies, such as reverse image search, to help detect deepfake images \cite{Aprin.2022, Qian.2022}. These tools can be effective, but are cognitively demanding, require time investment, and must be repeated for all potentially suspect images. Others have adapted media literacy tips -- originally developed for textual misinformation -- to the context of deepfake images \cite{Guo.2025, Chen.2025}  (see \cite{Kozyreva.2024} for an overview of media-literacy tips). However, they face a critical challenge that interventions improve deepfake image discernment while inadvertently increasing skepticism toward authentic content, a phenomenon also observed in fact-checking interventions for textual misinformation \cite{Hoes.2024, Altay.2024}. }

{Furthermore, research on digital literacy interventions for combating misinformation suggests that the \emph{medium} of literacy interventions may matter as much as their \emph{content} \cite{Hu.2023, Dan.2025}. Hence, how information is delivered, e.g., through text or visuals, could influence intervention effectiveness. However, one-off examinations of potential interventions often vary in both content and medium between experiments, making direct comparisons nearly impossible.}

{Building on these insights, we conduct a \emph{large-scale comparative evaluation} of five lightweight, accessible digital literacy interventions in varying formats: textual instruction, visual examples, gamified training, feedback-based learning, and knowledge about AI image generation. We independently vary modality while holding content and stimuli selection constant in a counterbalanced design. Critically, we assess each intervention's effect on both deepfake image and real image discernment to ensure our interventions improve discernment without eroding trust in authentic content. Moreover, research on misinformation interventions has mostly measured immediate effects and neglected long-term effectiveness \cite{Dan.2021}. We aim to fill this gap by re-testing the effectiveness of our interventions at a two-week follow-up.}

\begin{figure*}
    \centering
    \includegraphics[width=0.9\linewidth]{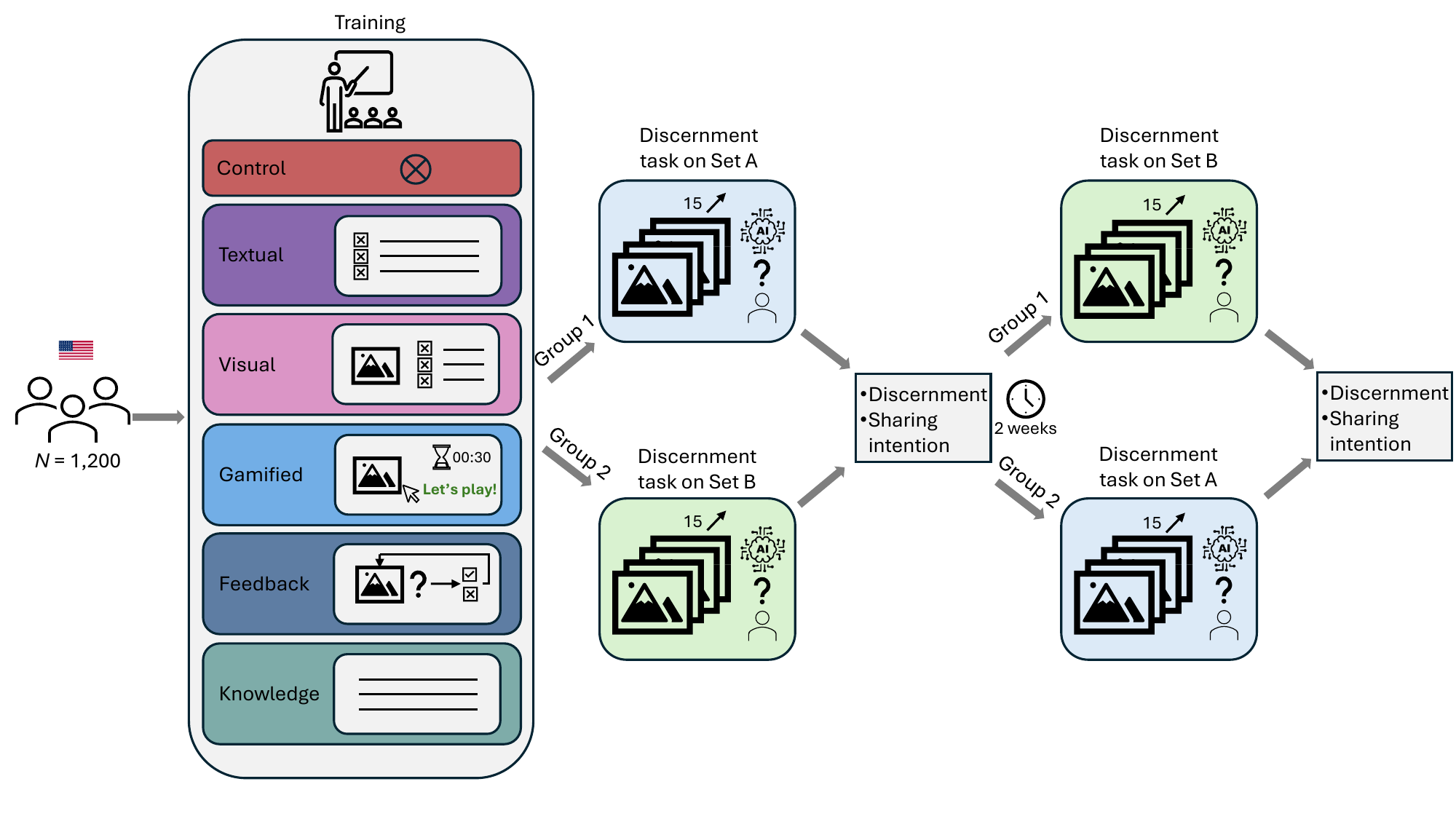}
    \caption{\textbf{Experimental flow.} We recruited participants from the U.S. for our online experiment. Participants were randomly assigned to either a control condition or one of five intervention conditions: (1)~a textual description of ways to spot errors commonly found in {deepfake images} (\emph{Textual}); (2)~the same description with illustrative {deepfake images} (\emph{Visual}); (3)~a gamified exercise to spot errors in {deepfake images} (\emph{Gamified}); (4)~a discernment task with feedback across multiple rounds (\emph{Feedback}); and (5)~an explanation of how {deepfake images} are generated (\emph{Knowledge}). Afterward, participants completed an image discernment task following a counter-balanced design (i.e., Group~1 sees image set A; Group~2 sees image set~B). In the image discernment task, participants were asked to (a)~classify 15 images as real or fake (using a four-point Likert scale) and (b)~indicate whether they would share them on social media (with answer options ``Yes'', ``No'', ``Don't know''). From these responses, we assess how the interventions affect the discernment, i.e., the accuracy in discerning real images and {deepfake images}, and the sharing intention for each participant. After two weeks, participants repeated the image discernment task with the alternate image set (i.e., Group~1 now sees image set~B; and Group~2 sees image set~A) to assess the long-term effects of the interventions.}
    \label{fig:overview_graphic}
    \Description[short explanation]{The figure shows a study-flow diagram for an online experiment and the counterbalanced test procedure. On the left, one cohort is depicted (N = 1,200 American people). Participants are randomly assigned to one of six training conditions stacked vertically: Control; Textual (a short checklist of deepfake cues); Visual (the same cues paired with an example image); Gamified (a 30-second “let’s play!” spot-the-error micro-task); Feedback (classify images and receive immediate feedback); and Knowledge (a brief explanatory text). From there, arrows split participants into Group 1 and Group 2. Group 1 first completes a discernment task on Image Set A; Group 2 first completes the same task on Image Set B. Each task contains 15 images and records two outcomes: (1) discernment (classifying real vs. fake) and (2) sharing intention. After two weeks, the groups switch image sets and repeat the task, yielding a counterbalanced follow-up with the same two outcomes.}
\end{figure*}

Figure~\ref{fig:overview_graphic} provides an overview of the experimental flow of our study. We ground our approach in AI literacy theory \cite{Long.2020}, specifically targeting the competency of recognizing AI-generated content while adhering to principles of low barrier to entry and minimal cognitive load \cite{Sweller.2011}. We test three specific hypotheses in our research:

\vspace{0.2cm}
\noindent
\emph{\textbf{H1:} People who receive a digital literacy intervention will have better discernment of {deepfake images}. }
\vspace{0.2cm}

\noindent
\emph{\textbf{H2:} People who receive a digital literacy intervention will not be more skeptical of real images. }
\vspace{0.2cm}

\noindent
\emph{\textbf{H3:} People who receive a digital literacy intervention will have better long-term discernment of {deepfake images}.}
\vspace{0.2cm}

Our results show that well-designed digital literacy interventions can boost {deepfake image discernment} while maintaining discernment of real images. Beyond testing these hypotheses, we also measure participants' intentions to share the images on social media to assess whether improved discernment translates to more responsible sharing behavior, which is a key mechanism through which misinformation spreads online \cite{Vosoughi.2018, Geissler.2023, Subbian.2017}.

Overall, we make the following contributions: \textbf{(1)}~We conduct a large-scale experiment with $N=1,200$ participants to systematically evaluate which digital literacy interventions are most effective for boosting {deepfake image discernment} and hence provide robust quantitative evidence. \textbf{(2)}~We explicitly test whether literacy interventions can improve {deepfake image discernment} without undermining trust in authentic content. \textbf{(3)}~We evaluate the persistence of intervention effects over a two-week period, providing insights into the long-term sustainability of digital literacy approaches.

\section{Related work}

\subsection{Design principles for effective interventions}
{Understanding why people struggle to detect deepfake images requires examining the cognitive limitations underlying visual perception and decision-making. Research in cognitive science demonstrates that people can fail to notice even conspicuous changes or anomalies when their attention is not properly directed \cite{Simons.1999, Rensink.1997, Neisser.1975}. Notably, even when images contain physically impossible inconsistencies (such as shadows inconsistent with the lighting source), people frequently fail to notice these detectable signs of manipulation \cite{Nightingale.2019}. However, attention alone is insufficient for successful detection; it also depends on knowing what to attend to and how to analyze it \cite{Nightingale.2022c}. Yet, previous work shows that only 2\% of participants spontaneously used a systematic attention strategy, such as carefully zooming in, to detect manipulations in an image, even though such strategies improved detection \cite{Nightingale.2022c}.}

Designing interventions for {deepfake image discernment} requires aligning with established principles from cognitive psychology and AI literacy research. {We first outline general design principles that guided our intervention development and then derive specific guidelines for the concrete formats we test.}

{From cognitive psychology, we draw on the source monitoring framework (SMF), which describes how people evaluate source information \cite{Johnson.1993}. SMF distinguishes between heuristic processes, which are fast, automatic, and rely on familiarity or qualitative characteristics (such as whether something ``feels real''), and systematic processes, which involve slower, more deliberate evaluation of specific features and logical reasoning about origins \cite{Johnson.1993, Mitchell.2003}. Empirical work shows that, in the context of deepfake images, people predominantly rely on heuristic cues and only rarely use systematic inspection strategies, even though such strategies could improve discernment \cite{Jin.2025, Groh.2022, Kasra.2018, Nightingale.2022c}. Building on the SMF framework, a first design principle is therefore to support a shift from heuristic toward more systematic source monitoring by making diagnostic visual cues explicit (e.g., lighting inconsistencies, anatomical implausibilities) and by encouraging inspection of these cues.}

Moreover, Cognitive Load Theory emphasizes that excessive mental effort can hinder learning and decision-making \cite{Sweller.2011}. {A second design principle is thus to manage the cognitive load of our interventions so that participants can focus on the core learning objective.} Instructions should be concise, present only a small number of deepfake errors at a time, and technical details should be avoided. This should help participants process and retain the key cues for deepfake image discernment without becoming overwhelmed.

Design guidance from the AI literacy framework further underscores the importance of accessibility. Long et al. \cite{Long.2020} highlight ``low barrier to entry'' as a core consideration and recommend interventions that reduce reliance on mathematical or computer science knowledge, connect to people's prior experiences, and minimize feelings of inadequacy. {A third design principle is therefore to ensure that materials are easy to understand and feel approachable.} Interventions can operationalize this principle by using plain language, familiar visual examples, and avoiding domain-specific jargon. This should allow participants to engage with the material without needing specialized background knowledge.

In the present work, we apply these principles by designing interventions that (a)~{teach visual cues for systematic evaluation}, (b)~focus on a small number of easily recognizable visual cues to keep cognitive demands low, and (c)~use easy-to-understand instructions that require no prior technical knowledge. {Research on digital literacy and inoculation interventions suggests that the \emph{medium} of a literacy intervention can shape its effectiveness \cite{Hu.2023, Dan.2025}. Given these general principles, we designed five intervention formats that operationalize them in different ways, drawing on distinct theoretical and empirical foundations:}

{\textbf{Textual intervention.} Text-based instructions provide explicit knowledge about the diagnostic features of deepfake images. Research on instructional design shows that clear verbal explanations can effectively convey conceptual knowledge and direct attention to relevant features \cite{Mayer.2002, Moreno.1999}. Textual formats allow for precise specification of what to look for (e.g., ``check for inconsistent lighting'') while minimizing extraneous cognitive load through concise presentation. Thus, our textual intervention supports systematic source monitoring by directing attention to the inaccuracies of deepfakes with no other modality as a distractor. This allows us to examine whether directed attention alone is sufficient to improve deepfake discernment. }

{\textbf{Visual intervention.} Visual examples complement verbal instruction by providing concrete instantiations of visual cues. Dual coding theory suggests that combining verbal and visual information enhances learning and retention \cite{Paivio.1990, Clark.2023}. By showing side-by-side textual instructions and visual example images, this format helps participants build mental models of what deepfake errors actually look like, making abstract concepts concrete. }

{\textbf{Gamified intervention.} Gamification incorporates game-like elements such as points, feedback, and interactive challenges, to make the learning process more engaging and motivate users to apply the learned skills beyond the intervention \cite{Hamari.2014, Kiryakova.2014}. By framing deepfake image discernment as a challenge with immediate scoring, this format may promote repeated practice and deeper processing of diagnostic features while maintaining user engagement. }

{\textbf{Feedback intervention.} Providing immediate corrective feedback after each judgment allows participants to calibrate their discernment strategies in real-time. Learning theory demonstrates that feedback is essential for skill acquisition, as it helps learners identify errors and adjust their approach \cite{Hattie.2007}. In the context of deepfake image discernment, we test feedback to see whether feedback and reinforcement alone, without any directed attention tips, is sufficient to boost discernment. Feedback can reinforce correct identification strategies and discourage reliance on misleading heuristics, and thereby support the shift from heuristic to systematic source monitoring. }

{\textbf{Knowledge intervention.} Explaining how AI image generation works provides participants with a conceptual understanding of the technical processes that generate deepfake images. Prior work suggests that knowledge about a topic can aid discernment performance \cite{Chein.2024}. This approach builds mental models of how AI image generation works, which potentially supports more principled evaluation strategies that transfer across different types of deepfake images.}

{By comparing these five formats, we aim to identify which approaches most effectively promote systematic source monitoring for deepfake image discernment while adhering to principles of low cognitive load and accessibility. }

\subsection{The challenge of {deepfake image} detection: {technical and cognitive barriers}}
The spread of {deepfake images} presents unique challenges for democratic discourse and public trust \cite{Feuerriegel.2023, Kupferschmidt.2024, DiResta.2024, Dobber.2019, CenterforCounteringDigitalHate.2024, Ternovski.2022, Goldstein.2023}. {Deepfake images} can impersonate individuals, fabricate news events, and generate persuasive disinformation at scale \cite{Goldstein.2024, Spitale.2023}. At the same time, when people think content is AI-generated, it can reduce their trust in all content, including authentic content \cite{Jakesch.2019}.

One proposed solution is the use of automated detection tools, for example, in the form of automated flagging of {deepfake images} on social media or through browser extensions. However, current technological approaches to deepfake detection face significant limitations {\cite{Wikum.2024} and struggle with generalization across contexts \cite{Groh.2022}}. On the one hand, automated detection tools are often inaccurate and many {deepfake images} lack watermarks or sufficient metadata for reliable detection \cite{Goldstein.2023}. On the other hand, automated detection tools rapidly become obsolete as generation methods evolve quickly \cite{Mirsky.2021, Shiohara.2022, Korshunov.2018, Almars.2021}. {Moreover, flagging images as AI-generated may not be sufficient to reduce their persuasive effects, as is the case for AI-generated texts \cite{Gallegos.2025}, fake news \cite{Ecker.2010}, and manipulated photographs \cite{Nash.2018}.} Another challenge arises when automated flagging of {deepfake images} is not available. Then, users would need to manually upload images for analysis, which is impractical, time-consuming, and cognitively demanding, limiting scalability \cite{Zhong.2024}. Thus, a more scalable solution requires designing interventions for {deepfake image discernment} that align with established principles from cognitive psychology and AI literacy research.

When people attempt to identify AI-generated content independently of automated detection tools, research demonstrates they frequently fail due to their reliance on flawed heuristics, such as associating first-person pronouns or familiar topics with human-created content \cite{Jakesch.2023, Kobis.2021}. {Similar shortcut-based reasoning occurs for deepfake images: people assess visual content based on perceived realism {\cite{Jin.2025}}, affective responses {\cite{Groh.2022}}, or prior beliefs, all of which are vulnerable to manipulation \cite{Clark.2021, Nightingale.2017}.} These heuristics also extend beyond the image itself. For example, people judge the credibility of {deepfake images} they encounter online based on the source, platform, or accompanying text, while visual cues such as inconsistencies in lighting and shadows are rarely considered \cite{Kasra.2018, Nightingale.2022c}. When people do evaluate visual content directly, they currently struggle to distinguish state-of-the-art {deepfake images} from authentic content, with detection rates often at chance levels \cite{Frank.2024, Nightingale.2022, Groh.2024, Jin.2025, Nightingale.2019}.

{Given that automated detection tools are unreliable, that people predominantly rely on flawed heuristics, and that they rarely employ systematic visual analysis strategies, there is a growing need for user-focused interventions that teach how to scrutinize the visual features of  {deepfake images}. Such interventions should redirect people away from purely heuristic judgments toward deliberate and systematic image analysis and equip them with transferable detection strategies that work across different contexts and evolving AI-generation methods.}

\subsection{Promises and pitfalls of digital literacy interventions}
A growing body of work in HCI and related fields has explored how digital literacy can help people resist misinformation online (see \cite{Kozyreva.2024} for a general overview of intervention tools against misinformation). {For textual misinformation, one common intervention are accuracy prompts.} Accuracy nudges prompt users to consider accuracy before sharing and have reduced belief in and sharing of false headlines \cite{Pennycook.2020b, Pennycook.2022, Epstein.2021, Arechar.2023}. {However, the effectiveness of these types of nudges is contested \cite{Roozenbeek.2021}, and their underlying mechanism, i.e., redirecting attention to accuracy, requires that people possess the ability to assess accuracy when prompted. For deepfake images, this prerequisite often does not hold as people struggle to reliably discern deepfake images \cite{Jakesch.2023, Kobis.2021, Kasra.2018, Frank.2024, Nightingale.2022, Groh.2024}.}

{Another approach involves inoculation-based interventions that prebunk misinformation by exposing people to weakened forms of misleading content or explaining manipulation techniques. While much inoculation research targets textual misinformation \cite{Badrinathan.2021, Basol.2020, Maertens.2025, Apuke.2023, Ali.2023}, some emerging approaches also target visual misinformation. For example, indicator-based cues on short-form videos aided adolescents in detecting misinformation tactics on TikTok \cite{Hartwig.2024}, AI art was used to counter misinformation \cite{Walker.2023}, and explaining how an image is manipulated reduced people's belief in the intended message \cite{Ruffin.2023}. However, inoculation interventions typically focus on understanding tactics and strategies used in misinformation rather than building perceptual detection skills. For deepfake images, understanding manipulation tactics alone may be insufficient to detect deepfake images.} 

{In this regard, other works focused on teaching verification strategies and lateral reading to improve misinformation discernment for text \cite{Panizza.2022, Lee.2024}. These are comparable to verification strategies proposed for visual misinformation, such as reverse image search \cite{Aprin.2022, Qian.2022}. However, these face similar challenges in that they are time-consuming, cognitively demanding, and impractical for everyday use. }

{One possible, light-weight solution to teach deepfake image discernment are digital literacy tips, which have been effective in teaching strategies to identify textual misinformation \cite{Guess.2020, Moore.2022, Arechar.2023, Ali.2023}. Recently, they have been adapted to the context of deepfake images, for example, early work taught people to recognize specific visual artifacts in synthetic faces \cite{Nightingale.2022}. While this approach led to reliable improvements in discernment accuracy, overall performance remained only slightly above chance. Newer work tested the effectiveness of general misinformation tips against deepfake-specific tips. While research on deepfake videos found that both types of literacy tips reduced credibility \cite{Hwang.2021}, prior work on deepfake images found that deepfake image-specific tips improved discernment more \cite{Guo.2025}. At the same time, the work has revealed a critical challenge: the intervention inadvertently also reduced discernment of real images. This challenge was also observed by recent work, which tested digital literacy tips for synthetic face detection including descriptions and example images \cite{Chen.2025}.}

{One possible influencing factor might be the format of the literacy intervention. Prior work tested both textual and video-based fact-checks and found that the latter was more effective in correcting misperceptions \cite{Dan.2025}. Other work on combating misinformation found that game-based inoculation strategies were more effective in reducing misinformation credibility than graphic-based inoculation strategies, while neither produced counter-effects on accurate information \cite{Hu.2023}. Yet, we lack a comparative evaluation of different formats of digital literacy interventions designed to boost deepfake image discernment. At the same time, the challenge of maintaining trust in real images remains.  }

\section{Methods}

\subsection{Overview}

We employed a between-subjects experiment with a counter-balanced design to analyze the effect of digital literacy interventions in boosting individuals' ability to discern between real images and {deepfake images}. Participants were randomly assigned to either the control condition or one of five intervention conditions: (1)~\emph{Textual}, (2)~\emph{Visual}, (3)~\emph{Gamified}, (4)~\emph{Feedback}, or (5)~\emph{Knowledge}, as described below. {Participants in the intervention groups received the corresponding training, while those in the control group received no training. Following this, all participants were given a discernment task in which they classified images as real or deepfakes.} After a period of two weeks, we conducted a follow-up where the image discernment task was repeated to analyze the long-term effects of our interventions.

\subsection{Sampling plan}

We recruited a sample of $N=1,200$ participants from the U.S. via the online platform Prolific (\url{prolific.com}). Participants can take part in the experiment if they reside in the U.S. and are over 18 years old. Participants received compensation equivalent to 12\$/hour, paid through Prolific. {We collected their unique Prolific IDs to invite them to the follow-up conducted two weeks later.} Data was collected through an online survey hosted on Qualtrics. 

To determine the sample size, we performed a power analysis for a one-tailed $t$-test for equally allocated groups. {We chose an $\alpha$ error probability of $0.05$ for this and aimed for a power of $0.8$ \cite{Haile.2023}. However, since we compared multiple intervention conditions to the same control condition, we adjusted the alpha with a Bonferroni correction and divided by 5 (5 comparisons for 5 intervention conditions), which gave us $\alpha_{\text{adj}} = 0.01$. } For the effect size, we found that, on average, the effect size of media literacy training studies is $d = 0.37$ \cite{Jeong.2012}, which would be a medium effect according to common interpretations of Cohen's $d$ \cite{Cohen.1988}. This gave a minimum sample size of {$N = 155$} per condition for equally allocated groups. Following best practices to account for potential attrition over the two-week interval \cite{Haile.2023}, {we increased the minimum sample size by 30\% to 201 participants per condition and rounded this to a final sample size $N=200$ per condition,} which is in line with previous work on literacy interventions \cite{Jeong.2012}. This gave us a final sample size of $N=1,200$. With this sample size, we are able to detect effect sizes above {$d=0.32$} with a power of $0.8$ according to a post-hoc sensitivity analysis.

\subsection{Interventions}

We tested five digital literacy interventions aimed at boosting people's ability to discern between real images and {deepfake images}, as detailed in the following. We built on the design principles laid out in the related work to guide our design choices.

(1)~\emph{Textual}: We presented participants with short textual descriptions of typical errors found in {deepfake images\footnote{{We used the simplified term "fake images" during the study, but participants received the more elaborate definition of ‘fake image (AI-generated)’ prior to the task.}}} (see Table~\ref{tab:intervention_textual} for the content of the intervention). Prior research has categorized such errors found in {deepfake images} into five different types: anatomical implausibilities  (e.g., distorted hands or unlikely alignment of teeth),  stylistic artifacts (e.g., unnaturally smooth skin or hyper-real details), functional implausibilities (e.g., dysfunctional objects or incomprehensible text), violations of physics (e.g., missing shadows of objects and people or inconsistent reflections), and sociocultural implausibilities (e.g., clothing or symbols that are culturally out of place) \cite{Kamali.2024}. We taught these to participants using so-called media literacy tips, i.e., tips that give people a list of strategies for identifying {deepfake images} \cite{Kozyreva.2024}. We focused on a small number of easy-to-understand, hands-on media literacy tips to minimize cognitive load \cite{Sweller.2011} and lower barriers to entry for participants \cite{Long.2020}. We selected a textual format for this intervention as it can be easily deployed across various channels, such as private messages, social media posts, or news articles. By presenting only text without images, the intervention may further avoid biasing participants toward specific visual cues and instead foster a more general understanding of how to detect the characteristics of {deepfake images}. To improve attention, each error was presented on a separate page, where the order was randomized to mitigate potential order effects.

\begin{table}[h]
    \centering
    \begin{spacing}{0.9}
    \begin{tabular}{p{\linewidth}}
    \toprule
        \emph{\textbf{Hint}: Fake images often have anatomical errors. For example, people often have missing, extra, or merged fingers, nonexistent fingernails, and unlikely hand proportions.} \\  
        \midrule
        \emph{\textbf{Hint}: Fake images often look a bit too perfect. People may have glossy, shiny skin, windswept hair, and they may look like a photo in a magazine or a scene from a movie. Sometimes, parts of an image have a different level of detail or vibrancy compared to the rest.} \\  
        \midrule
       \emph{\textbf{Hint}: Fake images often have functional errors. Image generators do not have a structural understanding of how objects in the world work and interact with each other. This can cause composition errors, dysfunctional objects, and atypical designs. Looking closely may also reveal distorted and unresolved details or incomprehensible text and logos.} \\ 
        \midrule
        \emph{\textbf{Hint}: Fake images can produce subtle artifacts that are inconsistent with the laws of physics. These include inaccurate shadows, reflections of alternative realities, and depth and perspective issues.} \\
        \midrule
        \emph{\textbf{Hint}: Fake images may have sociocultural errors. For example, the images may show scenarios that are inappropriate, unlikely to be seen in the real world, violate subtle rules specific to different cultures, or are historically inaccurate.} \\
    \bottomrule
    \end{tabular}
    \end{spacing}
    \vspace{0.2cm}
    \caption{The \emph{Textual} intervention presents common errors found in {deepfake images} through short, textual descriptions. Each hint explains one specific error.}
    \label{tab:intervention_textual}
\end{table}

(2)~\emph{Visual}: We presented participants with the same textual descriptions of common errors as in the \emph{Textual} intervention, but each description was accompanied by an illustrative deepfake image that exemplified the respective error. {The images were chosen to illustrate specific diagnostic cues rather than to exhaustively cover photorealistic styles.} For the sociocultural implausibilities, we used the deepfake image of Pope Francis in a puffer jacket. For the remaining errors, we generated example images using DALL-E 3 \cite{OpenAI.2023} via the ChatGPT website to show examples of the errors (see Table~\ref{tab:example_images} {in the Appendix} for the prompts and sources). Each image was supplemented with a short explanatory sentence that guides participants to the relevant error in the image (see Table~\ref{tab:intervention_visual} for the exact intervention content). We chose to show the textual information together with an example image to make the error more tangible and easier to recognize. Compared to the \emph{Textual} intervention, the visual format helps illustrate how these errors manifest in practice, which may support learning. This format is also well-suited for deployment in visually-driven environments, such as social media posts or digital ads. Again, each error was shown on a separate page in randomized order.

\begin{table}[h]
    \centering
    \begin{tabular}{p{0.3\linewidth}p{0.7\linewidth}}
    \toprule
        \textbf{Image} & \textbf{Error description} \\
    \midrule
        \raisebox{-0.9\height}{\includegraphics[width=0.15\textwidth]{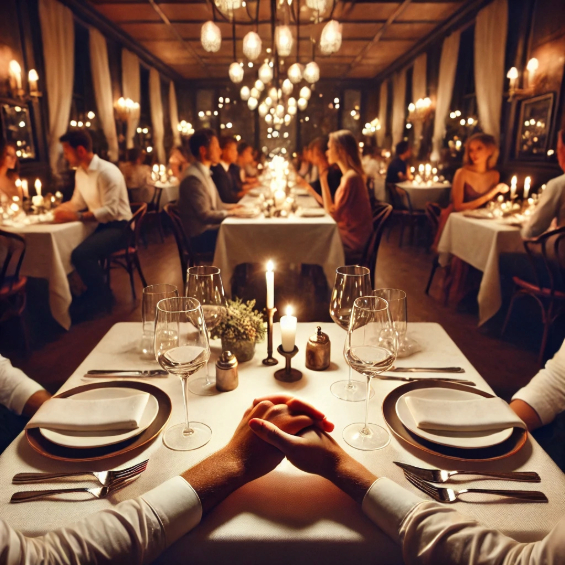}} & \emph{\textbf{Hint}: Fake images often have anatomical errors. For example, people often have missing, extra, or merged fingers, nonexistent fingernails, and unlikely hand proportions. \newline In the image below, you can see that the hands are merged together and that the hand on the right should have been a left hand, so the thumb should not be visible.} \\
    \midrule
       \raisebox{-0.9\height}{\includegraphics[width=0.15\textwidth]{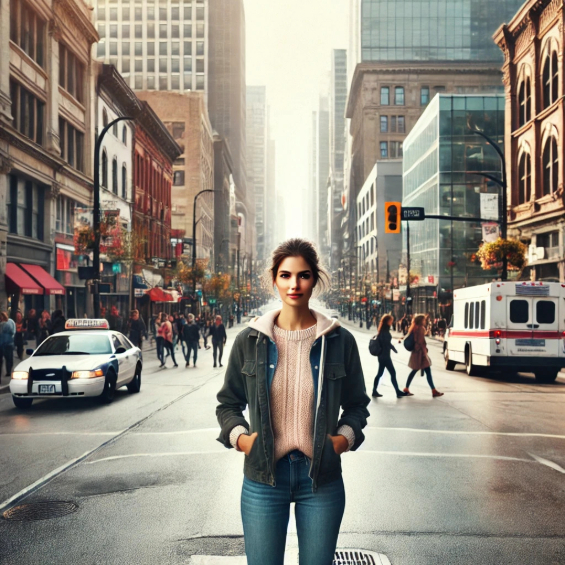}} &  \emph{\textbf{Hint}: Fake images images often look a bit too perfect. People may have glossy, shiny skin, windswept hair, and they may look like a photo in a magazine or a scene from a movie. Sometimes, parts of an image have a different level of detail or vibrancy compared to the rest. \newline In the image below, you can see that the face of the woman is not realistic.} \\
    \midrule
       \raisebox{-0.9\height}{\includegraphics[width=0.15\textwidth]{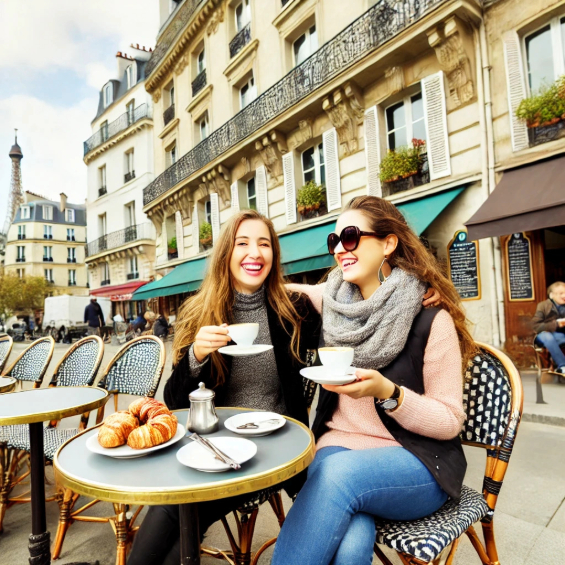}} &  \emph{\textbf{Hint}: Fake images often have functional errors. Image generators do not have a structural understanding of how objects in the world work and interact with each other. This can cause composition errors, dysfunctional objects, and atypical designs. Looking closely may also reveal distorted and unresolved details or incomprehensible text and logos. \newline In the image below, the saucer seems to be floating underneath the cup on the left with no hand holding it. }\\
    \midrule
       \raisebox{-0.9\height}{\includegraphics[width=0.15\textwidth]{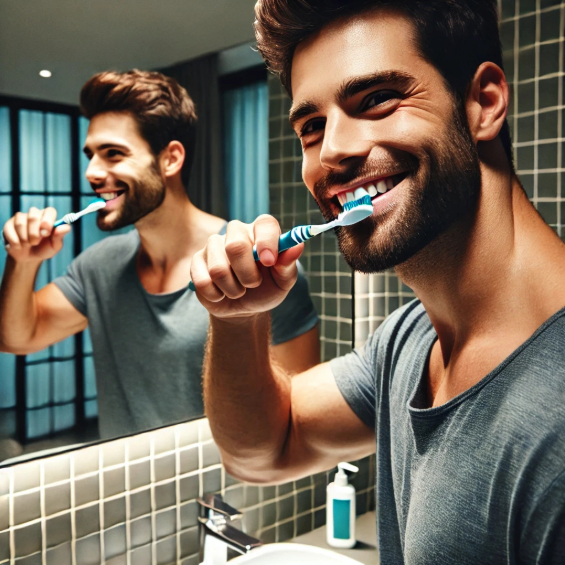}} &  \emph{\textbf{Hint}: Fake images can produce subtle artifacts that are inconsistent with the laws of physics. These include inaccurate shadows, reflections of alternative realities, and depth and perspective issues. \newline In the image below, the reflection of the man doesn't match the man in front of the mirror. The wrong arm is holding the toothbrush and the toothbrush should be facing a different direction.} \\
    \midrule
       \raisebox{-0.9\height}{\includegraphics[width=0.15\textwidth]{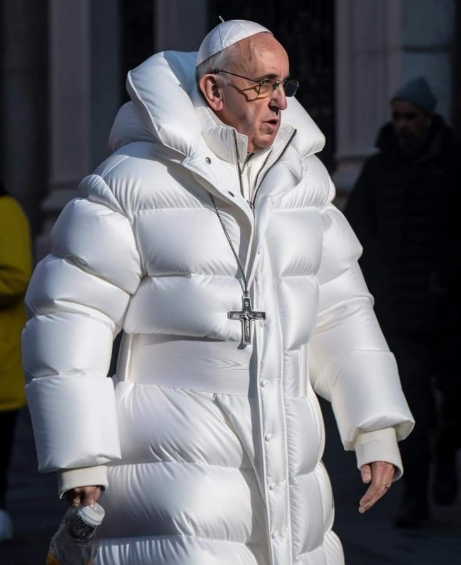}} &  \emph{\textbf{Hint}: Fake images may have sociocultural errors. For example, the images may show scenarios that are inappropriate, unlikely to be seen in the real world, violate subtle rules specific to different cultures, or are historically inaccurate. \newline In the image below, the scenario is highly unlikely.} \\
    \bottomrule
    \end{tabular}
    \caption{The \emph{Visual} intervention shows common errors found in {deepfake images} through a combination of an example image and an explanatory text.}
    \label{tab:intervention_visual}
\end{table}

(3)~\emph{Gamified}: This intervention draws upon elements of gamification to increase engagement, motivation, and learning in line with the design principles for effective interventions \cite{Fogg.2002}. Here, we first presented participants with brief instructions on how to play the game. The goal of the game was to locate the error in each deepfake image by clicking on the corresponding position in the image. For this, we showed participants the same example images and textual descriptions of how to recognize {deepfake images} as in the \emph{Visual} intervention and asked them to spot the error.  We employed various elements that are common in gamification \cite{Kiryakova.2014, Hamari.2014}, namely, a timer, rewards in the form of points, and real-time feedback. Participants had 30 seconds per image and were rewarded 10 points for each error they found. We showed their overall score throughout the game. To further motivate participants, participants received immediate feedback when clicking on an image: participants that identify the error correctly saw the error highlighted with a green circle and received a message in green font (\emph{\say{Correct! You found the error. You earned 10 points!}}); for all other clicks, the participants received a message in red font (\emph{\say{Not quite! Try again.}}). After finishing the game, participants saw their total score and a message that ranks their achievement (0--10 points: \emph{\say{Thanks for participating! Spotting errors in fake images can be challenging. With more exposure, you'll likely get better at it.}}; 20--30 points: \emph{\say{Good eye! You've shown a solid ability to spot errors in fake images.}}; 40--50 points: \emph{\say{Impressive results! You've demonstrated a keen ability to spot errors in fake images.}}). An example is shown in Table~\ref{tab:intervention_gamified}. Again, each error and image was presented on a separate page in randomized order.

\begin{table*}[h]
    \centering
    \begin{tabular}{p{0.3\linewidth}p{0.7\linewidth}}
    \toprule
        Points: 10 Time left: 20 seconds \newline 
        \raisebox{-0.6\height}{\includegraphics[width=0.2\textwidth]{images/holding_hands_at_dinner.jpg}} & \emph{\textbf{Hint}: Fake images often have anatomical errors. For example, people often have missing, extra, or merged fingers, nonexistent fingernails, and unlikely hand proportions. \newline Can you spot the error in the image? Click on it to earn points! \newline \newline
        \textcolor{red}{\textbf{Not quite! Try again.}} }
         \\
    \midrule
        Points: 10 Time left: 20 seconds \newline \raisebox{-0.6\height}{\includegraphics[width=0.2\textwidth]{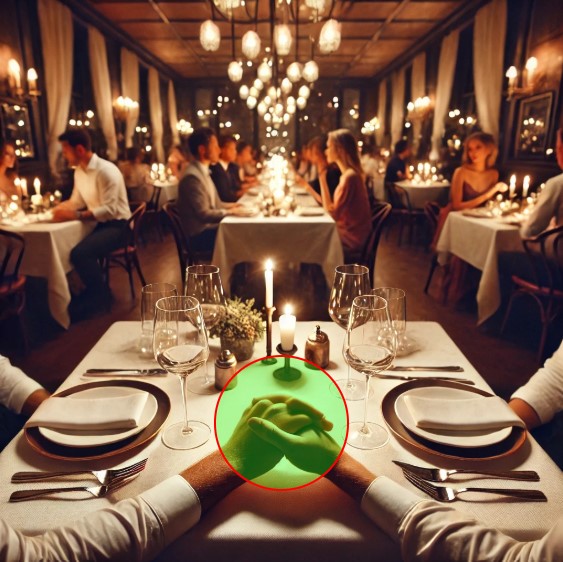}} & \emph{\textbf{Hint}: Fake images often have anatomical errors. For example, people often have missing, extra, or merged fingers, nonexistent fingernails, and unlikely hand proportions. \newline Can you spot the error in the image? Click on it to earn points! \newline \newline        
        \textcolor{DarkGreen}{\textbf{Correct! You found the error. You earned 10 points!}} }\\
        
    \bottomrule
        \end{tabular}
    \caption{The \emph{Gamified} intervention shows participants examples of deepfake images and explanatory texts in a gamified setting, which includes a timer, point-based rewards, and real-time feedback.}
    \label{tab:intervention_gamified}
\end{table*}

(4)~\emph{Feedback}: We presented participants with five real images and five {deepfake images} and asked them to complete ten rounds of image discernment with immediate feedback after each assessment. The {deepfake images} in this intervention are analogous to those in the \emph{Visual} and \emph{Gamified} intervention. The real images can be found in Table~\ref{tab:intervention_feedback} in the Appendix. Unlike the other interventions, no descriptions of errors are provided; this choice is informed by prior research suggesting that repeated exposure alone can support generalization to new stimuli \cite{Gordon.1983}. For each image, we asked participants whether they think the image is real or fake. After answering, they received immediate feedback indicating whether their answer was correct. If the participant misidentifies an image (e.g., by labeling a deepfake as real), a red-colored text box appears stating that their response was incorrect and that the image was fake. In addition, the answer button briefly ``shakes'' when an incorrect answer was given to reinforce the feedback in an engaging way. If the answer was correct, a green-colored text informs that the response was correct. Each image was again shown on a separate page in randomized order.

(5)~\emph{Knowledge}: We presented participants with a textual explanation of how AI technologies are used in image generators to create {deepfake images}. The explanation is written in simple, easy-to-understand language and presented as a brief paragraph (see Table~\ref{tab:intervention_knowledge} for the exact content). This intervention tests whether increased conceptual knowledge about AI image generation influences participants’ ability to discern {deepfake images}. This is inspired by prior research which suggests that familiarity with a topic can affect discernment performance \cite{Chein.2024}.

\begin{table}[h]
    \centering
    \begin{tabular}{p{\linewidth}}
    \toprule
        \emph{AI can learn how to generate realistic images by looking at millions of example pictures. After inputting the example images into the AI, they are transformed into representations that only computers can understand and that don't mean anything to humans. From those, AI then learns patterns of how the world, such as objects, people, and places look like. After training, humans can ask AI models to generate images through simple input text, this is called prompting. The possibilities for AI-generated images are endless, for example, AI can generate things that have never been seen before such as made-up scenarios or even attempt to imitate real people or places.} \\
    \bottomrule
    \end{tabular}
    \caption{The \emph{Knowledge} intervention provides a short explanation of how {deepfake images} are generated with the help of AI.}
    \label{tab:intervention_knowledge}
\end{table}

\subsection{Task and materials}
To test the discernment abilities, the participants completed an image discernment task in which they were shown a set of 15 images, presented one at a time in randomized order. The 15 images include 5 real images and 10 deepfake images. For each image, we asked the participants whether they think the image is real or fake using a four-point Likert scale (i.e., ``Definitely fake'', ``Probably fake'', ``Probably real'', ``Definitely real''). {We use the simplified terms ``real'' and ``fake'' as they are more accessible to a general audience. At the beginning of the survey, participants saw an explanation that real images refer to images shot by a human and fake images refer to AI-generated images. We included the ``probably'' options to lower the commitment cost of labeling an image when participants are unsure. This lets participants follow their gut when uncertain, which reduces conservative under-calls compared to a forced “definitely fake vs. definitely real” choice. At the same time, we omitted a neutral midpoint to discourage non-informative answers and require a directional stance (real vs. fake).}

{For each image, we also asked participants whether they would consider sharing the image on their social media (``Yes'', ``No'', ``Don't know''). We collected the sharing intention because previous research suggests that belief and sharing are distinct from each other \cite{Pennycook.2021b, Pennycook.2022}, but sharing is the behavior that propagates content online and hence acts as a behavioral proxy for diffusion \cite{Bashardoust.2024}. Participants were unaware of the ground truth label of the image when asked about their sharing intention, which mimics real-world scenarios, where people tend to share on social media without knowing whether an image is real or a deepfake.} A screenshot of the task interface is provided in {Appendix}~\ref{sec:example_discernment_task}.

Additionally, participants were given the option to tick a checkbox if they have seen the image before or if the image failed to load properly (e.g., due to bandwidth issues with their Internet). The responses for images where a participant checked one of the options (i.e., reported they had seen the image or that the image had not loaded properly) were excluded from the analysis, {while keeping the remaining responses for the other images where the checkbox was not ticked}. Later, we provide a robustness check with regard to people who have seen the images in {Appendix}~\ref{supp:robustness_checkbox_filtering}.

We presented participants with both real images and {deepfake images} that cover a broad range of themes, including everyday activities, potential disinformation scenarios, and portraits. In total, we collected 10 real images and 20 {deepfake images}, which we randomly split into two sets (Group A and B), each containing 15 images (see  {Appendix}~\ref{sec:images} for an overview of all images, including the source links and the collection date). To account for the potential heterogeneity of {deepfake images}, we included images from two sources: half of the images have gone viral and the other half have not. We thereby aim to account for that viral {deepfake images} may possess certain features that contribute to their widespread dissemination (e.g., such as being more convincing, more emotionally engaging, or crafted with greater technical sophistication). Including both types allows us to examine a broader and more representative spectrum of deepfake content. The viral {deepfake images} were sourced from global news outlets reporting on widely shared {deepfake images}. The non-viral {deepfake images} were collected from Midjourney's Discord channel \cite{Midjourney.Discord}, self-generated using DALL-E 3 \cite{OpenAI.2023} as provided by the ChatGPT website, or sourced from the Center for Countering Digital Hate \cite{CenterforCounteringDigitalHate.2024}. Real images were selected to match the themes of the {deepfake images} and were obtained from the public photo hosting service Flickr (\url{www.flickr.com}). We provide a robustness check in {Appendix}~\ref{supp:robustness_viral_nonviral}, where we test for differences in discernment abilities for viral and non-viral {deepfake images}.

\subsection{Procedure}

Participants first provided informed consent and completed an attention check {(see Appendix~\ref{sec:attention_checks} for the detailed questions of the attention check)}. They were then randomly assigned to either the control condition or one of five intervention conditions. {Participants in the intervention conditions received a digital literacy intervention; participants in the control condition did not. Then, all participants completed the image discernment task, while being aware that images may be real or deepfakes.} For this, participants were randomly assigned to one of two groups as part of a counter-balanced design. Group~1 completed image set A immediately after the intervention and image set B in the two-week follow-up; Group~2 received the sets in reverse order. This design controls for potential order and image-specific effects, ensuring that differences in the discernment task are attributable to the interventions rather than the sequence or content of the image sets.

Next, we collected covariates such as the participants' sociodemographics, social media use, and digital literacy levels. Participants were also asked to complete a cognitive reflection test (CRT) \cite{Frederick.2005} (see Table~\ref{tab:independent_variables} {in the Appendix} for details on questions and scales). At the end of the study, we also performed a second attention check and two honesty checks (i.e., asking whether participants responded randomly at any point and whether they searched for any of the images online such as, e.g., via Google) {to ensure participants relied on the information they learned from the interventions, in line with previous research \cite{Pennycook.2021b}}. Participants who failed both of the attention checks and/or both honesty checks were excluded from the analysis.\footnote{{When filtering out all participants who failed either one of the honesty checks ($N=40$), the main results of our paper remain robust.  } }

After two weeks, {we invited the same set of participants to complete a follow-up using their Prolific IDs. They were asked to repeat the image discernment task but this time with the previously unseen set of 15 images (i.e., image set A or B, depending on their group assignment).} Then, participants answered the same set of questions about their digital literacy abilities. The follow-up remained open for one week, after which it was closed for submissions. Finally, participants were debriefed and informed on which images were real and which were {deepfake images}. 

\subsection{Measures}

We collected the following dependent variables from our image discernment task, namely, (a)~\emph{Discernment}, and (b)~\emph{SharingIntention} as follows. After finishing the task, we also asked participants once for their overall (c)~\emph{Confidence} during the discernment task.
\begin{enumerate}[label=(\alph*)]
\item \emph{Discernment}: We asked participants to rate 15 images as real or fake on a four-point Likert scale (``Definitely fake'', ``Probably fake'', ``Probably real'', ``Definitely real''). {We mapped the responses to binary labels ``fake'' (responses ``Definitely fake'' and ``Probably fake'') and ``real'' (responses ``Definitely real'' and ``Probably real'').} Each response was coded as correct or incorrect based on the image's ground truth label, based on which we then computed two discernment measures for each participant, namely, their accuracy in detecting real images and in detecting {deepfake images}. The accuracy for real images is calculated as $TP / P$, where $TP$ denotes the true positives {(number of real images a participant correctly discerned as real)} and $P$ the total number of real images. The accuracy for {deepfake images} is calculated as $TN / N$, where $TN$ denotes true negatives {(number of deepfake images a participant correctly discerned as deepfake)} and $N$ the total number of {deepfake images}. 
\item \emph{SharingIntention}: For each image, we asked participants whether they would share it on their social media  (answer options: ``Yes'', ``No'', ``Don't know''). A ``Yes'' is counted as an intention to share. Again, we computed two measures for each participant: The sharing intention for real images, computed as $TP / P$, where $TP$ denotes the real images the participant is willing to share and $P$ the total number of real images; and the sharing intention for {deepfake images}, calculated as $TN / N$, where $TN$ denotes the {deepfake images} the participant is willing to share and $N$ the total number of {deepfake images}.
\item \emph{Confidence}: After completing the image discernment task, we asked participants about their confidence in identifying images as real and fake on a scale of 0 (not confident at all) to 100 (very confident). {We used a 0--100 scale as it provides an intuitive percentage-like metric that is familiar to participants, making it both interpretable and suitable for reporting individual differences in confidence.}
\end{enumerate}
Details are in Table~\ref{tab:dependent_variables} {in the Appendix}. 

To account for between-participant heterogeneity in our analyses, we further collected four sets of covariates. (1)~We asked participants about their sociodemographics such as age, ethnicity, gender, level of education, religion, political orientation, income, and subjective social status. {For example, age was a strong confounder of susceptibility to misinformation in previous work \cite{Jin.2025, Mitchell.2003}, with older adults particularly struggling to distinguish real from fake content \cite{Bashardoust.2024, Grinberg.2019}, and that tailored interventions can help improve their discernment abilities \cite{Moore.2022}.} (2)~We asked participants about their social media use, including the platforms they use, their sharing habits, and the amount of time they spend online. This helps capture the media environment participants are typically exposed to, which may influence their familiarity with manipulated content. (3)~We collected information about the participants' digital literacy, such as their knowledge of {deepfake images}, their experience in detecting {deepfake images}, as well as their experience with search engines, reverse image search, and generative AI. (4)~We collected information on the participants' analytical thinking skills through the use of CRT \cite{Frederick.2005}. This is informed by previous studies showing that psychological factors, such as partisanship \cite{Roozenbeek.2022} and analytical thinking \cite{Pennycook.2019, Pennycook.2021, Arechar.2023}, are important predictors of digital literacy. Details on all are provided in Table~\ref{tab:independent_variables} {in the Appendix}.

\subsection{Analysis plan}
\label{sec:analysis_plan}

First, we filtered the collected data as follows. Participants who failed both attention checks and/or both honesty checks were excluded from the analysis. We also removed participants whose completion time fell outside the 95\% interval of the sample distribution. In addition, we removed participants who gave the same answer for every image in the discernment task. Lastly, if a participant indicated that they had previously seen the image or experienced loading issues, we set these responses to missing values and computed the discernment and sharing intention without those responses.

After filtering, we evaluated the effectiveness of our digital literacy interventions on participants’ ability to discern {deepfake images} using a hypothesis-driven approach.

For \textbf{H1} (\emph{People who receive a digital literacy intervention will have better discernment of {deepfake images}}), we compared each intervention condition to the {same} control condition using Mann–Whitney $U$ tests {and apply Bonferroni correction for the many-to-one comparisons}. We selected this non-parametric test to account for potential non-normality in our data (for details, see {Appendix}~\ref{sec:normality_check}). A significant result is interpreted as evidence that the intervention improves {deepfake image discernment}.

For \textbf{H2} (\emph{People who receive a digital literacy intervention will not be more skeptical of real images}), we used equivalence tests \cite{Lakens.2018} to compare accuracy in discerning real images between each intervention condition and the control condition. We tested against an equivalence interval of ($-0.05$, $0.05$) {and applied Bonferroni correction for the many-to-one comparisons (per-comparison $\alpha_{\text{adj}}=0.01$)}. If our observed confidence interval is fully contained in the testing interval, we consider this as evidence for a null effect; otherwise, we consider the results inconclusive with respect to the null.

For \textbf{H3} (\emph{People who receive a digital literacy intervention will have better long-term discernment of {deepfake images}}), we conducted Mann–Whitney $U$ tests {(with Bonferroni correction)} comparing follow-up accuracy between intervention conditions and the control conditions two weeks after the intervention. {We treated the follow-up as a new between-subject comparison as we wanted to know whether participants who had received an intervention still outperformed comparable control participants two weeks later.} To account for attrition, the initial sample size included a 30\% buffer \cite{Haile.2023}. A significant result is interpreted as evidence that the intervention’s effect persists over time.

All analyses are conducted using Python (3.8) with \emph{numpy} (1.24.4), \emph{pandas} (2.0.3), \emph{scipy} (1.10.1), and \emph{statsmodels} (0.14.1). Data visualizations were created with \emph{matplotlib} (3.7.5), and \emph{seaborn} (0.13.2). All analysis scripts will be made publicly available. 

\subsection{Robustness checks}
\label{sec:robustness_checks}

We conducted a series of robustness checks to assess the validity and reliability of our findings. First, we repeated our analyses without excluding participants who failed both attention checks and/or honesty checks, to examine whether our effects remain robust. Second, we repeated our analysis without removing image-level responses flagged by participants as previously seen or not properly loaded. For both, we compared the difference in discernment ability for filtered and non-filtered responses using Mann-Whitney $U$ tests. 

Third, to account for individual-level heterogeneity, we estimated the effect of our interventions on participants' ability to discern {deepfake images} using an ordinary least squares (OLS) regression model via
\begin{equation}
    Y_i = \beta_0 + \beta_1 \emph{Condition}_i +  \beta_2 X_i + \varepsilon_{i},
    \label{eq:covariates}
\end{equation}
where $Y_i$ is the observed discernment ability for participant $i$, \emph{Condition}$_i$ is a binary variable equal to 1 for intervention and 0 for control, and $X_i$ is a vector of participant-level control variables.  

Fourth, we accounted for possible interaction effects with partici\-pant-specific control variables in our regression model via
\begin{equation}
    Y_i = \beta_0 + \beta_1 \emph{Condition}_i +  \beta_2 X_i + \beta_3 (\emph{Condition}_i \odot X_i) + \varepsilon_{i},
    \label{eq:covariates_interactions}
\end{equation}
where $\emph{Condition}_i \odot X_i$ is a two-way interaction term.

Fifth, to test for non-random attrition, we compared participants who completed the follow-up with those who dropped out using Mann-Whitney $U$ tests for differences in discernment abilities and sociodemographics.

Sixth, we controlled for the image set that participants saw in the image discernment task via
\begin{equation}
    Y_i = \beta_0 + \beta_1 \emph{Condition}_i +  \beta_2 Set_i + \beta_3 (\emph{Condition}_i \odot Set_i) + \varepsilon_{i},
    \label{eq:order_effect}
\end{equation}
where $Set_i$ is a binary variable which equals 0 if participant $i$ saw image set A in the first session and 1 if participant $i$ saw image set B, and where $\odot$ denotes element-wise multiplication.

Seventh, we assessed whether the interventions are differentially effective for viral versus non-viral {deepfake images} using a linear mixed-effects model
\begin{align}
\begin{split}
    Y_{ij} &= \beta_0 + \beta_1 \emph{Condition}_i + \beta_2 \emph{Source}_{j}  \label{eq:mixed_effects} \\ 
    &\quad + \beta_3 (\emph{Condition}_i \odot \emph{Source}_{j}) + u_i + \varepsilon_{ij},
\end{split}   
\end{align}
where $Y_{ij}$ is the observed discernment ability for participant $i$ for {deepfake images} set $j$, $\emph{Condition}_i$ is a categorical variable indicating the intervention condition for participant $i$ (with control as reference category), $\emph{Source}_{j}$ is a binary variable which equals 1 for {deepfake images} sets consisting of viral images and 0 for non-viral {deepfake images}, $\emph{Condition}_i \odot \emph{Source}_{j}$ is the interaction term, $u_i \sim \mathcal{N}(0, \sigma_u^2)$ are random intercepts accounting for individual differences, and $\varepsilon_{ij} \sim \mathcal{N}(0, \sigma_\varepsilon^2)$ is the residual error.

Eight, we conducted a robustness check where we evaluated the long-term effectiveness of our interventions using a linear mixed effects regression with participant-level random effects. Here, we estimate
\begin{align}
\begin{split}
    Y_{it} &= \beta_0 + \beta_1 \emph{Condition}_i + \beta_2 \emph{TimePoint}_{t} \label{eq:mixed_effects_long_term} \\
    &\quad + \beta_3 (\emph{Condition}_i \odot \emph{TimePoint}_{t}) + u_i + \varepsilon_{it},
    \end{split}
\end{align}
where $Y_{it}$ is the {deepfake image discernment} ability of participant $i$ at time point $t$, $\emph{Condition}_i$ is a binary variable indicating the intervention condition for participant $i$ (with control as reference category), $\emph{TimePoint}_{t}$ is a binary variable which equals 0 for the first session and 1 for the follow-up time, $\emph{Condition}_i \odot \emph{TimePoint}_{t}$ is the interaction term, $u_i \sim \mathcal{N}(0, \sigma_u^2)$ are random intercepts for individual differences, and $\varepsilon_{it} \sim \mathcal{N}(0, \sigma_\varepsilon^2)$ is the residual error.

Lastly, we conducted a robustness check to test for possible learning effects from the image discernment task itself (see {Appendix}~\ref{supp:learning_effect_robustness} for more information). We found that mere exposure to the discernment task did not create a learning effect for the control condition, suggesting that improvements in discernment performance cannot be explained by exposure to the task but must be attributed to our interventions.

\subsection{Validation data}
Our main study was exploratory and was not preregistered. To externally validate our findings, we conducted a preregistered validation study, where, informed by the above analysis, we specified the hypotheses, primary outcomes, and analysis plan (see OSF preregistration \url{https://osf.io/5au7p/?view_only=8c4cbf81ca0746debe3c6bac5c134229}). We collect a second set of data, which we refer to here as \emph{validation data}. For this, we collected $N=600$ participants from the U.S. to achieve approximately $N\approx 100$ participants per intervention condition across two image sets, while remaining within budgetary constraints. With this, we are able to detect effect sizes of Cohen's $d=0.46$ ($\alpha_{\text{adj}}=0.01$, power=0.8). The procedure, interventions, tasks, materials, and measures were the same as described above. We repeated the analyses with this validation data to replicate our findings.

\subsection{{Ethics information}}

This research complies with all relevant ethical regulations. Ethical approval was obtained from the Institutional Review Board (IRB) of the Faculty of Mathematics, Informatics, and Statistics at LMU Munich (EK-MIS-2024-319). 
All participants gave informed consent at the beginning of the experiment and could withdraw at any time. At the end of the first experiment, participants were debriefed that some of the images they viewed were deepfakes. At the end of the follow-up, they were informed which specific images were artificially generated. Data were collected and stored in accordance with local data privacy laws. Compensation was done via the Prolific platform in accordance with the fair‑wage policy of the platform. 

\begin{table}[h]
\centering
\begin{tabular}{lrr}
\hline
\textbf{Sociodemographics} & $N$ & \% \\
\hline
\textbf{Gender} & & \\
Female & 586 & 52.7 \\
Male & 515 & 46.3 \\
Non-binary/Third gender/Other & 10 & 0.9\\
Prefer not to say & 1 & 0.1 \\
\hline
\textbf{Age} & & \\
18--34 years & 410 & 36.9 \\
35--54 years & 506 & 45.5 \\
55+ years & 194 & 17.5 \\
Mean (SD) & \multicolumn{2}{r}{41.7 (13.3)} \\
Range & \multicolumn{2}{r}{18-82} \\
\hline
\textbf{Ethnicity} & & \\
White & 786 & 70.6 \\
blue or African American & 192 & 17.3 \\
Asian & 39 & 3.5 \\
Hispanic, Latino or Spanish origin & 26 & 2.3 \\
Other & 69 & 6.2 \\
\hline
\textbf{Education} & & \\
0-6 (up to primary School) & 1 & 0.1 \\
Up to high school (7--12 years) & 153 & 13.8 \\
College/Undergraduate (13--16 years) & 683 & 61.4 \\
Graduate/Professional (17+ years) & 269 & 24.2 \\
Prefer not to say & 6 & 0.5 \\
\hline
\textbf{Religious status} & & \\
Not religious & 617 & 55.5 \\
Religious & 444 & 39.9 \\
Prefer not to answer & 51 & 4.6 \\
\hline
\textbf{Income group} & & \\
Low & 164 & 14.8 \\
Medium & 422 & 38.1 \\
High & 503 & 45.2 \\
Prefer not to say & 23 & 2.1 \\
\hline
\textbf{Political orientation} & & \\
Very liberal (1--1.5) & 135 & 12.1 \\
Liberal (2--2.5) & 210 & 18.9 \\
Somewhat liberal (3--3.5) & 117 & 10.5 \\
Neutral (4) & 200 & 18.0 \\
Somewhat conservative (4.5--5) & 130 & 11.7 \\
Conservative (5.5--6) & 205 & 18.5 \\
Very conservative (6.5--7) & 115 & 10.3 \\
\hline
\multicolumn{3}{l}{SD: standard deviation}
\end{tabular}
\caption{Sociodemographics of participants after filtering ($N=1,112$).}
\label{tab:demographics}
\end{table}

\section{Results}
\label{supp:results}

We conducted our experiment with a sample of $N=1,200$ U.S. participants. Applying the above filtering resulted in a final sample $N=1,112$ participants. People were filtered roughly equally from each condition, with 187 people remaining in the \emph{Control} condition, 188 people in the \emph{Textual} condition, 183 in the \emph{Visual} condition, 184 in the \emph{Gamified} condition, 185 people in the \emph{Feedback} condition, and 185 people in the \emph{Knowledge} condition.  The response rate for the follow-up was: 72.19\% for \emph{Control} condition, 76.60\% for \emph{Textual}, 70.49\% for \emph{Visual}, 69.57\% for \emph{Gamified}, 76.76\% for \emph{Feedback}, and 71.51\% for \emph{Knowledge}. Hence, the number of participants in the follow-up was relatively balanced across conditions. In the following, we present our results. A significance level of {$\alpha_{\text{adj}} = 0.05$} was used for all statistical comparisons. The sociodemographics of our participants can be found in {Table}~\ref{tab:demographics} and the results of our robustness checks are in {Appendix}~\ref{supp:robustness_checks_results}.

\subsection{Effects of interventions on discernment abilities}

We analyze the effect of our digital literacy interventions on participants' abilities to discern between real images and {deepfake images} compared to the control condition {(see Table~\ref{tab:performance_across_condition} for an overview of the performance of participants across intervention conditions)}. We first assess in Section~\ref{sec:results_discerning_deepfakes} whether our interventions boosted participants' ability to correctly identify {deepfake images}, both immediately after the intervention and at follow-up (Hypothesis \textbf{H1} and \textbf{H3}, respectively). In Section~\ref{sec:results_discerning_real_images}, we analyze whether our interventions affected participants' trust in real images and made them more skeptical (Hypothesis \textbf{H2}). 

\begin{table*}[h]
\centering
\begin{tabular}{lrrrrrrrr}
\toprule
 & \multicolumn{4}{c}{\textbf{Main study}} & \multicolumn{4}{c}{\textbf{Follow-up}} \\
\cmidrule(lr){2-5}\cmidrule(lr){6-9}
\textbf{Condition} & \textbf{TN} & \textbf{TP} & \textbf{FPR} & \textbf{FNR} & \textbf{TN} & \textbf{TP} & \textbf{FPR} & \textbf{FNR} \\
\midrule
\emph{Control}   & 61.33 & 83.97 & 38.14 & 16.03 & 63.54 & 81.74 & 36.46 & 18.26 \\
\emph{Textual}   & 68.80 & 82.99 & 31.20 & 17.01 & 68.07 & 81.02 & 31.93 & 18.98 \\
\emph{Visual}    & \textbf{74.31} & 79.99 & 25.69 & 20.01 & \textbf{68.42} & 80.23 & 31.58 & 19.77 \\
\emph{Gamified}  & 65.70 & 81.36 & 34.30 & 18.64 & 61.53 & 82.66 & 38.47 & 17.34 \\
\emph{Feedback}  & 60.00 & \textbf{84.26} & 40.00 & 15.74 & 63.58 & \textbf{84.93} & 36.42 & 15.07 \\
\emph{Knowledge} & 64.48 & 84.00 & 35.52 & 16.00 & 63.96 & 81.39 & 36.04 & 18.61 \\
\bottomrule
\end{tabular}
\caption{{\textbf{Discernment performance by intervention (mean across participants, \%).} 
{TN} = Deepfake image discernment accuracy; {TP} = Real image discernment accuracy;
{FPR} = discerned deepfake images as real; {FNR} = discerned real images as deepfakes.}}
\label{tab:performance_across_condition}
\end{table*}

\subsubsection{Discerning {deepfake images}}
\label{sec:results_discerning_deepfakes}
\begin{figure*}[h]

\begin{subfigure}[t]{0.45\textwidth}
\llap{\textsf{\textbf{a}}\hspace{-10pt}}%
\includegraphics[width=\linewidth]{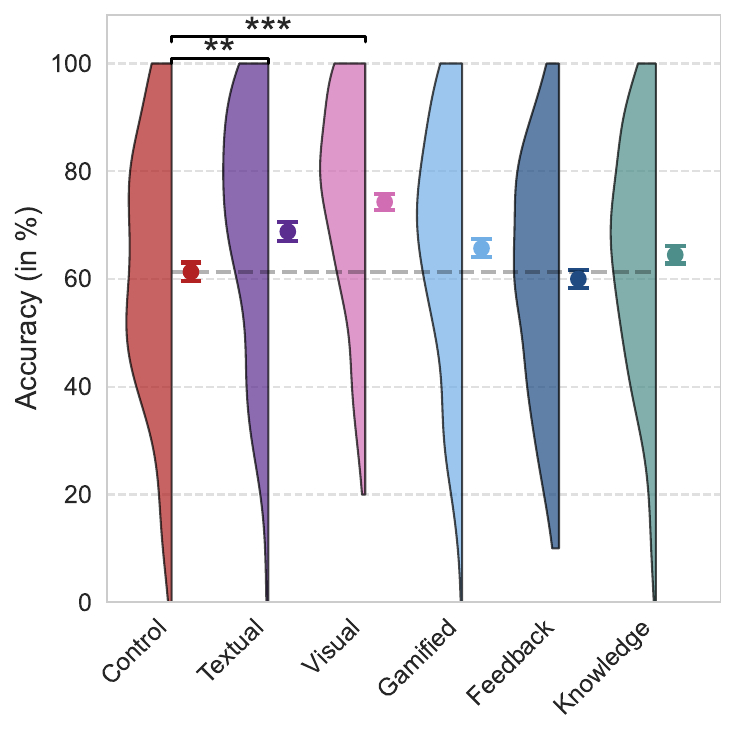} 
\label{fig:acc_fake_over_time_t1}
\end{subfigure}
\begin{subfigure}[t]{0.45\textwidth}
\llap{\textsf{\textbf{b}}\hspace{-10pt}}%
\includegraphics[width=\linewidth]{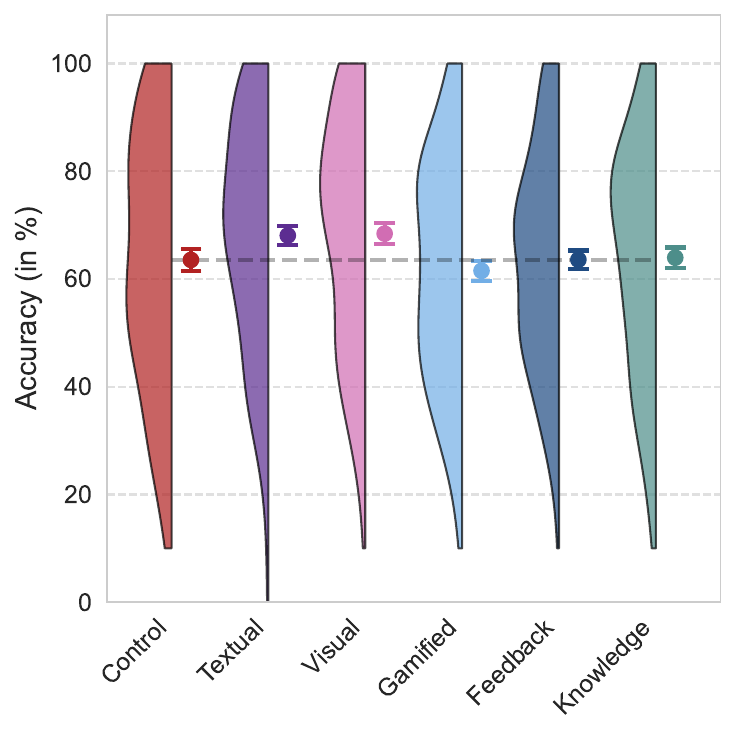}
\label{fig:acc_fake_over_time_t2}
\end{subfigure}
\vspace{-1cm}
\caption{\textbf{Discernment accuracy for {deepfake images} across conditions.} Shown is the participant-level accuracy in discerning {deepfake images}: \textsf{(\textbf{a})}~immediately after the intervention (time point T1) and \textsf{(\textbf{b})}~two weeks after the initial intervention (time point T2). The sample size was $N=1,112$ for time point T1 and $N=764$ for time point T2 after attrition. We hypothesized that participants who received a digital literacy interventions would show a better accuracy in discerning {deepfake images} compared to the control group. Evidently, at T1, the discernment accuracy was significantly higher for the \emph{Textual} and \emph{Visual} interventions. Each violin plot shows the distribution of the participant-level accuracy within each condition. Dots indicate mean values; whiskers indicate standard errors. Significance levels for Mann-Whitney $U$ tests: $^*{p_{\text{adj}}}<0.05$, $^{**}{p_{\text{adj}}}<0.01$, $^{***}{p_{\text{adj}}}<0.001$. }
    \label{fig:acc_fake_over_time}
    \Description[short explanation]{(a) The subfigure shows six vertical violin plots of participant accuracy for {deepfake image discernment} immediately after the intervention (y-axis labeled “Accuracy (in \%)”, 0–100\%) for Control, Textual, Visual, Gamified, Feedback, and Knowledge. The Textual and Visual violins are shifted higher than Control, and Gamified is also above Control, indicating better {deepfake image discernment} right after training. Asterisks next to Textual and Visual conditions (e.g., “”, “*”) mark statistically significant improvements versus Control. Feedback and Knowledge overlap more with Control, suggesting little or no immediate gain. (b) The subfigure shows the same six violin plots and axis as in (a), now for the follow-up session two weeks later. Distributions for Textual, Visual, and Gamified remain slightly higher than Control but show substantial overlap across conditions. No significance asterisks are displayed, indicating no statistically significant differences versus Control at follow-up.}
\end{figure*}

In our main session (Hypothesis \textbf{H1}), participants identified {deepfake images} with a mean accuracy of 61.3\% (see Figure~\ref{fig:acc_fake_over_time}\textsf{\textbf{a}}). {Deepfake image discernment} was boosted by 7.5 percentage points for participants in the \emph{Textual} condition ($\mu = 68.8\%$, Mann-Whitney $U=14302.0$, ${p_{\text{adj}}<0.01}$), and by 13 percentage points for participants in the \emph{Visual} intervention ($\mu = 74.3\%$, Mann-Whitney $U=11665.0$, ${p_{\text{adj}} < 0.001}$). Discernment for {deepfake images} {did not differ significantly from the control condition} for interventions \emph{Gamified} ($\mu = 65.7\%$,  Mann-Whitney $U=15292.5$, ${p_{\text{adj}}=0.310}$), \emph{Knowledge} ($\mu=64.5\%$, Mann-Whitney $U = 15964.0$, ${p_{\text{adj}} = 0.980}$), and \emph{Feedback} ($\mu = 60.0\%$, Mann-Whitney $U=17872.5$, ${p_{\text{adj}}=1.000}$).

To analyze long-term effects (Hypothesis \textbf{H3}), we conducted a follow-up two weeks after the initial intervention (see Figure~\ref{fig:acc_fake_over_time}\textsf{\textbf{b}}). We compared each intervention against the control condition using Mann-Whitney $U$ tests but found no significant difference in {deepfake image} discernment (\emph{Textual}: Mann-Whitney $U=8635.5$, ${p_{\text{adj}}=0.525}$; \emph{Visual}: $U=7724.0$, ${p_{\text{adj}}=0.550}$; \emph{Gamified};  $U=9132.5$, ${p_{\text{adj}}=1.000}$; \emph{Feedback}:  $U=9654.5$, ${p_{\text{adj}}=1.000}$; and \emph{Knowledge}: $U = 8915.5$, ${p_{\text{adj}}=1.000}$). The accuracy in detecting {deepfake images} in the control group was comparable and non-significantly different for the main session and the follow-up (Wilcoxon signed rank test $W = 2816.0$, $p=0.244$). {We also conducted a robustness check to control for possible learning effects of the task itself by comparing the control group in the follow-up to an independent robustness group and find no significant difference (see Appendix~\ref{supp:learning_effect_robustness}).}

\subsubsection{Discerning real images}
\label{sec:results_discerning_real_images}
\begin{figure*}[h]

\begin{subfigure}{0.45\textwidth}
\llap{\textsf{\textbf{a}}\hspace{-10pt}}
\includegraphics[width=\linewidth]{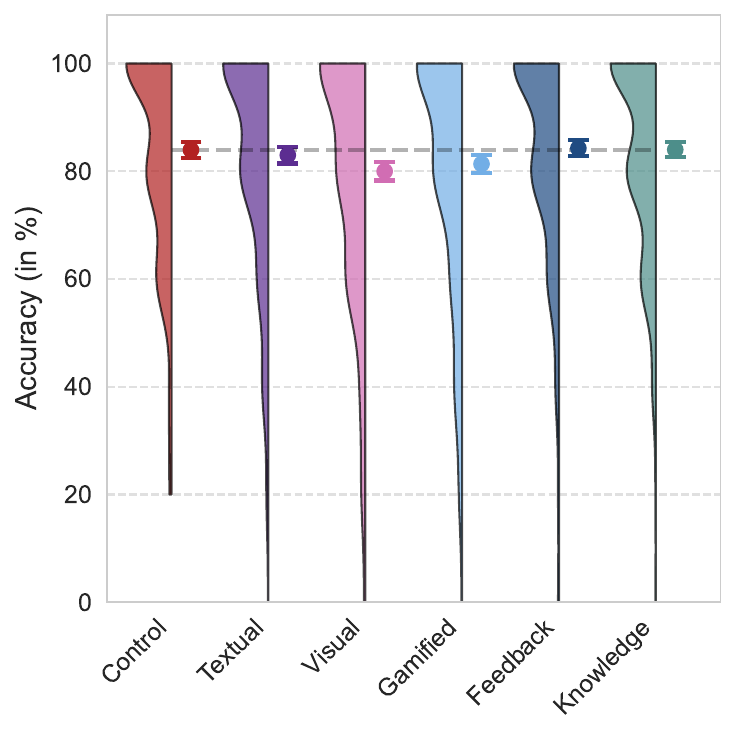} 
\label{fig:acc_real_over_time_t1}
\end{subfigure}
\begin{subfigure}{0.45\textwidth}
\llap{\textsf{\textbf{b}}\hspace{-10pt}}
\includegraphics[width=\linewidth]{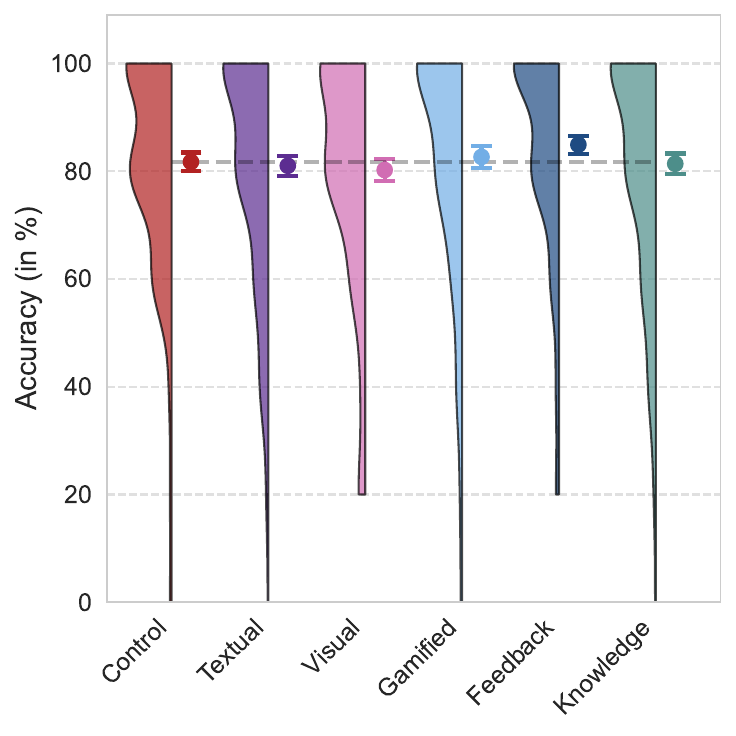}
\label{fig:acc_real_over_time_t2}
\end{subfigure}
\vspace{-1cm}
    \caption{\textbf{Discernment accuracy for real images across conditions.} Shown is the participant-level accuracy in discerning real images: \textsf{(\textbf{a})}~immediately after the intervention (time point T1) and \textsf{(\textbf{b})}~two weeks after the initial intervention (time point T2). The sample size was $N=1,112$ for time point T1 and $N=764$ for time point T2 after attrition. We hypothesized that participants who receive a digital literacy intervention would not become more skeptical of real images. Across both time points, the accuracy remained statistically indistinguishable from the control group. Each violin plot shows the distribution of the participant-level accuracy within each condition. Dots indicate mean values; whiskers indicate standard errors.}
    \label{fig:acc_real_over_time}
    \Description[short explanation]{(a) The figure shows six vertical violin plots of participants’ accuracy for real image discernment immediately after the intervention (y-axis “Accuracy (in \%)”, 0–100) for Control, Textual, Visual, Gamified, Feedback, and Knowledge. The distributions are similar in height and shape and largely overlap across conditions, indicating comparable ability to identify real images immediately after the intervention; no significance markers are displayed. (b) The figure shows the same set of six violin plots and axis as in (a), now for the follow-up session two weeks later. Distributions again overlap substantially across all conditions, suggesting real-image accuracy remained broadly comparable at follow-up; no significance markers are shown.}
\end{figure*}

Next, we examine whether our interventions had any unintended effects on participants’ ability to detect real images (Hypothesis \textbf{H2}). On average, participants correctly identified real images with a mean accuracy of 83.9\% (see Figure~\ref{fig:acc_real_over_time}\textsf{\textbf{a}}). {We conducted equivalence tests \cite{Lakens.2018} to compare each intervention to the control, using an equivalence region of $\pm 0.05$ (absolute accuracy points). To control the family-wise error across the five tests, we applied a Bonferroni correction (per-comparison $\alpha_{\text{adj}}=0.01$), which corresponds to reporting simultaneous two-sided $98\%$ confidence intervals.} We hypothesized that our intervention did not make participants more skeptical of real images. We find that the equivalence tests for each intervention fall in the undecided range (\emph{Textual}: $p_{\text{adj}}=1.000$, 98\% CI: $[-3.977,\, 5.934]$; \emph{Visual}: $p_{\text{adj}}=1.000$, 98\% CI: $[-1.287,\, 9.238]$; \emph{Gamified}: $p_{\text{adj}}=1.000$, 98\% CI: $[-2.569,\, 7.784]$; \emph{Feedback}: $p_{\text{adj}}=1.000$, 98\% CI: $[-5.139,\, 4.549]$; \emph{Knowledge}: $p_{\text{adj}}=1.000$, 98\% CI: $[-4.730,\, 4.663]$). This suggests that the results are statistically inconclusive with respect to a null effect and that our intervention did not increase skepticism towards real images.

Similarly, we compare the discernment of real images between the control condition and the intervention conditions in the follow-up (see Figure~\ref{fig:acc_real_over_time}\textsf{\textbf{b}}). For this, we conducted equivalence tests \cite{Lakens.2018} with threshold [$-0.05,+0.05$] for the smallest meaningful effect{, and applied a Bonferroni correction for five comparisons, which corresponds to 98\% confidence intervals}. We again found that the equivalence tests were undecided and hence inconclusive with respect to the null effect (\emph{Textual}: $p_{\text{adj}}=1.000$, 98\% CI: $[-5.219,\, 6.663]$; \emph{Visual}: $p_{\text{adj}}=1.000$, 98\% CI: $[-4.685,\, 7.702]$; \emph{Gamified}: $p_{\text{adj}}=1.000$, 98\% CI: $[-7.193,\, 5.362]$; \emph{Feedback}: $p_{\text{adj}}=1.000$, 98\% CI: $[-8.888,\, 2.510]$; \emph{Knowledge}: $p_{\text{adj}}=1.000$, 98\% CI: $[-5.722,\, 6.421]$). The accuracy in detecting real images in the control group was comparable and non-significantly different for the main session and the follow-up (Wilcoxon signed rank test $W = 1527.5$, $p=0.309$). Similarly, we conducted a robustness check to control for possible learning effects of the task itself and found no significant difference (see Appendix~\ref{supp:learning_effect_robustness}). Overall, the results indicate that none of the interventions decreased participants' ability to detect real images, neither immediately after the intervention nor at the follow-up.

\subsubsection{Summary of hypothesis testing}
Our results provide mixed support for our hypotheses. Hypothesis \textbf{H1} is supported: both the \emph{Textual} and \emph{Visual} interventions significantly improved deepfake detection accuracy compared to the control condition. Hypothesis \textbf{H2} is also supported: equivalence tests indicate that none of our interventions increased skepticism toward real images, with all results falling in the inconclusive range rather than showing decreased trust in authentic content. Hypothesis \textbf{H3} receives partial support: while participants in the intervention conditions were better at discerning than the control condition at the two-week follow-up, these differences were no longer statistically significant.

\subsection{{Effects of interventions on confidence in discernment ability}}
\label{supp_confidence}

{After completing the discernment task, participants reported their confidence in their ability to distinguish between real and deepfake images on a slider scale ranging from 0 (not confident at all) to 100 (very confident). We first analyzed the correlation between participants' self-reported confidence and their discernment accuracy. Confidence showed a significant positive correlation with both real image discernment (Pearson's $r = 0.16$, $p < .001$) and deepfake image discernment (Pearson's $r = 0.12$, $p < .001$). While statistically significant, these correlations are relatively modest: participants with higher confidence tended to perform somewhat better, but the relationship was weak.}

To test the effects of our interventions on discernment ability, we conducted Mann-Whitney $U$ tests with Bonferroni correction to compare confidence levels between the control condition and each intervention condition. Table~\ref{tab:confidence_by_intervention} presents mean confidence levels, discernment accuracy scores, and $p_{\text{adj}}$-values across intervention conditions. Notably, the interventions that improved fake image discernment (\emph{Visual} and \emph{Gamified}) also showed slight increases in confidence, while the \emph{Knowledge} intervention increased confidence without improving fake image accuracy. This pattern suggests that effective interventions may enhance both performance and participants' metacognitive awareness of their abilities, though the confidence effects did not reach statistical significance in our study.

\begin{table*}[h]
\centering
\begin{tabular}{lcccc}
\hline
Condition & Confidence & Discernment real images & Discernment deepfakes images & $p_{\text{adj}}$-value \\
\hline
\emph{Control} & 68.04 (21.53) & 83.97 (19.77) & 61.33 (23.67) & -- \\
\emph{Textual} & 72.80 (20.66) & 82.99 (21.28) & 68.80 (23.52) & 0.076 \\
\emph{Visual} & 72.36 (20.21) & 79.99 (23.36) & 74.31 (20.02) & 0.346 \\
\emph{Gamified} & 68.20 (19.71) & 81.36 (22.77) & 65.70 (23.00) & 1.000 \\
\emph{Feedback} & 74.22 (16.15) & 84.26 (20.21) & 60.00 (22.18) & 0.092 \\
\emph{Knowledge} & 67.76 (22.53) & 84.00 (19.00) & 64.48 (21.92) & 1.000 \\
\hline
\end{tabular}
\caption{{\textbf{Mean confidence and discernment accuracy by intervention condition.} Standard deviations in parentheses.}}
\label{tab:confidence_by_intervention}
\end{table*}
 
\subsection{Effect of interventions on sharing intention}
\label{sec:sharing_intention}

Beyond our hypotheses on the effect of our interventions on discernment accuracy, we also explored whether the interventions affected the sharing intention of participants, as this is a key mechanism for misinformation spread online. For this, we asked participants whether they would share the images they saw during the discernment task. {Note that participants were unaware whether an image was real or a deepfake. Table~\ref{tab:sharing_by_discernment} shows the results.}

{Overall, our \emph{Visual}, \emph{Gamified}, and \emph{Knowledge} interventions significantly reduced sharing intention for deepfake images that participants correctly discerned as deepfakes (TN) (\emph{Visual}: Mann-Whitney $U=19456.5$, $p_{\text{adj}}<0.05$; \emph{Gamified}: $U=19459.0$, $p_{\text{adj}}<0.05$; \emph{Knowledge}: $U=19707.5$, $p_{\text{adj}}<0.05$). Hence, our interventions can not only improve image discernment but also potentially help reduce the spread of deepfake images online by address both the cognitive ability to detect and the intention to avoid spreading misinformation. At the same time, sharing intention for real images was not reduced significantly. Sharing intention effects did not persist at the two-week follow-up. }

Our findings reveal an important connection between discernment and sharing intention: participants were less likely to share images they discerned as deepfakes than images they discerned as real. This pattern suggests that accurate detection is a prerequisite for appropriate sharing decisions since participants cannot avoid sharing deepfakes they fail to identify. Even when interventions successfully discourage sharing of identified deepfakes, misclassified deepfakes will still spread. This reinforces our focus on improving discernment accuracy as the primary goal, with sharing behavior as a secondary outcome that depends on accurate detection. However, the effects did not persist at follow-up, which highlights a critical limitation: while short-term interventions can temporarily suppress sharing of detected deepfakes, maintaining this behavioral change may require ongoing reinforcement or stronger attitudinal shifts.

\begin{table*}[t]
\centering
\begin{tabular}{lrrrrrrrr}
\toprule
 & \multicolumn{4}{c}{\textbf{Main study}} & \multicolumn{4}{c}{\textbf{Follow-up}} \\
\cmidrule(lr){2-5}\cmidrule(lr){6-9}
\textbf{Condition} & \textbf{TN} & \textbf{FP} & \textbf{FN} & \textbf{TP} & \textbf{TN} & \textbf{FP} & \textbf{FN} & \textbf{TP} \\
\midrule
\emph{Control}   & 14.72 & 24.97 & 5.35 & 28.89 & 11.10 & 22.60 & 3.52 & 22.43 \\
\emph{Textual}   &  8.87 & 23.32 & 6.38 & 30.53 & 10.88 & 21.02 & 3.70 & 24.38 \\
\emph{Visual}    &  6.34$^*$ & 22.76 & 2.05 & 27.34 &  5.27 & 18.08 & 2.33 & 21.03 \\
\emph{Gamified}  &  7.73$^*$ & 24.00 & 2.63 & 24.94 &  8.62 & 24.24 & 2.34 & 22.02 \\
\emph{Feedback}  &  9.71 & 24.68 & 3.96 & 24.82 &  9.87 & 23.56 & 3.87 & 25.18 \\
\emph{Knowledge} &  8.81$^*$ & 24.29 & 3.42 & 23.10 &  9.54 & 20.18 & 4.51 & 22.69 \\
\midrule
\textbf{Avg.} & \textbf{9.36} & \textbf{24.00} & \textbf{4.04} & \textbf{26.06} & \textbf{9.21} & \textbf{21.61} & \textbf{3.37} & \textbf{22.96} \\
\bottomrule
\end{tabular}
\caption{{\textbf{Sharing intention conditional on correctness (mean across participants, \%).} 
TN = deepfake image discerned as deepfake; FP = deepfake image discerned as real; FN = real image discerned as deepfake; TP = real image discerned as real. Significance levels for Mann-Whitney $U$ tests: $^*{p_{\text{adj}}}<0.05$, $^{**}{p_{\text{adj}}}<0.01$, $^{***}{p_{\text{adj}}}<0.001$.}}
\label{tab:sharing_by_discernment}
\end{table*}

\subsection{Validation}
We conducted a preregistered validation study with $N=600$ participants from the U.S. After applying filtering, a final sample of $N=561$ participants remained, with people being filtered roughly equally from each condition. The response rate was 76.8\% overall (80.0\% for \emph{Control}, 81.7\% for \emph{Textual}, 87.4\% for \emph{Visual}, 75.3\% for \emph{Gamified}, 85.7\% for \emph{Feedback}, and 77.4\% for \emph{Knowledge}). Hence, the number of participants in the follow-up was relatively balanced across conditions, and we found similar response rates to the main study. Again, a significance level of {$\alpha_{\text{adj}} = 0.05$} was used for all statistical comparisons.  {See Table~\ref{tab:performance_across_condition_val} for an overview of the performance of participants across intervention conditions in the validation data.}

\begin{table*}[h]
\centering
\begin{tabular}{lrrrrrrrr}
\toprule
 & \multicolumn{4}{c}{\textbf{Main study}} & \multicolumn{4}{c}{\textbf{Follow-up}} \\
\cmidrule(lr){2-5}\cmidrule(lr){6-9}
\textbf{Condition} & \textbf{TN} & \textbf{TP} & \textbf{FPR} & \textbf{FNR} & \textbf{TN} & \textbf{TP} & \textbf{FPR} & \textbf{FNR} \\
\midrule
\emph{Control}   & 68.49 & 83.32 & 31.51 & 16.68 & 68.91 & 78.68 & 31.09 & 21.32 \\
\emph{Textual}   & 79.76 & 77.69 & 20.24 & 22.31 & 75.76 & 75.99 & 24.24 & 24.01 \\
\emph{Visual}    & \textbf{81.06} & 83.11 & 18.94 & 16.89 & \textbf{77.51} & 75.66 & 22.49 & 24.34 \\
\emph{Gamified}  & 77.69 & 74.82 & 22.31 & 25.18 & 75.07 & 79.43 & 24.93 & 20.57 \\
\emph{Feedback}  & 69.21 & \textbf{85.00} & 30.79 & 15.00 & 74.88 & \textbf{81.46} & 25.12 & 18.54 \\
\emph{Knowledge} & 71.78 & 80.22 & 28.22 & 19.78 & 69.32 & 78.54 & 30.68 & 21.46 \\
\bottomrule
\end{tabular}
\caption{{\textbf{Discernment performance by intervention in the validation study (mean across participants, \%).} 
{TN} = Deepfake image discernment accuracy; {TP} = Real image discernment accuracy;
{FPR} = discerned deepfake images as real; {FNR} = discerned real images as deepfakes.}}
\label{tab:performance_across_condition_val}
\end{table*}

\subsubsection{Discerning {deepfake images}}
The results of our validation study (see Table~\ref{tab:discerning_deepfakes_val}) confirm our exploratory findings for Hypothesis H1 from the main experiment. Specifically, our interventions \emph{Textual}, \emph{Visual}, and \emph{Gamified} significantly boosted participants' accuracy in discerning {deepfake images} compared to the control condition. The \emph{Feedback} and \emph{Knowledge} interventions {did not show significantly different discernment of deepfake images in comparison to the control condition}, which is consistent with our exploratory findings from the main experiment. Thus, our validation provides robust evidence that some of our interventions effectively boost short-term discernment abilities.

\begin{table*}[h]
\centering
\begin{tabular}{lccccccc}
\toprule
\textbf{Intervention} & \textbf{Mean Accuracy} & \textbf{Effect size (Cohen's $d$)} & \textbf{Mann-Whitney $U$} & {\textbf{$p_{\text{adj}}$-value}} & \\
\midrule
\emph{Textual} & 79.8\% & 0.59 & 2906.5 & $< 0.001$ & \includegraphics[scale=0.008]{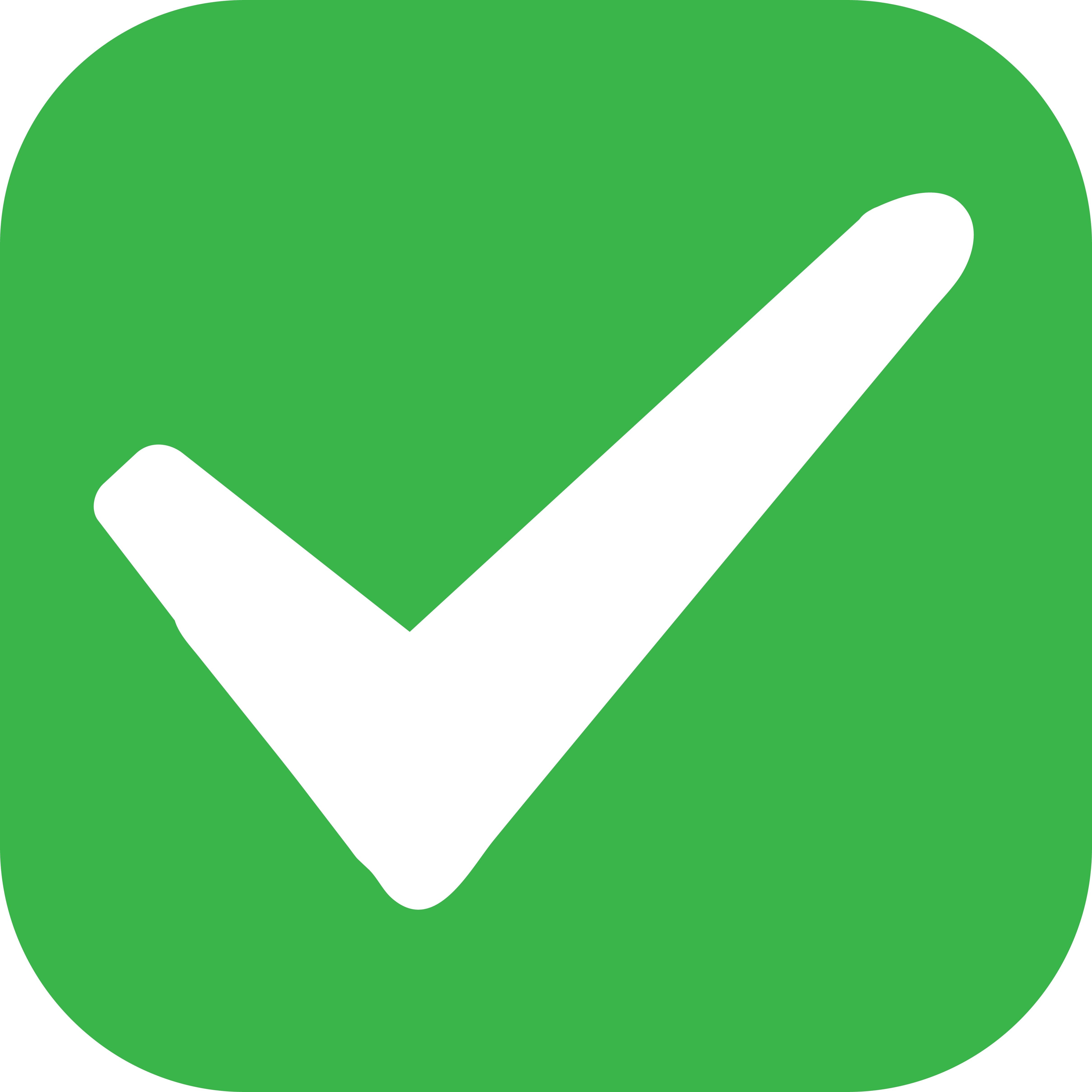} \\
\emph{Visual} & 81.1\% & 0.66 & 2625.5 & $< 0.001$ & \includegraphics[scale=0.008]{figures/tick-icon.png} \\
\emph{Gamified} & 77.7\% & 0.48 & 3175.5 & {$<0.01$}
& \includegraphics[scale=0.008]{figures/tick-icon.png} \\
\emph{Feedback} & 69.2\% & 0.03 & 4199.0 & {1.000}
& \includegraphics[scale=0.8]{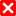} \\
\emph{Knowledge} & 71.8\% & 0.17 & 3943.5 & {1.000}
& \includegraphics[scale=0.8]{figures/x-mark.png} \\
\midrule
\multicolumn{4}{l}{\textbf{Control accuracy:} 68.5\%} \\
\bottomrule
\end{tabular}
\caption{\textbf{Results of the Mann-Whitney $U$ test comparing intervention conditions against the control condition at time point T1 (right after the intervention).} Checkmarks indicate significant improvement in discerning {deepfake images}. Significance levels for Mann-Whitney $U$ tests: {$p_{\text{adj}}<0.05$}.}
\label{tab:discerning_deepfakes_val}
\end{table*}

Similarly, the follow-up results of the validation study largely replicated the follow-up pattern observed in the main study for Hypothesis \textbf{H3} (see Table~\ref{tab:discerning_deepfakes_val_followup}). {The effects of the \emph{Textual}, \emph{Gamified}, \emph{Feedback} and \emph{Knowledge} interventions were not significantly different from the control condition in the follow up. This time, however, participants in the \emph{Visual} condition achieved significantly higher long-term accuracy than the control group.} Compared to the null findings of the main study, this suggests that the validation sample provided stronger evidence for lasting benefits of the \emph{Visual} intervention. One possible reason for this difference is variation in the sample composition between the studies, which may have allowed some effects to reach significance that did not in the main study.

\begin{table*}[h]
\centering
\begin{tabular}{lccccccc}
\toprule
\textbf{Intervention} & \textbf{Mean Accuracy} & \textbf{Effect size (Cohen's $d$)} & \textbf{Mann-Whitney $U$} & \textbf{$p$-value} & \\
\midrule
\emph{Textual} & 75.7\% & 0.41 & 2308.0 & {0.156}
& \includegraphics[scale=0.8]{figures/x-mark.png}  \\
\emph{Visual} & 77.5\% & 0.41 & 2407.0 & {$<0.05$}
& \includegraphics[scale=0.008]{figures/tick-icon.png} \\
\emph{Gamified} & 75.1\% & 0.29 & 2178.0 & {0.283}
& \includegraphics[scale=0.8]{figures/x-mark.png} \\
\emph{Feedback} & 74.9\% & 0.26 & 2514.5 & {0.386}
& \includegraphics[scale=0.8]{figures/x-mark.png} \\
\emph{Knowledge} & 69.3\% & 0.09 & 2652.5 & {1.000}
& \includegraphics[scale=0.8]{figures/x-mark.png} \\
\midrule
\multicolumn{4}{l}{\textbf{Control accuracy:} 68.9\%} \\
\bottomrule
\end{tabular}
\caption{\textbf{Results of the Mann-Whitney $U$ test comparing intervention conditions against the control condition at time point T2 (two weeks after the intervention).} Checkmarks indicate significant improvement in discerning {deepfake images}. Significance levels for Mann-Whitney $U$ tests: $p<0.05$.}
\label{tab:discerning_deepfakes_val_followup}
\end{table*}

\subsubsection{Discerning real images}
The results of our validation study (see Table~\ref{tab:discerning_real_images_val}) overall confirm our exploratory findings for Hypothesis \textbf{H2}.  The equivalence tests for the \emph{Textual}, \emph{Visual}, \emph{Feedback}, and \emph{Knowledge} interventions remained statistically indistinguishable from the control condition. This shows that participants in those conditions did not become more skeptical of real images which demonstrates the robustness of our interventions across both studies. Only the \emph{Gamified} intervention showed a significant difference from the exploratory results, with participants identifying real images less accurately in the validation sample. Taken together, these validation results broadly support our original conclusion that our digital literacy interventions do not make people more skeptical of real images.

\begin{table*}[h]
\centering
\setlength{\belowcaptionskip}{10pt}
\begin{tabular}{lccccccc}
\toprule
\textbf{Intervention} & \textbf{Mean Accuracy} & \textbf{Equivalence test} & \textbf{Confidence Interval ({98\%} CI)} & {\textbf{$p_{\text{adj}}$-value}} &   \\
\midrule
\emph{Textual} & 77.7\% & undecided & {[--1.889, 13.145]} & {1.000} & \includegraphics[scale=0.008]{figures/tick-icon.png} \\
\emph{Visual} & 83.1\% & undecided & {[--6.529, 6.951]} & {1.000} & \includegraphics[scale=0.008]{figures/tick-icon.png} \\
\emph{Gamified} & 74.8\% & significantly different & {[1.152, 15.837]} & {1.000} & \includegraphics[scale=0.8]{figures/x-mark.png} \\
\emph{Feedback} & 85.0\% & undecided & {[--8.347, 4.978]} & {1.000} & \includegraphics[scale=0.008]{figures/tick-icon.png} \\
\emph{Knowledge} & 80.2\% & undecided & {[--3.365, 9.567]} & {1.000} & \includegraphics[scale=0.008]{figures/tick-icon.png} \\
\midrule
\multicolumn{4}{l}{\textbf{Control accuracy:} 83.3\%} \\
\bottomrule
\end{tabular}
\caption{\textbf{Results of the equivalence tests comparing each intervention condition against the control condition at time point T1 (right after the intervention).} }
\label{tab:discerning_real_images_val}
\end{table*}

We also repeated the follow-up and the results of the validation study (see Table~\ref{tab:discerning_real_images_val_followup}) confirm our findings from the main study. Across all five interventions, the equivalence tests were again statistically undecided. This suggests that, consistent with the main study, our digital literacy interventions did not lead to increased skepticism toward authentic content over time. 

\begin{table*}[h]
\centering
\setlength{\belowcaptionskip}{10pt}
\begin{tabular}{lccccccc}
\toprule
\textbf{Intervention} & \textbf{Mean Accuracy} & \textbf{Equivalence test} & \textbf{Confidence Interval ({98\%} CI)} & {\textbf{$p_{\text{adj}}$-value}} &   \\
\midrule
\emph{Textual} & 75.9\% & undecided & {[--6.257, 11.652]} & {1.000} & \includegraphics[scale=0.008]{figures/tick-icon.png} \\
\emph{Visual} & 75.7\% & undecided & {[--5.775, 11.818]} & {1.000} & \includegraphics[scale=0.008]{figures/tick-icon.png} \\
\emph{Gamified} & 79.4\% & undecided & {[--9.002, 7.513]} & {1.000} & \includegraphics[scale=0.008]{figures/tick-icon.png} \\
\emph{Feedback} & 81.5\% & undecided & {[--10.844, 5.301]} & {1.000} & \includegraphics[scale=0.008]{figures/tick-icon.png} \\
\emph{Knowledge} & 78.5\% & undecided & {[--8.524, 8.809]} & {1.000} & \includegraphics[scale=0.008]{figures/tick-icon.png} \\
\midrule
\multicolumn{4}{l}{\textbf{Control accuracy:} 78.7\%} \\
\bottomrule
\end{tabular}
\caption{\textbf{Results of the equivalence tests comparing each intervention condition against the control condition at time point T2 (two weeks after the intervention).} }
\label{tab:discerning_real_images_val_followup}
\end{table*}

\subsubsection{{Sharing intention}}

{Our validation study (see Table~\ref{tab:sharing_by_discernment_validation}) confirms our exploratory findings for participants' sharing intention in that participants are more likely to share images they discerned as real than images they discerned as deepfakes. However, we did not find significant effects of our interventions on sharing intention. Notably, while the sharing intention rates from the participants in the validation data are overall lower than the sharing intention rates from our exploratory findings, participants in the \emph{Knowledge} intervention were more willing to share images than any other condition. Sharing intention was even significantly higher for deepfakes that were discerned as deepfakes (TN) in the follow-up (Mann-Whitney $U = 2274.0$, $p_{\text{adj}} < 0.05$). Since the procedure and materials were identical across the studies, this difference can most likely be attributed to between-sample differences. Additionally, we find that our interventions did not reduce sharing intention for real images.}

\begin{table*}[h]
\centering
\begin{tabular}{lrrrrrrrr}
\toprule
 & \multicolumn{4}{c}{\textbf{Main study}} & \multicolumn{4}{c}{\textbf{Follow-up}} \\
\cmidrule(lr){2-5}\cmidrule(lr){6-9}
\textbf{Condition} & \textbf{TN} & \textbf{FP} & \textbf{FN} & \textbf{TP} & \textbf{TN} & \textbf{FP} & \textbf{FN} & \textbf{TP} \\
\midrule
\emph{Control}   &  5.76 & 16.35 & 1.75 & 13.61 &  2.70 & 10.09 & 0.00 & 14.45 \\
\emph{Textual}   &  7.53 & 14.96 & 6.18 & 22.26 &  4.38 & 10.68 & 6.05 & 13.68 \\
\emph{Visual}    &  2.11 & 10.54 & 0.53 & 15.05 &  3.67 & 10.88 & 1.31 & 11.33 \\
\emph{Gamified}  &  2.47 &  9.64 & 0.00 & 13.96 &  4.42 & 11.19 & 4.29 & 11.31 \\
\emph{Feedback}  &  4.32 & 15.86 & 3.26 & 12.39 &  5.10 &  8.54 & 0.00 & 11.33 \\
\emph{Knowledge} &  9.83 & 25.20 & 6.09 & 23.37 & 11.74$^*$ & 24.59 & 4.17 & 25.44 \\
\midrule
\textbf{Average} &  \textbf{5.34} & \textbf{15.43} & \textbf{2.97} & \textbf{16.77} & \textbf{5.34} & \textbf{12.66} & \textbf{2.64} & \textbf{14.59} \\

\bottomrule
\end{tabular}
\caption{{\textbf{Sharing intention conditional on correctness (mean across participants, \%).} 
TN = deepfake image discerned as deepfake; FP = deepfake image discerned as real; FN = real image discerned as deepfake; TP = real image discerned as real. Significance levels for Mann-Whitney $U$ tests: $^*{p_{\text{adj}}}<0.05$, $^{**}{p_{\text{adj}}}<0.01$, $^{***}{p_{\text{adj}}}<0.001$.}}
\label{tab:sharing_by_discernment_validation}
\end{table*}

\section{Discussion}

We conducted an experiment with {$N=1,112$} participants and found that the \emph{Textual} and \emph{Visual} interventions significantly boosted {deepfake image discernment} by 7.5 percentage points and 13 percentage points, respectively (both {$p_{\text{adj}} < 0.001$}). {In contrast, the \emph{Knowledge}, \emph{Feedback}, and \emph{Gamified} interventions did not produce reliable improvements in immediate discernment. Intervention effects did not persist in the two-week follow-up.} At the same time, our interventions did not increase participants' skepticism toward real images{, an unintended effect of previous digital literacy tips interventions \cite{Guo.2025,Chen.2025}}. Our findings support Hypotheses \textbf{H1} and \textbf{H2}, with partial support for \textbf{H3}. Overall, our results suggest that our lightweight, easy-to-understand digital literacy interventions can boost {deepfake image discernment} without reducing trust in real images{, and that the way deepfake cues are presented (textual vs.\ visual vs.\ interactive) influences their effectiveness}.

\subsection{Implications for digital literacy intervention design}
Our findings provide several key insights for designing effective behavioral interventions against visual misinformation. {First, the stronger effect of the \emph{Visual} intervention compared to the \emph{Textual} intervention suggests that \emph{how} visual cues are explained matters. Both interventions introduced the same small set of deepfake image cues, but the \emph{Textual} intervention described them only with words (somewhat akin to an accuracy nudge), whereas the \emph{Visual} intervention paired the textual descriptions with concrete example images (somewhat akin to a pre-bunking intervention). The effects of the \emph{Visual} intervention, therefore, point to the added value of mapping textual cue descriptions onto concrete visual patterns, as was the case in prior HCI work on visual misinformation discernment \cite{Guo.2025} and synthetic faces \cite{Chen.2025}. Similarly, work on fact-checking also found that visual explanations were more effective in correcting misperceptions than textual explanations \cite{Dan.2025}. Our findings are consistent with multimedia learning research, which shows that combining verbal explanations with relevant visuals can facilitate understanding and application of new concepts \cite{Moreno.1999}. }

{Second, the effectiveness of the \emph{Textual} intervention demonstrates that explicit instruction about directed attention towards specific visual cues can still improve detection even without example images. This is important in light of current HCI work, where many interventions use example visuals \cite{Guo.2025, Hwang.2021} or even interactive interventions \cite{Chen.2025, Hu.2023}: our results show that clear, concise descriptions of what to look for in {deepfake images} already yield measurable benefits and can be delivered in low-friction formats such as short messages or posts. From a design perspective, this suggests that textual explanations should not be overlooked in favor of more technologically sophisticated interventions, especially when resources or attention are limited.}

{Third, the lack of reliable additional benefits for the \emph{Gamified}, \emph{Feedback}, and \emph{Knowledge} interventions within our study suggests that a small amount of practice, gamification, or general knowledge does not automatically translate into discernment gains. Our results on the \emph{Gamified} intervention are in contrast with previous work that found game-based inoculation more effective than visual-based inoculation \cite{Hu.2023}. This difference likely stems from fundamental differences in how gamification was implemented. Previous work (e.g., the ``Bad News'' game \cite{Roozenbeek.2019, Hu.2023}) used deep role-playing mechanics where participants actively simulated being misinformation spreaders, learning manipulation tactics through experiential engagement within an immersive game. In contrast, our \emph{Gamified} intervention added lightweight motivational elements (points, timer) to the same literacy tips from the \emph{Textual} and \emph{Visual} conditions that served primarily to sustain engagement during exposure rather than to teach through experience. This suggests that gamification used for motivation may produce different outcomes than gamification used as the core learning mechanism.}

{For the \emph{Feedback} intervention, previous work found that feedback trials on 128 images improved discernment of AI-synthesized faces; however, the resulting accuracy remained at chance-level \cite{Nightingale.2022}. Our feedback-based format exposed participants to only ten practice trials, which may be insufficient for error-based learning and calibration to take hold; more extensive or spaced practice might be required to see benefits. Our \emph{Knowledge} intervention showed that conceptual knowledge did not translate into improved discernment of deepfake images, which suggests a critical gap between understanding the generation process and possessing the perceptual skills needed to identify resulting artifacts. However, recent work found that implicit training, where participants observed the step-by-step transformation of real faces into AI-synthesized ones, improved detection accuracy \cite{Chen.2025}. This suggests that abstract conceptual knowledge about AI image generation is insufficient and can support discernment only when paired with concrete visual demonstrations of the artifacts that result from that process.}

{Taken together, these findings complement existing HCI work on misinformation interventions by highlighting the importance of both \emph{content} (clear, diagnostic cues) and \emph{medium} (text-only vs.\ visual examples vs.\ interactive formats). In our setting, the most effective design combined simple, easy-to-understand cue descriptions with concrete visual examples, while more complex interactive features did not yield additional short-term benefits.}

\subsection{Addressing the skepticism challenge in HCI}
A contribution of our work is demonstrating that digital literacy interventions can improve discernment without the unintended consequence of increased skepticism, a key challenge identified in prior HCI research on misinformation interventions \cite{Hoes.2024, Altay.2024, Guo.2025, Chen.2025}. {We used similarly lightweight media literacy tips as in recent work \cite{Guo.2025, Chen.2025}, but there are several important differences in how our interventions and settings are configured, which could explain why our intervention did not increase skepticism towards authentic content. First, the problem under study differs: Guo et al. \cite{Guo.2025} focus on belief in visual misinformation rather than image discernment. Their participants evaluate news headlines paired with images in a multimodal task, whereas we study deepfake image discernment in a single-medium, image-only classification task. This means that, in Guo et al., additional factors such as headline content and prior beliefs can shape skepticism, while our design isolates intervention effects on visual authenticity judgments. Second, the image domains differ: Chen et al. \cite{Chen.2025} focus exclusively on synthetic faces, whereas our images portray a broader range of everyday scenes, including cities, groups of people, and nature. Third, the intervention exposure differs: Chen et al. \cite{Chen.2025} repeatedly exposed participants to their artifact-focused intervention, which shifted people’s response bias toward judging more images as synthetic, while our study relies on a single, brief intervention without repeated training. Chen et al. \cite{Chen.2025} suggest that this bias is likely driven by the absence of real images in the training phase rather than by artifact instruction per se. Hence, future work should study how robust our effects are under repeated exposure, and whether this can be mitigated by including corresponding real images in the intervention materials. }

{Together, these differences in outcome measures, domain, and intervention design may help explain why our interventions did not produce the trust costs observed in previous work. More broadly, our results suggest that the skepticism side effect is not an unavoidable consequence of deepfake literacy tips; carefully targeting specific, diagnostically strong visual anomalies and presenting them in brief, image-focused interventions can improve detection without eroding trust in authentic images.}

\subsection{Practical deployment considerations}
Our interventions offer several practical advantages that support their real-world implementation. First, our toolbox of digital literacy interventions includes a variety of formats (i.e., textual, visual, gamified, etc.), which allows the interventions to be flexibly deployed and format-matched across different communication channels, such as private messages, social media posts, and advertising campaigns. {Textual tips can be embedded in low-friction formats such as push notifications, platform banners, or short social media posts, where space is limited and users are unlikely to engage with longer content. Visual formats that pair the same tips with annotated example images are well suited for contexts where images are already central, such as carousel posts, in-feed explainers, or illustrated modules in existing digital literacy curricula. }

{Second, all formats are lightweight and require no specialized technical infrastructure or access to proprietary detection systems. The interventions consist of short texts, static images, and simple interaction logic, making them easy to translate, adapt to different cultural contexts, and integrate into existing communication channels (e.g., school programs, NGO campaigns, or platform-hosted information hubs). This also means they can be updated as new characteristic errors of deepfake images emerge, without needing to retrain or redeploy automated detection models.}

{Third, the skills acquired through our interventions are under users’ direct control and can be applied even when platforms do not remove deepfake images or provide automated authenticity labels. In this sense, our interventions are complementary to platform-side detection efforts: they offer a user-focused layer of protection that can be deployed independently of platform cooperation and across different services and media environments. The lightweight nature of the interventions makes them suitable for integration into existing digital literacy programs, civic education, or social media platform features without substantial additional cost.}

\subsection{Limitations and future work}
As with any other study, ours is also subject to limitations. {First, our study focuses on deepfake images; however, images are just one part of AI-generated content, next to video, audio, and text, and image-specific training may not transfer to other modalities. This holds the potential risks that users may feel falsely secure about detecting other modalities or that they misapply image-based cues to other modalities, which can lead to missed detections or false skepticism. This leaves opportunities for future work to test the effectiveness of our interventions on other AI-generated content and whether simple guardrails could prevent misapplication. For example, image tips could be paired with explicit scope cues or pointers to use other methods, such as reverse image search, when users are unsure.

Second, we chose deepfake example images in our interventions to illustrate specific errors intended to teach generalizable cues; however, these examples were not fully photorealistic, which may have made them more obvious to participants and possibly impacted the learning effects. At the same time, the real images were photorealistic; this asymmetry may obscure effect estimates. Future work should balance photorealism across conditions or manipulate photorealism as an explicit factor.} 

Third, the experimental setup in our study does not fully replicate real-world online environments, where users are embedded in fast-paced and emotionally-charged contexts such as those of social media platforms. While such controlled experimental setups are common in research testing behavioral interventions designed for online settings \cite{Maertens.2025, Spampatti.2024}, future research could implement and evaluate our digital literacy interventions in field experiments. 

Fourth, our interventions are intentionally designed to be lightweight, minimally-disruptive, and scalable, which may have limited their effectiveness. Repeated exposure or multi-session training as in \cite{Maertens.2025} may lead to stronger or more persistent effects. {For example, future research should examine whether longer and more in-depth gamified trainings produce meaningful change in deepfake image detection. Moreover, our interventions differed in “dose” (time-on-task, cognitive load, and engagement), which may impact long-term effectiveness. Future work should equalize or control for exposure duration and cognitive load for comparisons across formats.} 

Fifth, while we assess the risk of increased skepticism toward real images, future research should more deeply examine possible downstream effects of deepfake literacy training on trust in institutions, media sources, or interpersonal communication.

Lastly, we measure whether people can be taught to better detect deepfake images, but we do not look into their reasoning for their image discernment choices. Knowing how people discern and what they base their decisions on could provide insights into participants' understanding of the interventions and verify the intervention effects. Future work should focus on this aspect by asking participants to explain their choices or by asking them to highlight the area of the image where they see AI-generated artifacts.

\subsection{Relevance of behavioral interventions in the face of technological advances}

Recent work has begun to explore AI literacy as a distinct competency area within the broader literacy framework \cite{Long.2020}. For example, some studies explore how users interact with AI-powered voice assistants \cite{Markus.2024}. Others show that short video trainings can improve self-efficacy when interacting with AI, although understanding and evaluation did not improve \cite{Cao.2025}. Again, others design interactive games to encourage reflection on prompting behaviors for generative AI \cite{Ma.2025}. While valuable for understanding user–AI interaction more generally, these studies focus on system-level knowledge and interaction patterns rather than the specific visual analysis skills needed to detect {deepfake images}.

Our research remains highly relevant, even as AI tools continue to advance and {deepfake images} become more realistic. First, the cognitive skills targeted by our interventions, such as visual discrimination, analytical judgment, and contextual reasoning, are not tied to any specific generation technique. Instead, they directly mitigate the potential for deception, which means that these skills remain applicable even as surface-level flaws in {deepfake images} diminish \cite{Diakopoulos.2021, Goldstein.2023}. Second, many disinformation campaigns do not strive for perfect realism; instead, in practice, malicious actors often use lower-quality models that are faster and cheaper to mislead in low-attention or emotionally charged environments \cite{Labuz.2024}. Third, there are growing regulatory efforts aimed at restricting access to advanced generative models or limiting the capabilities of these models \cite{RomeroMoreno.2024}. If such policies succeed, many real-world {deepfake images} may continue to be produced with mid-tier tools where human detection remains effective. Fourth, the intervention effects we study are not only useful for visual content; they may transfer to other forms of synthetic media, such as AI-generated audio or video, which are increasingly part of disinformation campaigns. Taken together, these points suggest that digital literacy, rather than automated detection tools alone, will play a crucial part in the long-term response to {deepfake images}.

\section{Conclusion}
AI-generated disinformation poses a serious threat to information integrity, such as eroding public trust, fueling disinformation campaigns, and enabling political manipulation \cite{Diakopoulos.2021}. In response, our study demonstrates that short, easy-to-understand digital literacy interventions can boost people’s ability to discern {deepfake images} without diminishing their trust in authentic content. Our findings offer actionable insights for researchers, educators, and platform designers seeking to counter {deepfake images} through evidence-based behavioral interventions. As generative AI technology continues to advance, empowering users with practical, transferable detection skills will be crucial for safeguarding public discourse and fostering resilient societies.

\section*{Acknowledgments}

Funding by the German Research Foundation (Grant: 543018872) is acknowledged.

\section*{Data availability}

The code, anonymized data, and QSF files for our analyses are available in a public GitHub repository: \url{https://github.com/DominiqueGeissler/digital_literacy_interventions_for_deepfake_discernment}.

\newpage
\bibliographystyle{ACM-Reference-Format}
\bibliography{literature}

\newpage
\appendix
\onecolumn

\setcounter{table}{0}
\renewcommand{\thetable}{S\arabic{table}}

\setcounter{figure}{0}
\renewcommand{\thefigure}{S\arabic{figure}}

\section{Variables}

\begin{spacing}{0.9}
\begin{table}[ht]
    \centering
    \begin{scriptsize}
    \begin{tabular}{p{0.2\linewidth}p{0.4\linewidth}p{0.4\linewidth}}
    \toprule
        \textbf{Variable} & \textbf{Question} &\textbf{Options} \\
    \midrule
        \emph{Discernment}$^\dagger$ & \emph{\say{Is this image real or fake?}}& definitely fake / probably fake / probably real / definitely real \\
        \emph{SharingIntention} & \emph{\say{Would you share this image on your social media?}}& yes (1) / no / don't know (0) \\
        \emph{Confidence} & \emph{\say{How confident are you about your fake image detection choices?}}& not confident at all (0) / ... / very confident (100) \\
    \bottomrule
    \multicolumn{3}{l}{$^\dagger$Encoded as discernment accuracy for each participant in the analysis}
    \end{tabular}
    \caption{Dependent variables.}
    \label{tab:dependent_variables}
    \end{scriptsize}
\end{table}
\end{spacing}

\newpage

\begin{scriptsize}
\begin{spacing}{1}
    \begin{longtable}{@{}p{0.2\linewidth}p{0.4\linewidth}p{0.4\linewidth}@{}}

\toprule
\textbf{Variable} & \textbf{Question} & \textbf{Options} \\
\midrule
\endfirsthead

\multicolumn{3}{@{}l}{\tablename\ \thetable{} -- continued from previous page} \\
\toprule
\textbf{Variable} & \textbf{Question} & \textbf{Options} \\
\midrule
\endhead

\midrule
\multicolumn{3}{r@{}}{Continued on next page} \\
\endfoot

\bottomrule
\multicolumn{3}{p{\linewidth}}{$^*$We grouped ethnicities with lower counts into one group.}\\
\multicolumn{3}{p{\linewidth}}{$^{**}$ In our study, only 10 subjects identified themselves as being different from male or female and 1 subject preferred not to answer. Therefore, in our analysis, we encoded Gender as a binary variable and removed other observations from the analysis due to the low statistical power.}\\
\multicolumn{3}{p{\linewidth}}{$^{***}$ In our study, only 6 subjects did not answer and 1 subject had less than 7 years of formal education. Therefore, in our analysis, we removed these observations.}\\
\multicolumn{3}{p{\linewidth}}{$^\dagger$We encoded religion as a binary variable and removed observations where subjects preferred not to answer.}\\
\multicolumn{3}{p{\linewidth}}{$\ddagger$The variables were averaged into a single score that represents political leaning encoded as very liberal (1) / liberal (2) / somewhat liberal (3) / neutral (4) / somewhat conservative (5) / conservative (6) / very conservative (7).} \\
\multicolumn{3}{p{\linewidth}}{$\S$We transform our income values into three categories of incomes: low, middle, and high. This coding is based on \cite{Bennett.2021}, and we use \$80,000 as the threshold between middle and high. We removed observations where subjects preferred not to answer.}\\
\caption{Covariates.}
\label{tab:independent_variables}\\
\endlastfoot

        \emph{Sociodemographics} && \\
    \midrule
        Age & \emph{\say{How old are you? (in years)}} & Continuous variable \\
        Ethnicity$^*$ & \emph{\say{Select all that apply to you}} & White (0) / blue or African American (1) / Hispanic, Latino or Spanish origin (2) / Asian (3) / American Indian or Alaska Native / Middle Eastern or North African / Native Hawaiian or Other Pacific Islander / some other race, ethnicity, or origin (4) \\
         Gender$^{**}$ & \emph{\say{What is your gender?}} & male (0) / female (1) / non-binary / third gender / other / prefer not to say \\
         Education$^{***}$ & \emph{\say{How many years of formal education have you completed?}} &  7--12 (up to high school) (0) / 13--16 (college / undergraduate university / certificate training) (1) / more than 17 years (doctorate degree, medical degree, etc.) (2) / 0--6 (up to primary school) / prefer not to say \\
         Religion$^\dagger$ & \emph{\say{What describes you best?}} & a religious person (0) / not a religious person (1) / prefer not to say \\
         Social orientation$^\ddagger$ & \emph{\say{What is your political orientation for social issues? For example in health care, education, etc\newline Here: \say{liberal} means classically left-wing; \say{conservative} means classically right-wing}} & very liberal (1) / liberal (2) / somewhat liberal (3) / neutral (4) / somewhat conservative (5) / conservative (6) / very conservative (7)\\
         Economic orientation$^\ddagger$ & \emph{\say{What is your political orientation for economic issues? For example in taxes, etc\newline Here: \say{liberal} means classically left-wing; \say{conservative} means classically right-wing}} & very liberal (1) / liberal (2) / somewhat liberal (3) / neutral (4) / somewhat conservative (5) / conservative (6) / very conservative (7)\\
         Income$^\S$ & \emph{\say{What is your total yearly family/household income?}} &  [\$10,000, \$19,999] / [\$20,000, \$29,999] / [\$30,000, \$39,999] / [\$40,000, \$49,999] / [\$50,000, \$59,999] / [\$60,000, \$69,999] / [\$70,000, \$79,999] / [\$80,000, \$89,999] / [\$90,000, \$99,999] / [\$100,000, \$109,999] / [\$110,000, \$119,999] / [\$120,000, \$129,999] / [\$130,000, \$139,999] / [\$140,000, \$149,999] / [\$150,000 or more]\newline encoded as [\$10,000, \$29,999]: low (0) / [\$30,000, \$79,999]: middle (1) / $\geq$ \$80,000: high (2)\\
        Status & \emph{\say{Think of this ladder as representing where people stand in your country. \newline Where would you place yourself on this ladder? (10 equals highest, 1 equals lowest)}} & 1 (0) / \ldots / 10 (9) \\
    \midrule
        \emph{Social media use} &&\\
    \midrule
        PlatformCount & \emph{\say{Which social media platforms do you use (if any)?}} & Facebook / Twitter/X / Snapchat / Instagram / WhatsApp / TikTok / other / none; \newline encoded as count \\
         SharingCount & \emph{\say{Which type of content would you consider sharing on social media (if any)?}} & political news / sports news / celebrity news / science/technology news / business news / personal content / other / none; \newline encoded as count \\
         TimeOnline & \emph{\say{How much time (in hours) on average do you spend on social media daily?}} & Continuous variable \\
    \midrule
        \emph{Digital literacy} &&\\
    \midrule
        KnowledgeDeepfakes & \emph{\say{How do you rate your knowledge about deepfakes?}} & no knowledge at all (0) / \ldots / very knowledgeable (7)\\
        ExpDetecting & \emph{\say{How do you rate your experience with detecting deepfakes?}} & no experience at all (0) / \ldots / very experienced (7) \\
        ExpSearchEngines & \emph{\say{How do you rate your experience with using online search engines? For example with Google, Bing, etc}} & no experience at all (0) / \ldots / very experienced (7) \\ \\
        ExpImageSearch & \emph{\say{How do you rate your experience with using reverse image search?}} & no experience at all (0) / \ldots / very experienced (7) \\
        ExpGenAI &  \emph{\say{How do you rate your experience with using AI image generators. For example with DALL-E, Midjourney, etc}} & no experience at all (0) / \ldots / very experienced (7) \\
    \midrule
        CRT & \emph{\say{The ages of Mark and Adam add up to 28 years total. Mark is 20 years older than Adam. How old is Adam?}\newline \say{If it takes 10 seconds for 10 printers to print out 10 pages of paper, how many seconds will it take 50 printers to print out 50 pages of paper?}\newline \say{On a loaf of bread, there is a patch of mold. Every day, the patch doubles in size. If it takes 40 days for the patch to cover the entire loaf of bread, how many days would it take for the patch to cover half of the loaf of bread?}} & all incorrect (0) / \ldots / all correct (3)\\
    \end{longtable}
\end{spacing}\end{scriptsize}

\newpage
\section{Example discernment task}
\label{sec:example_discernment_task}

For the image discernment task, we show 15 images to each participant and ask them about the image's veracity as well as their willingness to share the image. In addition, we include an optional question where participants can tick a checkbox if they had trouble seeing the image or if they have seen the image before. A screenshot of the question setup can be found in Figure~\ref{fig:example_discernment_questions} {in the Appendix}.

\begin{figure}[H]
    \centering
    \fbox{\includegraphics[width=0.5\linewidth]{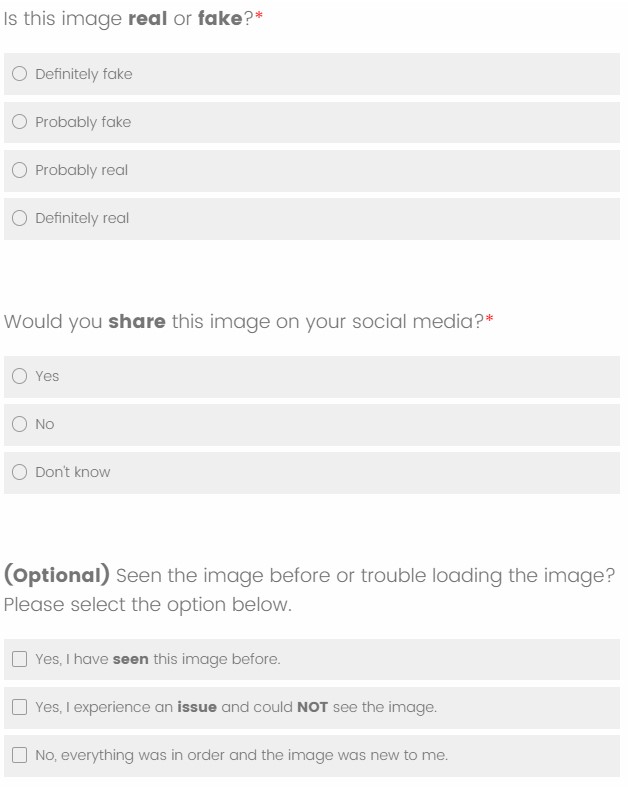}}
    \caption{\textbf{Screenshot of the question setup for each image rating task.}}
    \label{fig:example_discernment_questions}
    \Description[short explanation]{The figure shows a screenshot of the image-level survey screen used in the discernment task. At the top, participants answer “Is this image real or fake?” using four single-choice radio options: Definitely fake, Probably fake, Probably real, Definitely real (question marked with an asterisk to indicate it is required). Below, a second required question asks, “Would you share this image on your social media?” with answer options Yes, No, Don’t know. A final optional checklist records context/technical issues with three checkboxes: Yes, I have seen this image before; Yes, I experienced an issue and could NOT see the image; No, everything was in order and the image was new to me.}
\end{figure}

\newpage
\section{{Attention checks}}
\label{sec:attention_checks}

\begin{spacing}{0.9}
\begin{table}[ht]
    \centering
    \begin{scriptsize}
    \begin{tabular}{p{0.7\linewidth}p{0.3\linewidth}}
    \toprule
        \textbf{Question} &\textbf{Options} \\
    \midrule
        \emph{\say{Participants in surveys have a wide variety of interests and participate in a wide variety of sports. We appreciate your help with this survey and are interested in whether you take enough time to read the survey directions and questions carefully before you provide your answers. In order to demonstrate that you have read these instructions carefully, choose the I do not play any sports answer. Which sports have you been asked to select?}}& Basketball / Hockey / Baseball / Volleyball / Soccer / Tennis / I do not play any sports \\
    \midrule
        \emph{\say{Most modern theories of decision making recognize that decisions do not take place in a vacuum. Individual preferences and knowledge, along with situational variables can greatly impact the decision process. To demonstrate that you've read this much, just go ahead and select both red and green among the alternatives below, no matter what your favorite color is. What color are you asked to select?}} & White / blue / Red / Pink / Green / blue\\
    \midrule
        \emph{\say{Environmental issues are increasingly becoming a global concern, and many people are more mindful of how their actions impact the planet. We appreciate your help with this survey and are interested in whether you take enough time to read the survey directions and questions carefully before you provide your answers. To confirm that you're reading the instructions, please choose both 'Recycling' and 'Composting' from the options below, even if you do not participate in either. What actions are you asked to select?}}& Eating less meat / Donating clothes / Reducing water usage / Using public transportation / Composting / Recycling / I do not take any specific actions\\
    \midrule
        \emph{\say{Reading is a great way to gain knowledge, relax, and explore new ideas. As you proceed with this survey, it's important to carefully read and follow instructions. To demonstrate that you’re paying attention, please select the option 'I prefer audiobooks' in response to the following question. What should you choose to indicate you’ve read these instructions carefully?}} & Fiction / Non-fiction / Comics / Magazines / I prefer audiobooks / I don't enjoy reading \\
    \bottomrule
    \end{tabular}
    \caption{{Attention checks.}}
    \label{tab:attention checks}
    \end{scriptsize}
\end{table}
\end{spacing}

\newpage 

\section{Normality check}
\label{sec:normality_check}

To assess whether parametric statistical tests are appropriate, we conducted Shapiro-Wilk tests for normality on all outcome variables across experimental conditions. Table~\ref{tab:normality_tests} shows that none of the outcome measures (real image discernment, {deepfake image discernment}, real image sharing intention, deepfake sharing intention) were normally distributed in any condition (all $p < 0.001$). Specifically, this was the case for the control condition as well as all intervention conditions: \emph{Textual}, \emph{Visual}, \emph{Gamified}, \emph{Feedback}, and \emph{Knowledge}. Given these violations of the normality assumption, we used the Mann-Whitney $U$ test as a non-parametric alternative to the independent samples $t$-test in order to compare differences between conditions.

\begin{table}[htbp]
\centering
\begin{scriptsize}
\begin{tabular}{llcc}
\toprule
\textbf{Condition} & \textbf{Variable} & \textbf{$W$-statistic} & \textbf{$p$-value} \\
\midrule
\emph{Control}        & Real image discernment       & 0.7756 & < 0.001 \\
\emph{Textual}        & Real image discernment       & 0.7756 & < 0.001 \\
\emph{Visual}         & Real image discernment       & 0.8038 & < 0.001 \\
\emph{Gamified}       & Real image discernment       & 0.7887 & < 0.001 \\
\emph{Feedback}       & Real image discernment       & 0.7583 & < 0.001 \\
\emph{Knowledge}      & Real image discernment       & 0.7863 & < 0.001 \\
\addlinespace
\emph{Control}        & {Deepfake image discernment}       & 0.9666 & < 0.001 \\
\emph{Textual}        & {Deepfake image discernment}       & 0.9368 & < 0.001 \\
\emph{Visual}         & {Deepfake image discernment}       & 0.9280 & < 0.001 \\
\emph{Gamified}       & {Deepfake image discernment}       & 0.9540 & < 0.001 \\
\emph{Feedback}       & {Deepfake image discernment}       & 0.9667 & < 0.001 \\
\emph{Knowledge}      & {Deepfake image discernment}       & 0.9641 & < 0.001 \\
\addlinespace
\emph{Control}        & Real image sharing intention   & 0.7472 & < 0.001 \\
\emph{Textual}        & Real image sharing intention   & 0.7551 & < 0.001 \\
\emph{Visual}         & Real image sharing intention   & 0.7144 & < 0.001 \\
\emph{Gamified}       & Real image sharing intention   & 0.7011 & < 0.001 \\
\emph{Feedback}       & Real image sharing intention   & 0.7022 & < 0.001 \\
\emph{Knowledge}      & Real image sharing intention   & 0.6932 & < 0.001 \\
\addlinespace
\emph{Control}        & Deepfake sharing intention   & 0.7416 & < 0.001 \\
\emph{Textual}        & Deepfake sharing intention   & 0.6958 & < 0.001 \\
\emph{Visual}         & Deepfake sharing intention   & 0.6778 & < 0.001 \\
\emph{Gamified}       & Deepfake sharing intention   & 0.7187 & < 0.001 \\
\emph{Feedback}       & Deepfake sharing intention   & 0.7065 & < 0.001 \\
\emph{Knowledge}      & Deepfake sharing intention   & 0.6923 & < 0.001 \\
\bottomrule
\end{tabular}
\caption{Shapiro-Wilk normality test results for all outcome variables across conditions.}
\label{tab:normality_tests}
\end{scriptsize}
\end{table}

\newpage
\section{Robustness checks}
\label{supp:robustness_checks_results}

\subsection{Robustness check including participants failing attention and honesty checks}
\label{supp:robustness_filtering}

We repeated the analyses from Section~\ref{sec:analysis_plan} while including the participants who failed both attention and/or both honesty checks. For {deepfake image discernment}, we find robust evidence of our findings. In particular, the effects of our \emph{Textual} and \emph{Visual} interventions remain significant {(\emph{Textual}: Mann-Whitney $U = 14306.5$, $p_{\text{adj}} < 0.01$; \emph{Visual}: Mann-Whitney $U = 11696.0$, $p_{\text{adj}} < 0.001$; \emph{Gamified}: Mann-Whitney $U = 15298.5$, $p_{\text{adj}} = 0.264$; \emph{Feedback}: Mann-Whitney $U = 17880.0$, $p_{\text{adj}} = 1.000$; \emph{Knowledge}: Mann-Whitney $U = 15970.5$, $p_{\text{adj}} = 0.853$)}.

We also repeated the equivalence tests to check for meaningful differences in the discernment ability for real images using a threshold of $[-0.05, +0.05]$. {To control family-wise error across the five comparisons, we applied a Bonferroni correction (per-comparison $\alpha_{\text{adj}}=0.01$) and report simultaneous two-sided $98\%$ CIs.} The equivalence tests for each intervention were undecided {(\emph{Textual}: $p_{\text{adj}}=1.000$, 98\% CI: $[-3.883,\, 6.011]$; \emph{Visual}: $p_{\text{adj}}=1.000$, 98\% CI: $[-0.723,\, 9.925]$; \emph{Gamified}: $p_{\text{adj}}=1.000$, 98\% CI: $[-2.476,\, 7.862]$; \emph{Feedback}: $p_{\text{adj}}=1.000$, 98\% CI: $[-5.045,\, 4.626]$; \emph{Knowledge}: $p_{\text{adj}}=1.000$, 98\% CI: $[-4.636,\, 4.739]$). Thus, we find no evidence that any intervention altered discernment of authentic images relative to control after multiplicity adjustment.} This suggests that our findings are largely robust. 

To check for robustness of our follow-up results, we also repeat the Mann-Whitney $U$ tests for {deepfake image discernment} in the follow-up without excluding participants who failed the attention and honesty checks. We find robust results of our findings {(\emph{Textual}: Mann–Whitney $U=8635.5$, $p_{\text{adj}}=0.527$; \emph{Visual}: $U=7724.0$, $p_{\text{adj}}=0.554$; \emph{Gamified}: $U=9132.5$, $p_{\text{adj}}=1.000$; \emph{Feedback}: $U=9654.5$, $p_{\text{adj}}=1.000$; \emph{Knowledge}: $U=8915.5$, $p_{\text{adj}}=1.000$).}

\newpage
\subsection{Discernment accuracy for previously seen images or loading issues}
\label{supp:robustness_checkbox_filtering}
In the main analysis, we excluded {individual} responses where participants indicated they had previously seen the image or experienced loading issues{, but retained participants' remaining responses for other images.}. {On average, 18 responses per image were excluded (14 for real images and 20 for fake images). As each image was rated by 600 participants in the main study, this represents an almost negligible exclusion rate of only 2--3\% per image.} Here, we repeated the analysis without excluding these responses and re-computed discernment accuracy for both real images and {deepfake images}. We again test for differences in discernment of {deepfake images} for each intervention against the control group using Mann-Whitney $U$ tests {and adjusted $p$-values}. We find that the effects of our \emph{Textual} and \emph{Visual} interventions remain significant {(\emph{Textual}: Mann-Whitney $U=58160.0$, $p_{\text{adj}} < 0.001^{***}$;  \emph{Visual}: $U = 47543.5$, $p_{\text{adj}} < 0.001^{***}$). } Further, our findings remain robust ({\emph{Gamified}: Mann-Whitney $U = 62189.0$, $p_{\text{adj}} =0.068$; \emph{Feedback}: $U = 73192.5$, $p_{\text{adj}} = 1.000$; and \emph{Knowledge}: $U = 64471.0$, $p_{\text{adj}} = 0.407$).}

For the discernment of real images, we again repeat the equivalence tests using a threshold of $[-0.05, +0.05]$ {with a Bonferroni correction (per-comparison $\alpha_{\text{adj}}=0.01$) and report simultaneous two-sided $98\%$ CIs.}. Our findings are largely robust as we find that the equivalence tests for each intervention were undecided ({\emph{Textual}: $p_{\text{adj}} = 1.000$, 98\% CI: [$-2.463, 4.416]$; \emph{Gamified}: $p_{\text{adj}} = 1.000$, 98\% CI: $[-1.081, 6.115]$; \emph{Feedback}: $p_{\text{adj}} = 1.000$, 98\% CI: $[-3.503, 3.231]$; \emph{Knowledge}: $p_{\text{adj}} = 1.000$, 98\% CI: $[-3.272, 3.193]$}). Only, intervention \emph{Visual} showed a significant difference ({$p_{\text{adj}} = 1.000$, 98\% CI: $[0.198, 7.491]$}). 

We also repeat the Mann-Whitney $U$ tests for the discernment of {deepfake images} in the follow-up while keeping the responses where participants indicated they had previously seen the image of experienced loading issues. {We find results that are robust with our main study (\emph{Textual}: Mann–Whitney $U=34853.5$, $p_{\text{adj}}=0.166$; \emph{Visual}: $U=31064.0$, $p_{\text{adj}}=0.152$; \emph{Gamified}: $U=36761.5$, $p_{\text{adj}}=1.000$; \emph{Feedback}: $U=38855.5$, $p_{\text{adj}}=1.000$; \emph{Knowledge}: $U=35870.0$, $p_{\text{adj}}=1.000$).}

\newpage
\subsection{Regression analysis with sociodemographic controls}

We now conduct a regression analysis where we control for sociodemographic covariates, namely, gender, age, ethnicity, education, religion, political orientation, and income (as defined in Table~\ref{tab:independent_variables} {in the Appendix}). We removed observations with low-frequency categories or missing responses (e.g., because participants answered \say{prefer not to say}). That is, we removed 10 observations where participants identified as non-binary/Third gender/Other due to the limited sample size in the study and 1 participant who preferred not to report their gender. We also excluded observations with missing data on education (6 people) and those with less than 7 years of formal education (1 person). Similarly, we removed participants who preferred not to answer on their income (23 people) and religion (51 people), as well as 2 people who reported to be younger than 18 (see Table~\ref{tab:demographics} for more details on the distribution of sociodemographic in our experiment). After removing these observations, $N=1034$ participants remained for the regression analysis.

Table~\ref{tab:regression_demo} reports the results from the OLS regression model (as defined in Equation~\ref{eq:covariates}) for the different interventions. Model (1)~\textsc{Overall} estimates the effect of each of our interventions (all included in the model as a binary variable). In models (2)~\textsc{Textual}, (3)~\textsc{Visual}, (4)~\textsc{Gamified}, (5)~\textsc{Feedback}, and (6)~\textsc{Knowledge}, we only include individuals from the respective intervention condition and the control condition to isolate the effect of our intervention on the outcome. Model (7)~\textsc{Control} estimates direct effects of sociodemographics on the outcome within the control condition.  We find that the estimated coefficients for the interventions remain robust. We find that religion tends to play a significant role in the discernment ability of participants and that non-religious people tend to be better at discerning {deepfake images}.

\begin{spacing}{0.9}
\setlength{\belowcaptionskip}{10pt}
\begin{scriptsize}
\begin{longtable}{lcccccc|c}
\hline
 & \textsc{Overall} & \textsc{Textual} & \textsc{Visual} & \textsc{Gamified} & \textsc{Feedback} & \textsc{Knowledge} & \textsc{Control} \\
\hline
\endfirsthead

\multicolumn{8}{c}{\tablename\ \thetable{} -- continued from previous page} \\
\hline
 & Overall & Textual & Visual & Gamified & Feedback & Knowledge & Control \\
\hline
\endhead

\hline \multicolumn{8}{r}{Continued on next page} \\
\endfoot

\hline

\caption{\textbf{Regression results for discernment of {deepfake images} while controlling for sociodemographics.} Parentheses report standard errors. Variance inflation factor (VIF) values for the predictors in the models did not exceed 10, indicating that multicollinearity is not a major concern \cite{Obrien.2007} and that intervention effects remain reliable. Significance levels:  $^*p<0.05$, $^{**}p<0.01$, $^{***}p<0.001$.} \\
\label{tab:regression_demo}
\endlastfoot

Intercept & 63.400$^{***}$ & 67.954$^{***}$ & 66.526$^{***}$ & 55.851$^{***}$ & 64.323$^{***}$ & 67.370$^{***}$ & 66.622$^{***}$ \\
 & (3.866) & (6.761) & (6.124) & (6.845) & (6.215) & (6.279) & (9.254) \\
\emph{Textual} & 8.112$^{***}$ & 8.089$^{**}$ &  &  &  &  &  \\
 & (2.423) & (2.597) &  &  &  &  &  \\
\emph{Visual} & 13.491$^{***}$ &  & 13.493$^{***}$ &  &  &  &  \\
 & (2.443) &  & (2.402) &  &  &  &  \\
\emph{Gamified} & 5.670$^{*}$ &  &  & 5.344$^{*}$ &  &  &  \\
 & (2.436) &  &  & (2.517) &  &  &  \\
\emph{Feedback} & --0.832 &  &  &  & --1.245 &  &  \\
 & (2.409) &  &  &  & (2.488) &  &  \\
\emph{Knowledge} & 3.438 &  &  &  &  & 3.326 &  \\
 & (2.436) &  &  &  &  & (2.480) &  \\
\emph{Gender} & --0.172 & 1.316 & --4.571 & --2.813 & --0.533 & --3.540 & --4.495 \\
 & (1.418) & (2.645) & (2.423) & (2.558) & (2.549) & (2.503) & (3.742) \\
\emph{Age} & --0.024 & --0.128 & --0.167 & 0.050 & --0.154 & --0.126 & --0.224 \\
 & (0.054) & (0.101) & (0.093) & (0.096) & (0.092) & (0.087) & (0.132) \\
\emph{Ethnicity} & 0.243 & --0.883 & --2.249$^{*}$ & --1.335 & --0.112 & --0.559 & --3.141 \\
 & (0.657) & (1.272) & (1.089) & (1.233) & (1.184) & (1.226) & (1.813) \\
\emph{Education} & --1.178 & 0.075 & 1.828 & 2.206 & 0.647 & 2.836 & 4.415 \\
 & (1.228) & (2.197) & (2.044) & (2.160) & (2.068) & (2.189) & (3.020) \\
\emph{Religion} & 5.387$^{***}$ & 5.005 & 6.141$^{*}$ & 7.581$^{**}$ & 6.292$^{*}$ & 7.061$^{*}$ & 8.021 \\
 & (1.561) & (2.954) & (2.646) & (2.897) & (2.796) & (2.757) & (4.148) \\
\emph{PoliticalOrientation} & --0.186 & --0.031 & 0.686 & 0.242 & 0.738 & --0.017 & 1.217 \\
 & (0.410) & (0.788) & (0.685) & (0.750) & (0.743) & (0.727) & (1.089) \\
\emph{Income} & --1.726 & --3.446 & --1.919 & --1.338 & --2.452 & --3.944$^{*}$ & --4.103 \\
 & (1.047) & (1.924) & (1.755) & (1.865) & (1.820) & (1.880) & (2.664) \\
\midrule
$R^2$ & 0.068 & 0.058 & 0.128 & 0.046 & 0.025 & 0.054 & 0.070 \\
Adj. $R^2$ & 0.057 & 0.035 & 0.107 & 0.023 & 0.002 & 0.032 & 0.029 \\
\midrule
$N$ & 1034 & 345 & 338 & 339 & 348 & 340 & 169 \\
\end{longtable}
\end{scriptsize}
\end{spacing}

\newpage 
\subsection{Regression analyzing the varying effectiveness of interventions across sociodemographics}

We now control for sociodemographic covariates and their possible interaction effects with our interventions. Table~\ref{tab:regression_demo_interactions} shows the results of our OLS regression models with interaction terms for participant-specific sociodemographic controls as defined in Equation~\ref{eq:covariates_interactions}. We find that the estimated coefficients for the interventions remain robust. {The interaction terms provide exploratory evidence on which groups benefit more (or less) from specific intervention formats.}  Overall, we again find evidence that our interventions are more effective for non-religious participants. {Beyond this general pattern, several interventions appear to be differentially effective across groups. The \emph{Textual} intervention is significantly more effective for female participants and less effective for participants with higher educational attainment, suggesting that concise, text-based tips may be particularly useful for groups that are not already highly educated in related domains. The \emph{Gamified} intervention is more effective for older participants, indicating that adding light game elements can help close an age-related gap by making practice with the cues more engaging for older users. For the \emph{Feedback} and \emph{Knowledge} interventions, interaction terms with ethnicity are positive and significant, suggesting that these formats may be relatively more beneficial for non-white participants.}

{These interaction effects suggest that intervention effectiveness is moderated by individual characteristics, indicating that one-size-fits-all approaches may be suboptimal. Future work could explore personalized or adapted interventions that tailor content and delivery format based on user demographics.}

\begin{spacing}{0.9}
\setlength{\belowcaptionskip}{10pt}
\begin{scriptsize}
\begin{longtable}{lccccc}
\hline
 & \textsc{Textual} & \textsc{Visual} & \textsc{Gamified} & \textsc{Feedback} & \textsc{Knowledge} \\
\hline
\endfirsthead

\multicolumn{6}{c}{\tablename\ \thetable{} -- continued from previous page} \\
\hline
& \textsc{Textual} & \textsc{Visual} & \textsc{Gamified} & \textsc{Feedback} & \textsc{Knowledge} \\
\hline
\endhead

\hline \multicolumn{6}{r}{Continued on next page} \\
\endfoot

\hline
\caption{\textbf{Regression results for discernment of {deepfake images} while controlling for sociodemographics and their interactions with the interventions.} Parentheses report standard errors. Variance inflation factor (VIF) values for the predictors in the models did not exceed 10, indicating that multicollinearity is not a major concern \cite{Obrien.2007} and that intervention effects remain reliable. Significance levels: $^*p<0.05$, $^{**}p<0.01$, $^{***}p<0.001$.} \\
\label{tab:regression_demo_interactions}
\endlastfoot

Intercept & 66.622$^{***}$ & 66.622$^{***}$ & 66.622$^{***}$ & 66.622$^{***}$ & 66.622$^{***}$ \\
 & (9.247) & (8.695) & (9.029) & (9.025) & (8.914) \\
Intervention & 11.926 & 13.528 & --19.719 & --6.676 & 5.287 \\
 & (13.374) & (12.110) & (13.524) & (12.193) & (12.402) \\
\emph{Gender} & --4.495 & --4.495 & --4.495 & --4.495 & --4.495 \\
 & (3.739) & (3.516) & (3.651) & (3.650) & (3.605) \\
Intervention $\times$ \emph{Gender} & 11.982$^{*}$ & 0.403 & 4.211 & 7.775 & 1.644 \\
 & (5.287) & (4.939) & (5.113) & (5.110) & (5.033) \\
\emph{Age} & --0.224 & --0.224 & --0.224 & --0.224 & --0.224 \\
 & (0.132) & (0.124) & (0.129) & (0.129) & (0.127) \\
Intervention $\times$ \emph{Age} & 0.222 & 0.095 & 0.592$^{**}$ & 0.149 & 0.198 \\
 & (0.210) & (0.192) & (0.193) & (0.183) & (0.175) \\
\emph{Ethnicity} & --3.141 & --3.141 & --3.141 & --3.141 & --3.141 \\
 & (1.812) & (1.704) & (1.769) & (1.768) & (1.747) \\
Intervention $\times$ \emph{Ethnicity} & 4.735 & 1.802 & 3.219 & 5.785$^{*}$ & 5.288$^{*}$ \\
 & (2.519) & (2.234) & (2.455) & (2.378) & (2.454) \\
\emph{Education} & 4.415 & 4.415 & 4.415 & 4.415 & 4.415 \\
 & (3.017) & (2.837) & (2.946) & (2.945) & (2.909) \\
Intervention $\times$ \emph{Education} & --11.181$^{*}$ & --6.000 & --5.093 & --8.258$^{*}$ & --4.681 \\
 & (4.391) & (4.176) & (4.327) & (4.138) & (4.452) \\
\emph{Religion} & 8.021 & 8.021$^{*}$ & 8.021$^{*}$ & 8.021$^{*}$ & 8.021$^{*}$ \\
 & (4.145) & (3.897) & (4.047) & (4.045) & (3.995) \\
Intervention $\times$ \emph{Religion} & --6.210 & --3.574 & 0.845 & --2.596 & --1.554 \\
 & (5.870) & (5.375) & (5.785) & (5.610) & (5.511) \\
\emph{PoliticalOrientation} & 1.217 & 1.217 & 1.217 & 1.217 & 1.217 \\
 & (1.089) & (1.024) & (1.063) & (1.062) & (1.049) \\
Intervention $\times$ \emph{PoliticalOrientation} & --2.249 & --0.844 & --1.296 & --0.499 & --2.386 \\
 & (1.573) & (1.393) & (1.496) & (1.481) & (1.452) \\
\emph{Income} & --4.103 & --4.103 & --4.103 & --4.103 & --4.103 \\
 & (2.662) & (2.503) & (2.599) & (2.598) & (2.566) \\
Intervention $\times$ \emph{Income} & 1.327 & 4.472 & 5.008 & 3.124 & 0.734 \\
 & (3.842) & (3.599) & (3.728) & (3.631) & (3.799) \\
 \midrule
$R^2$ & 0.107 & 0.138 & 0.086 & 0.060 & 0.082 \\
Adj. $R^2$ & 0.067 & 0.098 & 0.043 & 0.018 & 0.039 \\
\midrule
$N$ & 345 & 338 & 339 & 348 & 340 \\
\end{longtable}
\end{scriptsize}
\end{spacing}

\newpage
\subsection{Regression analysis controlling for self-reported digital literacy of participants}

We also control for the digital literacy of participants in our robustness checks. For this, we collected the participants' knowledge about {deepfake images} and their experience in detecting {deepfake images}, using online search engines, using reverse image search, and using AI image generators (see Table~\ref{tab:independent_variables} {in the Appendix} for details). All covariates were collected on a 7-point Likert scale.

Table~\ref{tab:regression_media_literacy} {in the Appendix} shows the results of the OLS regression model as defined in Equation~\ref{eq:covariates} where we control for digital literacy. Again, we find that the estimated coefficients for the interventions remain robust. We also find that previous experience with search engines (covariate \emph{ExpSearchEngine}) has a significant direct effect on people's {deepfake image discernment} abilities. This suggests that people who have more experience with search engines tend to do better at discerning {deepfake images}.

\begin{spacing}{0.9}
\setlength{\belowcaptionskip}{10pt}
\begin{scriptsize}
\begin{longtable}{lcccccc|c}
\hline
 & \textsc{Overall} & \textsc{Textual} & \textsc{Visual} & \textsc{Gamified} & \textsc{Feedback} & \textsc{Knowledge} & \textsc{Control} \\
\hline
\endfirsthead

\multicolumn{8}{c}{\tablename\ \thetable{} -- continued from previous page} \\
\hline
 & \textsc{Overall} & \textsc{Textual} & \textsc{Visual} & \textsc{Gamified} & \textsc{Feedback} & \textsc{Knowledge} & \textsc{Control} \\
\hline
\endhead

\hline \multicolumn{8}{r}{Continued on next page} \\
\endfoot

\hline
\caption{\textbf{Regression results for discernment of {deepfake images} while controlling for digital literacy.}  Parentheses report standard errors. Variance inflation factor (VIF) values for the predictors in all models exceeded 10 for \emph{KnowledgeDeepfakes}, \emph{ExpDetecting}, and \emph{ExpSearchEngine}, which means that we can only interpret the coefficients but not the standard errors or significance levels \cite{Obrien.2007}. Significance levels:  $^*p<0.05$, $^{**}p<0.01$, $^{***}p<0.001$.} \\
\label{tab:regression_media_literacy}
\endlastfoot

Intercept & 50.600$^{***}$ & 44.784$^{***}$ & 48.190$^{***}$ & 63.109$^{***}$ & 54.784$^{***}$ & 54.955$^{***}$ & 56.434$^{***}$ \\
 & (3.977) & (6.571) & (6.122) & (6.693) & (6.373) & (6.484) & (8.867) \\
\emph{Textual} & 7.270$^{**}$ & 7.190$^{**}$ &  &  &  &  &  \\
 & (2.313) & (2.431) &  &  &  &  &  \\
\emph{Visual} & 12.701$^{***}$ &  & 12.982$^{***}$ &  &  &  &  \\
 & (2.340) &  & (2.311) &  &  &  &  \\
\emph{Gamified} & 4.335 &  &  & 4.351 &  &  &  \\
 & (2.326) &  &  & (2.442) &  &  &  \\
\emph{Feedback} & --1.438 &  &  &  & --1.180 &  &  \\
 & (2.334) &  &  &  & (2.389) &  &  \\
\emph{Knowledge} & 2.854 &  &  &  &  & 3.046 &  \\
 & (2.326) &  &  &  &  & (2.389) &  \\
\emph{KnowledgeDeepfakes} & --0.496 & --0.168 & 0.180 & --2.088 & --2.836 & --0.981 & --2.122 \\
 & (0.872) & (1.664) & (1.597) & (1.475) & (1.483) & (1.537) & (2.246) \\
\emph{ExpDetecting} & 0.655 & 0.479 & --0.322 & 1.193 & 1.095 & --0.291 & --0.067 \\
 & (0.794) & (1.521) & (1.464) & (1.339) & (1.431) & (1.389) & (2.120) \\
\emph{ExpSearchEngine} & 2.188$^{**}$ & 3.030$^{*}$ & 1.565 & 0.323 & 1.422 & 1.642 & 0.896 \\
 & (0.740) & (1.289) & (1.243) & (1.261) & (1.250) & (1.293) & (1.769) \\
\emph{ExpImageSearch} & 0.024 & 0.750 & 1.077 & --0.448 & 1.086 & 0.026 & 0.978 \\
 & (0.476) & (0.857) & (0.808) & (0.841) & (0.844) & (0.818) & (1.193) \\
\emph{ExpGenAI} & --0.305 & --0.864 & 0.751 & 0.611 & 0.823 & 0.872 & 1.793 \\
 & (0.475) & (0.910) & (0.816) & (0.927) & (0.851) & (0.860) & (1.355) \\
 \midrule
$R^2$ & 0.053 & 0.047 & 0.104 & 0.015 & 0.020 & 0.013 & 0.024 \\
Adj. $R^2$ & 0.044 & 0.031 & 0.089 & --0.001 & 0.004 & --0.004 & --0.003 \\
\midrule
$N$ & 1112 & 375 & 370 & 371 & 372 & 372 & 187 \\
\end{longtable}
\end{scriptsize}
\end{spacing}

\newpage
\subsection{Regression analyzing the varying effectiveness of interventions across self-reported digital literacy}
We also control for interactions between the digital literacy covariates and our interventions with an OLS regression model as defined in Equation~\ref{eq:covariates_interactions}. Table~\ref{tab:regression_media_literacy_interactions} shows the results of models \textsc{Textual}, \textsc{Visual}, \textsc{Gamified}, \textsc{Feedback}, \textsc{Knowledge}, where we only include participants from the respective intervention condition and the control condition. We find that the estimated coefficients for the interventions remain robust. Moreover, we find that the  \emph{Textual} intervention were less effective for participants who are more experienced in AI image generation. This might be the explained by the fact that participants with experience in AI image generation already know about the typical errors found in {deepfake images}, which may have led to a ceiling effect, limiting their potential for further improvement. Another possibility is that they were overconfident in their expertise which might have lead to reduced attention to the intervention content.

\begin{spacing}{0.9}
\setlength{\belowcaptionskip}{10pt}
\begin{scriptsize}
\begin{longtable}{lccccc}
\hline
 & \textsc{Textual} & \textsc{Visual} & \textsc{Gamified} & \textsc{Feedback} & \textsc{Knowledge} \\
\hline
\endfirsthead

\multicolumn{6}{c}{\tablename\ \thetable{} -- continued from previous page} \\
\hline
  & \textsc{Textual} & \textsc{Visual} & \textsc{Gamified} & \textsc{Feedback} & \textsc{Knowledge} \\
\hline
\endhead

\hline \multicolumn{6}{r}{Continued on next page} \\
\endfoot

\hline
\caption{\textbf{Regression results for discernment of {deepfake images} while controlling for self-reported digital literacy and the interactions with the interventions.} Parentheses report standard errors.  Variance inflation factor (VIF) values for the predictors in all models exceeded 10 for \emph{KnowledgeDeepfakes}, \emph{ExpDetecting}, and \emph{ExpSearchEngine}, which means that we can only interpret the coefficients but not the standard errors or significance levels \cite{Obrien.2007}. Significance levels: $^*p<0.05$, $^{**}p<0.01$, $^{***}p<0.001$.} \\
\label{tab:regression_media_literacy_interactions}
\endlastfoot

Intercept & 56.434$^{***}$ & 56.434$^{***}$ & 56.434$^{***}$ & 56.434$^{***}$ & 56.434$^{***}$ \\
& (8.673) & (8.158) & (8.730) & (8.599) & (8.571) \\
Intervention & --19.022 & --3.892 & 18.568 & --5.560 & --1.501 \\
 & (12.925) & (12.272) & (13.355) & (12.876) & (13.119) \\
\emph{KnowledgeDeepfakes} & --2.122 & --2.122 & --2.122 & --2.122 & --2.122 \\
 & (2.197) & (2.067) & (2.211) & (2.178) & (2.171) \\
Intervention $\times$ \emph{KnowledgeDeepfakes} & 4.324 & 5.327 & 0.693 & --1.305 & 2.423 \\
 & (3.326) & (3.280) & (2.984) & (3.019) & (3.081) \\
\emph{ExpDetecting} & --0.067 & --0.067 & --0.067 & --0.067 & --0.067 \\
 & (2.074) & (1.951) & (2.087) & (2.056) & (2.050) \\
Intervention $\times$ \emph{ExpDetecting} & 0.101 & --1.537 & 1.737 & 2.089 & --0.741 \\
 & (3.046) & (3.005) & (2.735) & (2.923) & (2.821) \\
\emph{ExpSearchEngine} & 0.896 & 0.896 & 0.896 & 0.896 & 0.896 \\
 & (1.730) & (1.627) & (1.741) & (1.715) & (1.710) \\
Intervention $\times$ \emph{ExpSearchEngine} & 4.887 & 1.624 & --1.332 & 1.179 & 1.636 \\
 & (2.562) & (2.526) & (2.530) & (2.525) & (2.624) \\
\emph{ExpImageSearch} & 0.978 & 0.978 & 0.978 & 0.978 & 0.978 \\
 & (1.167) & (1.098) & (1.175) & (1.157) & (1.153) \\
Intervention $\times$ \emph{ExpImageSearch} & --0.376 & --0.112 & --2.917 & 0.146 & --1.798 \\
 & (1.696) & (1.634) & (1.678) & (1.717) & (1.657) \\
\emph{ExpGenAI} & 1.793 & 1.793 & 1.793 & 1.793 & 1.793 \\
 & (1.325) & (1.246) & (1.334) & (1.314) & (1.309) \\
Intervention $\times$ \emph{ExpGenAI} & --4.926$^{**}$ & --1.796 & --2.205 & --1.644 & --1.626 \\
 & (1.811) & (1.654) & (1.867) & (1.732) & (1.759) \\
 \midrule
$R^2$ & 0.084 & 0.117 & 0.036 & 0.024 & 0.024 \\
Adj. $R^2$ & 0.056 & 0.090 & 0.007 & --0.005 & --0.006 \\
\midrule
$N$ & 375 & 370 & 371 & 372 & 372 \\
\end{longtable}
\end{scriptsize}
\end{spacing}

\newpage
\subsection{Regression analysis controlling for social media use of participants}

Further, we control for the social media use of the participants when estimating the effects of our interventions on {deepfake image discernment}. For this, we collected the number of platforms that participants use, what type of content they share, and how much time they spend on social media daily (see Table~\ref{tab:independent_variables} {in the Appendix} for details). 

Table~\ref{tab:regression_social_media} reports the result for the OLS regression model as defined in Equation~\ref{eq:covariates}. Again, we find robust results for the estimated coefficients. We also find indications that people who tend to share more content on social media and people who tend to spend more time online are less good at discerning {deepfake images}.

\begin{spacing}{0.9}
\setlength{\belowcaptionskip}{10pt}
\begin{scriptsize}
\begin{longtable}{lcccccc|c}
\hline
 & \textsc{Overall} & \textsc{Textual} & \textsc{Visual} & \textsc{Gamified} & \textsc{Feedback} & \textsc{Knowledge} & \textsc{Control} \\
\hline
\endfirsthead

\multicolumn{8}{c}{\tablename\ \thetable{} -- continued from previous page} \\
\hline
 & \textsc{Overall} & \textsc{Textual} & \textsc{Visual} & \textsc{Gamified} & \textsc{Feedback} & \textsc{Knowledge} & \textsc{Control} \\
\hline
\endhead

\hline \multicolumn{8}{r}{Continued on next page} \\
\endfoot

\hline
\caption{\textbf{Regression results for discernment of {deepfake images} while controlling for participants' social media use.} Parentheses report standard errors. Variance inflation factor (VIF) values for the predictors in the models did not exceed 10, indicating that multicollinearity is not a major concern \cite{Obrien.2007} and that intervention effects remain
reliable. Significance levels:  $^*p<0.05$, $^{**}p<0.01$, $^{***}p<0.001$.} \\
\label{tab:regression_social_media}
\endlastfoot

Intercept & 64.708$^{***}$ & 63.507$^{***}$ & 62.633$^{***}$ & 63.973$^{***}$ & 65.825$^{***}$ & 62.768$^{***}$ & 62.326$^{***}$ \\
 & (2.253) & (3.310) & (3.164) & (3.227) & (3.181) & (3.215) & (4.393) \\
\emph{Textual} & 7.702$^{***}$ & 7.761$^{**}$ &  &  &  &  &  \\
 & (2.299) & (2.414) &  &  &  &  &  \\
\emph{Visual} & 13.056$^{***}$ &  & 13.136$^{***}$ &  &  &  &  \\
 & (2.315) &  & (2.285) &  &  &  &  \\
\emph{Gamified} & 4.650$^{*}$ &  &  & 4.537 &  &  &  \\
 & (2.311) &  &  & (2.417) &  &  &  \\
\emph{Feedback} & --1.391 &  &  &  & --1.443 &  &  \\
 & (2.310) &  &  &  & (2.365) &  &  \\
\emph{Knowledge} & 2.939 &  &  &  &  & 2.902 &  \\
 & (2.308) &  &  &  &  & (2.356) &  \\
\emph{PlatformCount} & 0.521 & 1.296 & 0.763 & 0.860 & 0.759 & 1.447 & 1.690 \\
 & (0.465) & (0.811) & (0.820) & (0.828) & (0.843) & (0.845) & (1.219) \\
\emph{SharingCount} & --1.493$^{**}$ & --1.992$^{*}$ & --0.986 & --2.132$^{*}$ & --2.106$^{*}$ & --1.579 & --2.242 \\
 & (0.476) & (0.861) & (0.820) & (0.900) & (0.863) & (0.839) & (1.279) \\
\emph{TimeOnline} & --0.496$^{**}$ & --0.603$^{*}$ & --0.520 & --0.119 & --0.639 & --0.885 & --0.468 \\
 & (0.170) & (0.296) & (0.361) & (0.319) & (0.367) & (0.580) & (0.792) \\
 \midrule
$R^2$ & 0.061 & 0.052 & 0.091 & 0.025 & 0.030 & 0.023 & 0.021 \\
Adj. $R^2$ & 0.054 & 0.041 & 0.081 & 0.014 & 0.019 & 0.013 & 0.005 \\
\midrule
$N$ & 1112 & 375 & 370 & 371 & 372 & 372 & 187 \\
\end{longtable}
\end{scriptsize}
\end{spacing}

\newpage
\subsection{Regression analyzing the varying effectiveness of interventions across participants social media use}

To account for possible interaction effects, we run OLS regression models as defined in Equation~\ref{eq:covariates_interactions} with social media covariates. Tables~\ref{tab:regression_social_media_interactions} show the results. Here, the models \textsc{Textual}, \textsc{Visual}, \textsc{Gamified}, \textsc{Feedback}, \textsc{Knowledge} only include individuals from the respective intervention condition and the control condition to isolate the effect of our intervention on the outcome. We find that the estimated coefficients for the interventions remain robust.

\begin{spacing}{0.9}
\setlength{\belowcaptionskip}{10pt}
\begin{scriptsize}
\begin{longtable}{lccccc}
\hline
 & \textsc{Textual} & \textsc{Visual} & \textsc{Gamified} & \textsc{Feedback} & \textsc{Knowledge} \\
\hline
\endfirsthead

\multicolumn{6}{c}{\tablename\ \thetable{} -- continued from previous page} \\
\hline
 & Textual & Visual & Gamified & Feedback & Knowledge \\
\hline
\endhead

\hline \multicolumn{6}{r}{Continued on next page} \\
\endfoot

\hline
\caption{\textbf{Regression results for discernment of {deepfake images} while controlling for participants' social media use and the interactions with the interventions.} Parentheses report standard errors. Variance inflation factor (VIF) values for the predictors in the models did not exceed 10, indicating that multicollinearity is not a major concern \cite{Obrien.2007} and that intervention effects remain reliable.  Significance levels:  $^*p<0.05$, $^{**}p<0.01$, $^{***}p<0.001$.} \\
\label{tab:regression_social_media_interactions}
\endlastfoot

Intercept & 62.326$^{***}$ & 62.326$^{***}$ & 62.326$^{***}$ & 62.326$^{***}$ & 62.326$^{***}$ \\
 & (4.362) & (4.081) & (4.337) & (4.230) & (4.234) \\
Intervention & 10.036 & 13.915$^{*}$ & 8.031 & 5.703 & 3.952 \\
 & (6.257) & (5.909) & (6.112) & (5.974) & (5.970) \\
\emph{PlatformCount} & 1.690 & 1.690 & 1.690 & 1.690 & 1.690 \\
 & (1.211) & (1.133) & (1.204) & (1.174) & (1.175) \\
Intervention $\times$ \emph{PlatformCount} & --0.783 & --1.858 & --1.529 & --2.033 & --0.370 \\
 & (1.650) & (1.653) & (1.674) & (1.697) & (1.700) \\
\emph{SharingCount} & --2.242 & --2.242 & --2.242 & --2.242 & --2.242 \\
 & (1.270) & (1.189) & (1.263) & (1.232) & (1.233) \\
Intervention $\times$ \emph{SharingCount} & 0.400 & 2.384 & 0.302 & 0.301 & 1.269 \\
 & (1.740) & (1.647) & (1.810) & (1.730) & (1.687) \\
\emph{TimeOnline} & --0.468 & --0.468 & --0.468 & --0.468 & --0.468 \\
 & (0.787) & (0.736) & (0.782) & (0.763) & (0.764) \\
Intervention $\times$ \emph{TimeOnline} & --0.161 & --0.051 & 0.410 & --0.190 & --0.976 \\
 & (0.850) & (0.845) & (0.857) & (0.872) & (1.180) \\
 \midrule
$R^2$ & 0.052 & 0.097 & 0.027 & 0.035 & 0.026 \\
Adj. $R^2$ & 0.034 & 0.079 & 0.008 & 0.016 & 0.008 \\
\midrule
$N$ & 375 & 370 & 371 & 372 & 372 \\
\end{longtable}
\end{scriptsize}
\end{spacing}

\newpage
\subsection{Differences in sociodemographics and discernment for attrition group}

The response rate for our follow-up was 72.93\% of our participants. Here, we check whether there are significant differences between the participants that returned for the follow-up (respondents) and those who did not (dropouts) using Mann-Whitney $U$ tests. We find that both groups show similar discernment abilities for real images and {deepfake images}. Also, the sociodemographics are largely similar, except for age, where participants who dropped out were slightly younger on average than participants who returned for the follow-up (see Table~\ref{tab:attrition_differences}).

\begin{spacing}{.9}
\begin{table}[H]
\centering
\setlength{\belowcaptionskip}{10pt}
\begin{scriptsize}
\begin{tabular}{lccrr}
\hline
\textbf{Variable} & \textbf{Mean value (respondents)} & \textbf{Mean value (dropouts)} & \textbf{$U$-statistic} & \textbf{$p$-value} \\
\hline
{Gender} & 0.54 & 0.49 & 123252.00 & 0.133 \\
{Age} & 43.11 & 38.54 & 105464.00 & $<0.000$\\
{Ethnicity} & 0.57 & 0.56 & 135005.00 & 0.603 \\
{Education} & 1.08 & 1.13 & 136166.00 & 0.214 \\
{Religion} & 0.43 & 0.37 & 115084.00 & 0.081 \\
{PoliticalOrientation} & 3.95 & 3.90 & 130480.00 & 0.618 \\
{Income} & 1.31 & 1.30 & 128194.00 & 0.954 \\
\hline
{Real image discernment} & 83.26\% & 82.54\% & 152803.00 & 0.514 \\
{{Deepfake image discernment}} & 66.66\% & 65.33\% & 152803.00 & 0.288 \\
\hline
\end{tabular}
\caption{\textbf{Differences of sociodemographic and discernment abilities for participants who completed the follow-up (respondents) and those who did not (dropouts) using Mann-Whitney $U$ tests.} }
\label{tab:attrition_differences}
\end{scriptsize}
\end{table}
\end{spacing}

\newpage
\subsection{Regression with image set covariates}
\label{supp:robustness_order_effects}

To test for possible effects of the images that were used for the image discernment task, we conducted a robustness check which includes the image set as a control (as defined in Equation~\ref{eq:order_effect}). Table~\ref{tab:order_effects_deepfakes} shows the results for {deepfake image discernment}. We find that, while participants were better at discerning image set B, there were no interaction effects with our interventions. This shows that the effectiveness of our interventions is not influenced by the image set. This suggests that our interventions are effective, regardless of which images participants see. This implies that our findings can be generalized to a large variety of {deepfake images}.

\begin{spacing}{.9}
\begin{table}[H]
\centering
\setlength{\belowcaptionskip}{10pt}
\begin{scriptsize}
\begin{tabular}{l rc rrr}
\toprule
 & \textbf{Coef.} & \textbf{s.e.} & \textbf{Lower CI} & \textbf{Upper CI} & \textbf{$p$-value} \\
\midrule
{Intercept} & 57.65 & (2.260) & 53.22 & 62.09 & $< 0.001$ \\
\emph{Textual} & 4.74 & (3.188) & --1.51 & 10.99 & 0.137 \\
\emph{Visual} & 13.94 & (3.214) & 7.63 & 20.25 & $< 0.001$ \\
\emph{Gamified} & 0.93 & (3.196) & --5.33 & 7.20 & 0.770 \\
\emph{Feedback} & --5.09 & (3.196) & --11.36 & 1.18 & 0.112 \\
\emph{Knowledge} & 3.25 & (3.188) & --2.99 & 9.51 & 0.307 \\
Set & 7.30 & (3.188) & 1.04 & 13.55 & $< 0.05$ \\ 
\emph{Textual} $\times$ Set & 5.50 & (4.502) & --3.33 & 14.33 & 0.222 \\
\emph{Visual} $\times$ Set & --1.92 & (4.533) & --10.81 & 6.97 & 0.672 \\
\emph{Gamified} $\times$ Set & 7.06 & (4.527) & --1.81 & 15.94 & 0.119 \\
\emph{Feedback} $\times$ Set & 7.64 & (4.521) & --1.22 & 16.51 & 0.091 \\
\emph{Knowledge} $\times$ Set & --0.05 & (4.521) & --8.92 & 8.81 & 0.990 \\

\midrule
\multicolumn{5}{l}{Observations: 1112} \\ 
\bottomrule
\end{tabular}
\caption{\textbf{OLS regression results for intervention effect on discernment of {deepfake images} while controlling for the image set.} Reported are 95\% confidence intervals (CIs). Standard errors (s.e.) are reported in parentheses. }
\label{tab:order_effects_deepfakes}
\end{scriptsize}
\end{table}
\end{spacing}

\noindent
Similarly, for discernment of real images, we find that participants were better at discerning image set B, but there was no interaction effect with our interventions (see Table~\ref{tab:order_effects_real_images}). Hence, the effectiveness of our intervention was independent of the image set that participants were exposed to in the image discernment task. 

\begin{spacing}{.9}
\begin{table}[H]
\centering
\setlength{\belowcaptionskip}{10pt}
\begin{scriptsize}
\begin{tabular}{l rc rr r}
\toprule
 & \textbf{Coef.} & \textbf{s.e.} & \textbf{Lower CI} & \textbf{Upper CI} & \textbf{$p$-value} \\
\midrule
{Intercept} & 77.74 & (2.127) & 73.56 & 81.91 & $< 0.001$ \\
\emph{Textual} & 2.22 & (2.999) & --3.66 & 8.10 & 0.459 \\
\emph{Visual} & --1.77 & (3.024) & --7.70 & 4.16 & 0.558 \\
\emph{Gamified} & --2.84 & (3.007) & --8.75 & 3.05 & 0.344 \\
\emph{Feedback} & 1.00 & (3.007) & --4.89 & 6.90 & 0.739 \\
\emph{Knowledge} & 1.03 & (2.999) & --4.85 & 6.92 & 0.730 \\
Set & 12.38 & (2.999) & 6.49 & 18.26 & $< 0.001$ \\
\emph{Textual} $\times$ Set & --6.33 & (4.236) & --14.64 & 1.97 & 0.135 \\
\emph{Visual} $\times$ Set & --4.38 & (4.265) & --12.75 & 3.98 & 0.304 \\
\emph{Gamified} $\times$ Set & 0.69 & (4.259) & --7.66 & 9.04 & 0.871 \\
\emph{Feedback} $\times$ Set & --1.29 & (4.253) & --9.63 & 7.05 & 0.762 \\
\emph{Knowledge} $\times$ Set & --1.76 & (4.253) & --10.10 & 6.58 & 0.679 \\

\midrule
\multicolumn{5}{l}{Observations: 1112} \\ 
\bottomrule
\end{tabular}
\caption{\textbf{OLS regression results for intervention effect on discernment of real images while controlling for the image set.} Reported are 95\% confidence intervals (CIs). Standard errors (s.e.) are reported in parentheses.}
\label{tab:order_effects_real_images}
\end{scriptsize}
\end{table}
\end{spacing}

\newpage
\subsection{Regression for differences in discernment for viral and non-viral {deepfake images}}
\label{supp:robustness_viral_nonviral}

{Deepfake images} that went viral may be structurally different from those that did not. To analyze the heterogeneity of our effects across both types of {deepfake images}, we estimated  a linear mixed effects regression as defined in Equation~\ref{eq:mixed_effects}. Table~\ref{tab:mixed_effects_viral_deepfakes} shows that participants were 12.55\% better at detecting viral {deepfake images} than non-viral {deepfake images}. Moreover, our \emph{Textual} and \emph{Gamified} interventions were less effective for viral {deepfake images} compared to non-viral {deepfake images}.

\begin{spacing}{.9}
\begin{table}[H]
\centering
\setlength{\belowcaptionskip}{10pt}
\begin{scriptsize}
\begin{tabular}{lrcrrr}
\toprule
 & \textbf{Coef.} & \textbf{s.e.} & \textbf{Lower CI} & \textbf{Upper CI} & \textbf{$p$-value} \\
\midrule
{Intercept} & 55.00 & (1.957) & 51.17 & 58.84 & $< 0.001$ \\
\emph{Textual} & 10.87 & (2.764) & 5.46 & 16.29 & $< 0.001$ \\
\emph{Visual} & 15.47 & (2.783) & 10.02 & 20.92 & $< 0.001$ \\
\emph{Gamified} & 8.02 & (2.779) & 2.57 & 13.47 & $< 0.01$ \\ 
\emph{Feedback} & --1.91 & (2.775) & --7.35 & 3.52 & 0.489 \\
\emph{Knowledge} & 2.30 & (2.775) & --3.13 & 7.74 & 0.406 \\
Source & 12.55 & (2.126) & 8.39 & 16.72 & $< 0.001$ \\
\emph{Textual} $\times$ Source & --6.19 & (3.003) & --12.07 & --0.30 & $< 0.05$ \\ 
\emph{Visual} $\times$ Source & --4.88 & (3.023) & --10.81 & 1.03 & 0.106 \\
\emph{Gamified} $\times$ Source & --7.25 & (3.019) & --13.16 & --1.33 & $< 0.05$ \\ 
\emph{Feedback} $\times$ Source & 1.56 & (3.015) & --4.34 & 7.47 & 0.603 \\
\emph{Knowledge} $\times$ Source & 2.10 & (3.015) & --3.81 & 8.00 & 0.486 \\
\midrule
\multicolumn{5}{l}{Observations: 2224} \\ 
\bottomrule
\end{tabular}
\caption{\textbf{Linear mixed effects regression results for intervention effect on discernment of viral vs. non-viral {deepfake images}.} Reported are 95\% confidence intervals (CIs). Standard errors (s.e.) are reported in parentheses. Standard errors are reported in parentheses.}
\label{tab:mixed_effects_viral_deepfakes}
\end{scriptsize}
\end{table}
\end{spacing}

\newpage
\subsection{Regression analysis for long-term effects}

To test the long-term effectiveness of our interventions, we estimated a linear mixed effects regression where we model {deepfake image discernment} of each participant at time points T1 and T2. The direct effects capture the effectiveness of our intervention at time point T1 (immediately after), while the interaction effects capture how each intervention's effectiveness changed over time, i.e., whether the effect differs in the follow-up. Table~\ref{tab:mixed_effects_deepfakes_over_time} shows that our main intervention effects remain robust immediately after the intervention, with the \emph{Textual} and \emph{Visual} intervention boosting participants significantly in their discernment. Over time, the effect of both the \emph{Textual} and \emph{Knowledge} interventions decreased, although not significantly, while the effects of the \emph{Visual} and \emph{Gamified} intervention decreased significantly over time. Only the effect of the \emph{Feedback} intervention increased over time, however, this cancels out with the negative direct effect. This confirms our results regarding Hypotheses H1 and H3 from Section~\ref{sec:results_discerning_deepfakes} in that both the \emph{Textual} and \emph{Visual} interventions were able to boost discernment immediately after the intervention but not long-term.

\begin{spacing}{.9}
\begin{table}[H]
\centering
\setlength{\belowcaptionskip}{10pt}
\begin{scriptsize}
\begin{tabular}{lrcrrr}
\toprule
 & \textbf{Coef.} & \textbf{s.e.} & \textbf{Lower CI} & \textbf{Upper CI} & \textbf{$p$-value} \\
\midrule
{Intercept} & 61.40 & (1.95) & 57.58 & 65.21 & $< 0.001$ \\
\emph{Textual} & 7.53 & (2.71) & 2.21 & 12.84 & 0.006 \\
\emph{Visual} & 12.17 & (2.78) & 6.72 & 17.62 & $< 0.001$ \\
\emph{Gamified} & 3.50 & (2.78) & --1.94 & 8.94 & 0.207 \\
\emph{Feedback} & --1.62 & (2.72) & --6.95 & 3.72 & 0.553 \\
\emph{Knowledge} & 2.38 & (2.77) & --3.05 & 7.81 & 0.391 \\
Time Point & 2.98 & (2.19) & --1.31 & 7.27 & 0.173 \\
\emph{Textual} $\times$ Time Point & --3.45 & (3.05) & --9.42 & 2.53 & 0.258 \\
\emph{Visual} $\times$ Time Point & --6.89 & (3.13) & --13.02 & --0.76 & 0.028 \\
\emph{Gamified} $\times$ Time Point & --6.35 & (3.12) & --12.46 & --0.23 & 0.042 \\
\emph{Feedback} $\times$ Time Point & 1.14 & (3.06) & --4.86 & 7.13 & 0.710 \\
\emph{Knowledge} $\times$ Time Point & --2.49 & (3.11) & --8.59 & 3.61 & 0.424 \\
\midrule
\multicolumn{5}{l}{Observations: 2224} \\ 
\bottomrule
\end{tabular}
\caption{\textbf{Linear mixed effects regression results for intervention effect on discernment of {deepfake images} over time.} Reported are 95\% confidence intervals (CIs). Standard errors are reported in parentheses.}
\label{tab:mixed_effects_deepfakes_over_time}
\end{scriptsize}
\end{table}
\end{spacing}

\newpage
\section{Learning effect of image discernment task}
\label{supp:learning_effect_robustness}

To check for possible learning effects from the task itself, we recruited another round of $N=200$ participants (in addition to $N=1200$ participants from the main study), who first received an unrelated emotion discernment task at time point T1 and then the image discernment task only in the follow-up at time point T2. Here, half of the participants saw image set A for the discernment task, while the other half saw image set B. None of the participants received an intervention. We then compared the results from the discernment task of the new robustness condition with the results from the follow-up of the control conditions. For this, we use Mann-Whitney $U$ tests for difference in means for the images sets. This means that we compare the follow-up results of the robustness condition on image set A with the follow-up results of the control condition on image set A (the same procedure is applied to image set B). By doing so, we can see whether exposure to the task itself (=the {deepfake image discernment}) has had a significant effect on the performance of participants. 

We find no statistically significant difference between the results of the control condition (average discernment of real images $\mu=78.05\%$, average discernment of {deepfake images} $\mu=56.69\%$) and the robustness condition (average discernment of real images $\mu = 72.93\%$, average discernment of {deepfake images} $\mu=53.17\%$) that saw image discernment set A (Mann-Whitney $U = 2304.5$, $p=0.147$ for real images and Mann-Whitney $U = 2158.5$, $p=0.491$ for {deepfake images}) and the control condition (average discernment of real images $\mu=85.07\%$, average discernment of {deepfake images} $\mu=69.71\%$) and robustness condition (average discernment of real images $\mu=87.93\%$, average discernment of {deepfake images} $\mu=67.59\%$) that saw image discernment set B (Mann-Whitney $U = 1853.5$, $p=0.282$ for real images and Mann-Whitney $U = 2169.0$, $p=0.601$ for {deepfake images}). Hence, mere exposure to the task did not lead to a learning effect.

\newpage
\section{Images}
\label{sec:images}

\begin{longtable}{>{\raggedright}p{0.18\textwidth}p{0.47\textwidth}p{0.1\textwidth}p{0.1\textwidth}}
\toprule
    \textbf{Image} & \textbf{Prompt} & \textbf{Model} & \textbf{Date} \\
\midrule
\endfirsthead

\toprule
    \textbf{Image} & \textbf{Prompt} & \textbf{Model} & \textbf{Date} \\
\midrule
\endhead

\midrule
\endfoot

\midrule
\caption{Example image sources used in the interventions \emph{Visual}, \emph{Gamified}, and \emph{Feedback}.}\label{tab:example_images}\\
\endlastfoot

        \raisebox{-0.6\height}{\includegraphics[width=0.2\textwidth]{images/holding_hands_at_dinner.jpg}} & A photo of a couple holding hands at dinner in a restaurant & DALL-E 3  & 17.09.2024 \\
    \midrule
        \raisebox{-0.6\height}{\includegraphics[width=0.2\textwidth]{images/woman_on_street.jpg}} & A photo of a woman on a street & DALL-E 3 & 17.09.2024  \\
    \midrule
        \raisebox{-0.6\height}{\includegraphics[width=0.2\textwidth]{images/friends_drinking_coffee.jpg}} & A photo of two friends having coffee in Paris & DALL-E 3 & 17.09.2024 \\
    \midrule
        \raisebox{-0.6\height}{\includegraphics[width=0.2\textwidth]{images/man_brushing_teeth.jpg}} & A photo of a man brushing his teeth in his bathroom & DALL-E 3 & 17.09.2024\\
    \midrule
        \raisebox{-0.6\height}{\includegraphics[width=0.2\textwidth]{images/pope-wearing-balenciaga-jacket.jpg}} & \url{https://piktochart.com/blog/viral-ai-images/} & & 17.09.2024 \\ 
\end{longtable}

\newpage

\begin{longtable}{>{\raggedright}p{0.25\textwidth}p{0.47\textwidth}p{0.1\textwidth}}
\toprule
    \textbf{Image} & \textbf{Source} & \textbf{Date} \\
\midrule
\endfirsthead

\toprule
    \textbf{Image} & \textbf{Source} &  \textbf{Date} \\
\midrule
\endhead

\midrule
\endfoot

\midrule
\caption{Example real image sources used in the interventions \emph{Feedback}.}\label{tab:intervention_feedback}\\
\endlastfoot

        Girl posing in front of flowers & \url{https://www.flickr.com/photos/126108832@N06/48249053611/} & 17.09.2024 \\
    \midrule
        Police guarding the U.S. Capitol & \url{www.flickr.com/photos/schweitn/5459838053} & 17.09.2024  \\
    \midrule
        Portrait of a man on the street & \url{https://www.flickr.com/photos/benedictflett/27196027047/} & 17.09.2024 \\
    \midrule
        Portrait of a Mexican woman & \url{https://www.flickr.com/photos/radargeek/54019285321/} & 17.09.2024\\
    \midrule
        A man and a woman walking in the rain & \url{https://www.flickr.com/photos/radargeek/54031248344/} & 17.09.2024 \\ 
\end{longtable}

\newpage
\begin{longtable}{p{0.25\textwidth}p{0.55\textwidth}p{0.1\textwidth}p{0.1\linewidth}}
\toprule
    \textbf{Image} & \textbf{Source} & \textbf{Date} & \textbf{Image Set} \\
\midrule
\endfirsthead

\toprule
    \textbf{Image} & \textbf{Source} & \textbf{Date}  & \textbf{Image Set}\\
\midrule
\endhead

\midrule
\endfoot

\midrule
\caption{Real images used in the experiment.}\label{tab:real_images_main}\\
\endlastfoot

        Three boys walking on a beach & \url{https://www.flickr.com/photos/t\_riel/5675863222} & 16.09.2024 & A \\
    \midrule
        Police man taking of helmet during a protest & \url{https://www.flickr.com/photos/benedictflett/19672183508} & 17.09.2024 & A \\
    \midrule
        Three boys smiling into the camera & \url{https://www.flickr.com/photos/theopendoor/26187763682} & 16.09.2024 & A\\
    \midrule
         Man looking at a street & \url{https://www.flickr.com/photos/46924752@N03/4920054002} & 16.09.2024 & A\\
    \midrule
        Woman walking on street & \url{https://www.flickr.com/photos/148042613@N02/54000168419} & 16.09.2024 & A\\ 
    \midrule
        Cheerleader posing for camera & \url{https://www.flickr.com/photos/126108832@N06/48638756656} & 16.09.2024 & B \\
    \midrule
        Group gathering & \url{https://www.flickr.com/photos/17101115@N00/2497641135} & 16.09.2024 & B \\   
    \midrule
        Couple posing & \url{https://www.flickr.com/photos/radargeek/35769697725} & 16.09.2024 & B\\
    \midrule
        Woman giving speech & \url{https://www.flickr.com/photos/americanfarmschool/52470077293} & 16.09.2024 & B\\
    \midrule
        Women signing papers at a desk & \url{https://www.flickr.com/photos/kgs/180105927} & 17.09.2024 & B\\
\end{longtable}

\newpage
\begin{longtable}{>{\raggedright}p{0.18\textwidth}p{0.4\textwidth}p{0.17\textwidth}p{0.1\textwidth}p{0.1\textwidth}}
\toprule
    \textbf{Image} & \textbf{Prompt} & \textbf{Model} & \textbf{Date} & \textbf{Image Set}\\
\midrule
\endfirsthead

\toprule
    \textbf{Image} & \textbf{Prompt} & \textbf{Model} & \textbf{Date} & \textbf{Image Set} \\
\midrule
\endhead

\midrule
\endfoot

\midrule
\caption{Deepfake images that are not viral we used in the experiment. $^*$Image from the Center for Countering Digital Hate (CCDH) \cite{CenterforCounteringDigitalHate.2024}.}\label{tab:fake_images_main}\\
\endlastfoot

        \raisebox{-0.6\height}{\includegraphics[width=0.2\textwidth]{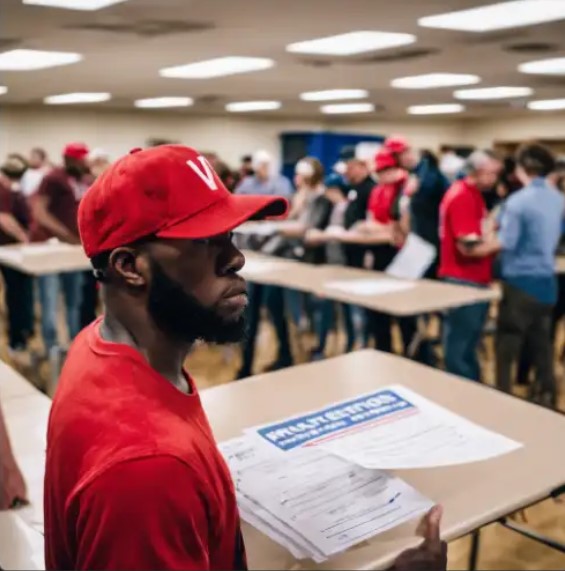}} & A photo of an angry protester in a red baseball cap inside a polling place, voting booths are visible in the background & DreamStudio$^*$ & 22.02.2024 & A\\
    \midrule
        \raisebox{-0.6\height}{\includegraphics[width=0.2\textwidth]{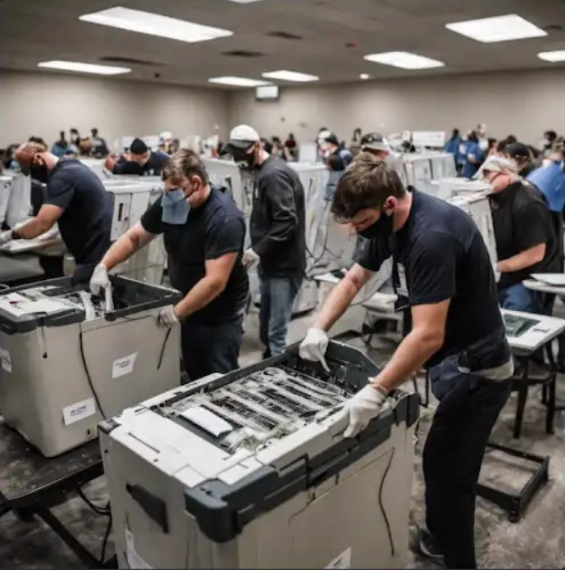}} & A photo of election workers damaging the machinery of voting machines & DreamStudio$^*$ & 06.02.2024 & A\\
    \midrule
        \raisebox{-0.6\height}{\includegraphics[width=0.2\textwidth]{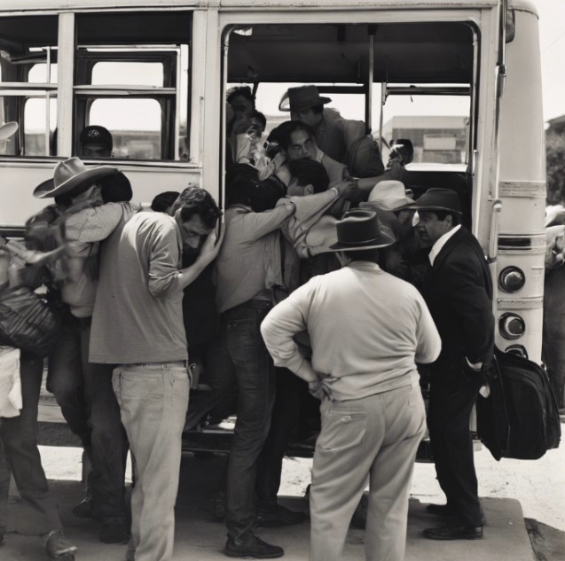}} & A photo of Mexican immigrants getting off of a bus outside a polling place & DreamStudio$^*$ & 06.02.2024 & A \\
    \midrule
        \raisebox{-0.6\height}{\includegraphics[width=0.2\textwidth]{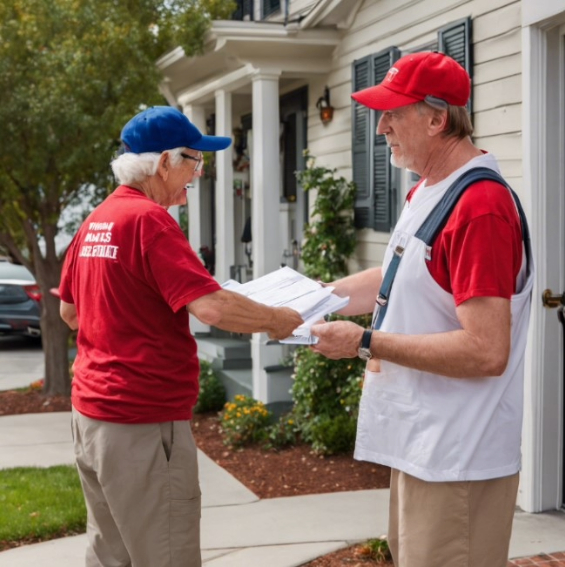}} & A photo of a white man going door to door in a red baseball cap picking up mail in ballots, talking to an old lady, ballot harvesting & DreamStudio$^*$ & 06.02.2024 & A \\
    \midrule
        \raisebox{-0.6\height}{\includegraphics[width=0.2\textwidth]{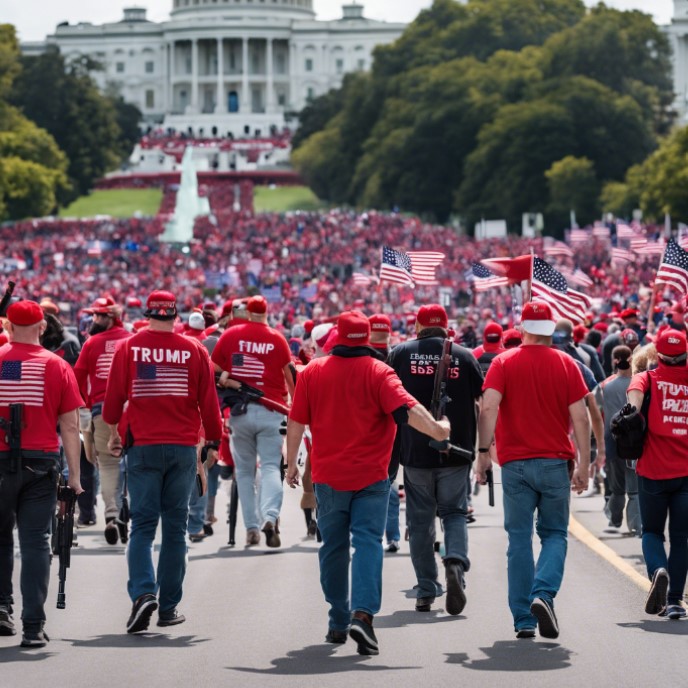}} & A photo of Trump supporters in red baseball hats, holding guns marching towards the United states capitol & DreamStudio$^*$ & 06.02.2024 & A \\
    \midrule
        \raisebox{-0.6\height}{\includegraphics[width=0.2\textwidth]{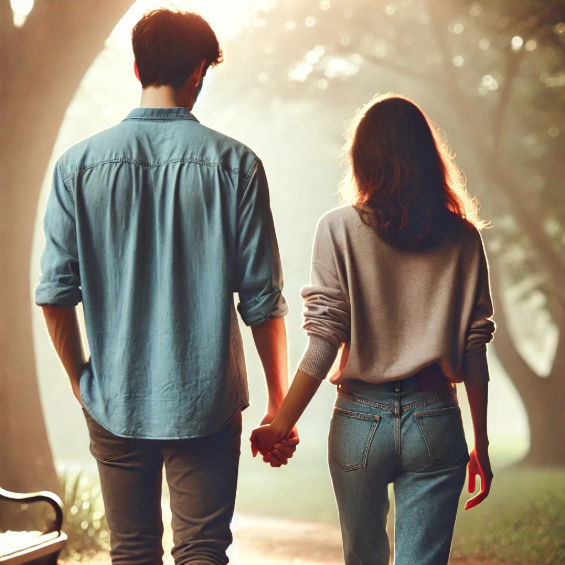}} & A couple holding hands & DALL-E 3 & 17.09.2024 & B\\
    \midrule
        \raisebox{-0.6\height}{\includegraphics[width=0.2\textwidth]{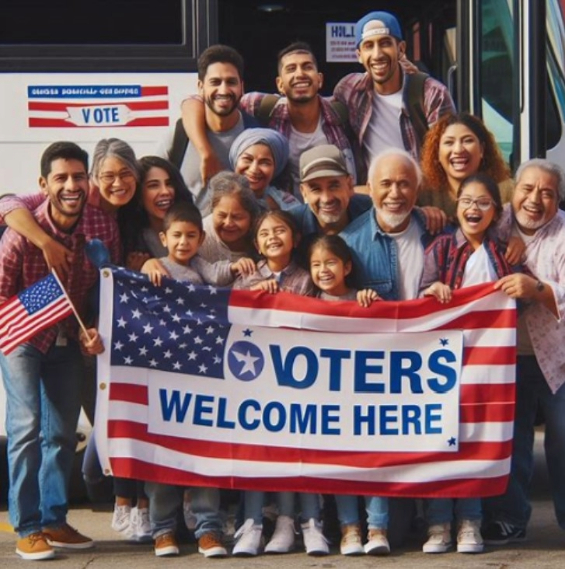}} & A photo of Mexican immigrants getting off of a bus outside a polling place & Image Creator (Bing)$^*$ & 06.02.2024 & B\\        
    \midrule
        \raisebox{-0.6\height}{\includegraphics[width=0.2\textwidth]{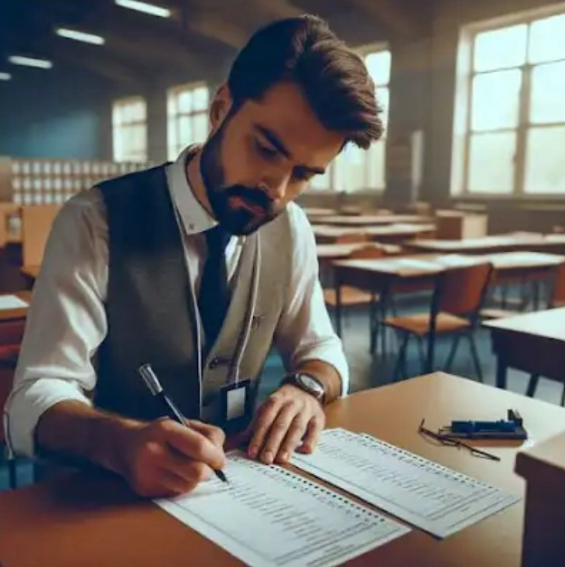}} & A photo of an election worker marking ballots with pen in a mundane looking office & Image Creator (Bing)$^*$ & 06.02.2024 & B\\
    \midrule
        \raisebox{-0.6\height}{\includegraphics[width=0.2\textwidth]{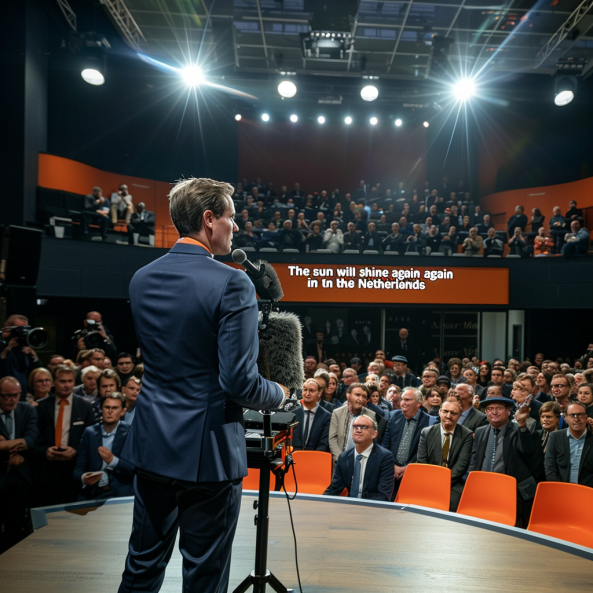}} & a politician in front of an audience says "The sun will shine again in the Netherlands" & Midjourney$^*$ & 06.02.2024 & B \\
    \midrule
        \raisebox{-0.6\height}{\includegraphics[width=0.2\textwidth]{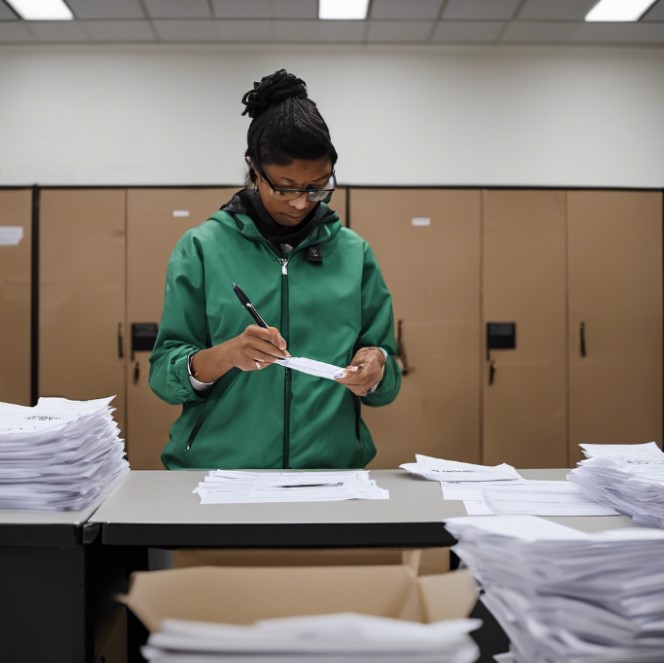}} & A photo of an election worker marking ballots with pen in a mundane looking office & DreamStudio$^*$ & 06.02.2024 & B\\
\end{longtable}

\newpage
\begin{longtable}{>{\raggedright}p{0.18\textwidth}p{0.5\textwidth}p{0.12\textwidth}p{0.1\textwidth}}
\toprule
    \textbf{Image} & \textbf{Source} & \textbf{Date}  & \textbf{Image Set}\\
\midrule
\endfirsthead

\toprule
    \textbf{Image} & \textbf{Source} & \textbf{Date}  & \textbf{Image Set}\\
\midrule
\endhead

\midrule
\endfoot

\midrule
\caption{Viral {deepfake images} used in the experiment.}\label{tab:viral_images_main}\\
\endlastfoot

        Black Lives Matter protest & Image with caption \say{This AI-generated fake photo, labeled by The Washington Post, can be found on Adobe Stock when users search for “BLM protests.” (Adobe)} on \url{https://www.washingtonpost.com/technology/2023/11/23/stock-photos-ai-images-controversy/} & 17.09.2024 & A \\
    \midrule
        Musk and Barra hand in hand & Image from \url{https://piktochart.com/blog/viral-ai-images/} & 17.09.2024 & A\\
    \midrule
        AI Model on field & Image under headline \say{AI Model with More Followers Than You and Me} from \url{https://piktochart.com/blog/viral-ai-images/} & 17.09.2024 & A \\
    \midrule
        Pentagon in (fake) flames & Image from \url{https://piktochart.com/blog/viral-ai-images/} & 17.09.2024 & A \\
    \midrule
        Refugees panicked in water & Image with caption \say{AI image by Michael Christopher Brown that depicts the perilous 90-mile journey from Havana to Florida that some Cubans attempt.} from \url{https://petapixel.com/2023/12/28/the-ai-images-that-shook-the-photography-world-in-2023/} & 17.09.2024 & A\\
    \midrule
        Two women hugging & Image under headline \say{AI Wins Prestigious Sony Award} from \url{https://piktochart.com/blog/viral-ai-images/} & 17.09.2024 & B \\
    \midrule
        Girl holding a teddy & Image with caption: \say{a girl holding his teddy bear with destructive civilian area during war time, sorrow scenery of war victims, idea for support children's right , especially Ukrainian, Generative Ai.} on \url{https://www.washingtonpost.com/technology/2023/11/23/stock-photos-ai-images-controversy/} & 17.09.2024 & B \\
    \midrule
        Girl taking selfie with a bear & Image from \url{https://twitter.com/heyBarsee/status/1641746873210814467} & 17.09.2024 & B\\
    \midrule
        Paris covered in trash & Image from \url{https://piktochart.com/blog/viral-ai-images/} & 17.09.2024 & B\\
    \midrule
        Trump being arrested & Image from \url{https://piktochart.com/blog/viral-ai-images/} & 17.09.2024 & B\\
\end{longtable}

\end{document}